\documentclass[fleqn,usenatbib]{mnras}

\usepackage{newtxtext,newtxmath}

\usepackage[T1]{fontenc}

\DeclareRobustCommand{\VAN}[3]{#2}
\let\VANthebibliography\thebibliography
\def\thebibliography{\DeclareRobustCommand{\VAN}[3]{##3}\VANthebibliography}

\usepackage{graphicx}	
\usepackage{amsmath}	
\usepackage{comment}
\usepackage{sansmath}

\usepackage[export]{adjustbox} 

\newcommand{\HI}{H\textsc{i}\ }
\newcommand{\hinospace}{\textrm{H\textsc{i}}}
\newcommand{\size}[2]{{\fontsize{#1}{0}\selectfont#2}}

\newcommand{\fastica}{F\size{7.5}{AST}ICA}
\newcommand{\Nfg}{$N_{\rm fg}$\,}
\newcommand{\secref}[1]{\hyperref[#1]{Section~\ref*{#1}}}
\newcommand{\appref}[1]{\hyperref[#1]{Appendix~\ref*{#1}}}

\title[Blind Foreground Subtraction Challenge]{SKAO \HI Intensity Mapping: Blind Foreground Subtraction Challenge}

\author[M. Spinelli, I.P. Carucci et al.]{
Marta Spinelli,$^{1,2,3,4}$\thanks{E-mail: marta.spinelli@inaf.it}
Isabella P. Carucci,$^{5,6,7}$\thanks{E-mail: isabellapaola.carucci@unito.it}
Steven Cunnington,$^{8}$
Stuart E. Harper,$^{9}$
Melis O. Irfan,$^{4,8}$
\newauthor
Jos\'e Fonseca,$^{8,4,10,11}$
Alkistis Pourtsidou,$^{8,4}$
Laura Wolz$^{9}$
\\
$^{1}$Institute for Particle Physics and Astrophysics,
ETH Z{\"u}rich, Wolfgang Pauli Strasse 27, 8093 Z{\"u}rich, Switzerland\\
$^{2}$INAF-Osservatorio Astronomico di Trieste, Via G.B. Tiepolo 11, 34143 Trieste, Italy\\
$^{3}$ IFPU - Institute for Fundamental Physics of the Universe, Via Beirut 2, 34014 Trieste, Italy\\
$^{4}$ Department of Physics and Astronomy, University of the Western Cape, Robert Sobukhwe Road, Bellville, 7535, South Africa\\
$^{5}$Dipartimento di Fisica, Universit\`a degli Studi di Torino, via P.\ Giuria 1, 10125 Torino, Italy\\
$^{6}$INFN -- Istituto Nazionale di Fisica Nucleare, Sezione di Torino, via P.\ Giuria 1, 10125 Torino, Italy\\
$^{7}$AIM, CEA, CNRS, Université Paris-Saclay, Université Paris Diderot, Sorbonne Paris Cité, F-91191 Gif-sur-Yvette, France\\
$^{8}$Astronomy Unit, School of Physics and Astronomy, Queen Mary University of London,
Mile End Road, London, E1 4NS, UK\\
$^{9}$Jodrell Bank Centre for Astrophysics, Alan Turing Building, Department of Physics \& Astronomy, School of Natural Sciences, The University of Manchester, \\
Oxford Road, Manchester, M13 9PL, UK\\
$^{10}$Dipartimento di Fisica ``G. Galilei'', Universit\`a degli Studi di Padova, Via Marzolo 8, 35131 Padova, Italy\\
$^{11}$INFN -- Istituto Nazionale di Fisica Nucleare, Sezione di Padova, Via Marzolo 8, 35131 Padova, Italy
}

\date{Accepted XXX. Received YYY; in original form ZZZ}

\pubyear{2021}

\begin{document}
\label{firstpage}
\pagerange{\pageref{firstpage}--\pageref{lastpage}}
\maketitle

\begin{abstract}
Neutral Hydrogen Intensity Mapping (\HI IM) surveys will be a powerful new probe of cosmology. However, strong astrophysical foregrounds contaminate the signal and their coupling with instrumental systematics further increases the data cleaning complexity.
In this work, we simulate a realistic single-dish \HI IM survey of a $5000$~deg$^2$ patch in the $950 - 1400$ MHz range, with both the MID telescope of the SKA Observatory (SKAO) and MeerKAT, its precursor. We include a state-of-the-art \HI simulations and explore different foreground models and instrumental effects such as non-homogeneous thermal noise and beam side-lobes. 
We perform the first Blind Foreground Subtraction Challenge for \HI IM on these synthetic data-cubes, aiming to characterise the performance of available foreground cleaning methods with no prior knowledge of the sky components and noise level. Nine foreground cleaning pipelines joined the Challenge, based on statistical source separation algorithms, blind polynomial fitting, and an astrophysical-informed parametric fit to foregrounds.
We devise metrics to compare the pipeline performances quantitatively. In general, they can recover the input maps' 2-point statistics within 20 per cent in the range of scales least affected by the telescope beam. 
However, spurious artefacts appear in the cleaned maps due to interactions between the foreground structure and the beam side-lobes. We conclude that it is fundamental to develop accurate beam deconvolution algorithms and test data post-processing steps carefully before cleaning. This study was performed as part of SKAO preparatory work by the \HI IM Focus Group of the SKA Cosmology Science Working Group.
\end{abstract}




\section{Introduction}

Over the next decades, a new generation of radio telescopes will revolutionise our understanding of cosmology via observations of the radio continuum emission and the 21-cm line emission from neutral hydrogen gas (\hinospace). Most notably, the telescope arrays of the SKA Observatory (SKAO) will conduct large radio surveys in order to test the standard cosmological model \citep{RedBook2018}. 

With the combined frequency range of both the SKAO-LOW and SKAO-MID telescopes, from $1.4$ GHz  down to $50$ MHz, the \HI surveys can trace the matter distribution from the present time to the epoch of reionization and beyond. 
\HI gas, as the first and most abundant element in the Universe, is an excellent tracer of the large-scale structure and its evolution. However, due to the weakness of its emission, the highest redshift at which a galaxy has been observed thanks to its 21-cm line is $z\sim0.376$ \citep{Fernandez2016},
and even the forthcoming SKAO surveys can only detect statistically significant samples for cosmology up $z \sim 0.4$ \citep{RedBook2018}.

Intensity Mapping (IM) is a relatively recent technique to circumvent detection limitations by observing the integrated \HI line emission from unresolved sources in large volume elements of the sky \citep{Bharadwaj2001,Battye2004,chang2008,Peterson2009,Wyithe2009,Seo2010}. \HI IM surveys are very time-efficient compared to traditional galaxy surveys as the low spatial resolution and large redshift range allow us to observe immense cosmic volumes within relatively short observation times.
The resulting \HI maps trace the largest scales of the matter distribution of the underlying dark matter field with excellent redshift resolution due to the telescope's fine frequency channelisation. Even though the original idea for \HI IM stems from using large single-dish telescopes such as the Green Bank Telescope (GBT) \citep{Chang2010, Switzer2013, Masui2013}, surveys can be conducted by a range of instrumental settings such as compact interferometric arrays or arrays of smaller dish telescopes.

For the SKAO Project, the planned cosmological IM surveys will be conducted in the so-called single-dish mode: each dish operates as a single telescope, and maps are co-added \citep{battyeps,Bull2015}. The resulting angular resolution of about one degree at $z\sim 0.4$ is very low, but the scanning is fast, and a large area coverage of $\sim 30,000\, \deg^2$ can be achieved using a few thousand hours \citep{RedBook2018}. The single-dish observations can be complemented by deep interferometric surveys which access the smaller scales beyond the primary beam of the telescope. Additionally, it has been shown that a large amount of small-scale information can be retrieved from the line-of-sight modes with its high redshift resolution \citep{Villaescusa2017}. Present and forthcoming instruments with planned \HI IM surveys are BINGO \citep{Battye2016}, CHIME \citep{bandura2014}, FAST \citep{Hu2020}, HIRAX \citep{Newburgh2016}, Tianlai \citep{Tianlai} and uGMRT \citep{Chakraborty2021}. Most importantly for this work, the SKAO pathfinder MeerKAT in South Africa is already taking pilot data for its MeerKLASS survey \citep{Santos:2017qgq,Wang2021} and the $64$ MeerKAT dishes will eventually be incorporated into the SKAO-MID telescope array when it will commences operation in the late 2020's. 

However, the detection of the \HI IM has proven to be observationally challenging. Since its first application by \citet{Chang2010} with GBT data more than a decade ago, few other studies have claimed detection of the signal, and always in cross-correlation with a galaxy catalogue \citep{Masui2013, Anderson2018, Wolz:2021ofa}. The main obstacle to detecting the \HI signal comes from the presence of astrophysical foregrounds orders of magnitude stronger than the \HI signal. While astrophysical foregrounds, predominantly due to synchrotron and free-free emission at the relevant (around $1$ GHz) frequencies, have a known spatial distribution and frequency correlation, their convolution with instrumental systematics and other observational effects can render signal separation a very challenging task. 

In recent years, many studies have addressed the problem in the context of single-dish \HI IM and investigated the quality of foreground removal methods on data \citep{Switzer:2015ria, Wolz:2015lwa} as well as simulations \citep[e.g.,][]{2012A&A...540A.129A, Wolz:2013wna, Alonso:2014dhk, Shaw2015, Olivari2016, Carucci:2020enz,  2021JCAP...04..081M, 2021MNRAS.504.5231Y, 2021MNRAS.504..267F, Soares2021}, where blind and non-parametric methods such as Principal Component Analysis (PCA), Independent Component Analysis (ICA), and Generalised Morphological Component Analysis (GMCA) have proven most powerful. In addition, many studies set particular focus on individual observational systematics such as primary beam effects \citep{Matshawule2020}, polarisation leakage \citep{Shaw2015, Spinelli2018,Carucci:2020enz, Cunnington:2020njn}, $1/f$ noise \citep{2018MNRAS.478.2416H, 2020MNRAS.491.4254C, 2021MNRAS.501.4344L} and radio frequency interference due to satellites \citep{Harper2018}. 
Findings of these studies point to the fact that all observational effects sensitively depend on the individual instrument and survey design, making end-to-end simulations a crucial requirement towards a valid detection of the \HI IM signal in auto-correlation. 

In this study, we present a detailed study of foreground removal methods for MeerKAT and future SKAO-MID \HI IM surveys implemented by the \HI Intensity Mapping Focus Group of the SKA Cosmology Science Working Group (SWG). For the first time, we conduct a Blind Foreground Subtraction Challenge where participants are presented with simulated data-cubes of unknown foregrounds, \HI signal, and instrumental specifics such as the beam and noise level.
We implement a realistic scanning strategy for a $\sim5,000\deg^2$ survey resulting in anisotropic noise, as well as a more sophisticated beam model with chromatic side-lobes for the SKAO-MID and MeerKAT dishes in addition to the conventional Gaussian beam approximation. We use two different implementations of the astrophysical foregrounds in order to investigate the impact of foreground models on the separation techniques.
The true level of \HI signal and noise in the mocks was not known to the participants of the Blind Challenge and the submitted results have not been adjusted or modified after unblinding the submissions. The participants used a total of nine different pipelines to clean the data-cubes, ranging from different kinds of blind (PCA, \fastica\ and GMCA) to non-blind (parametric fitting) source separation algorithms.
We stress that, although the cleaning techniques employed in this work have been proven powerful when applied to less realistic simulations, we do not expect them to perform perfectly facing these new complexities. We are thus equally interested in their absolute and relative performances to understand weaknesses and strengths.

The paper is structured as follows. In \secref{sec:sim} we describe the end-to-end simulation and properties of the mock data-cubes. In \secref{sec:methods} we outline the cleaning methods and in \secref{sec:estimators} the statistical estimators we use for the comparison among cleaned residuals and input maps. In \secref{sec:blind} we describe how we run the Challenge. Results are presented in \secref{sec:results}, followed by a broad discussion in \secref{sec:discussion}. We draw our conclusions and give future perspectives in \secref{sec:conclusions}.

\begin{figure}
  \centering
  \includegraphics[width=\columnwidth]{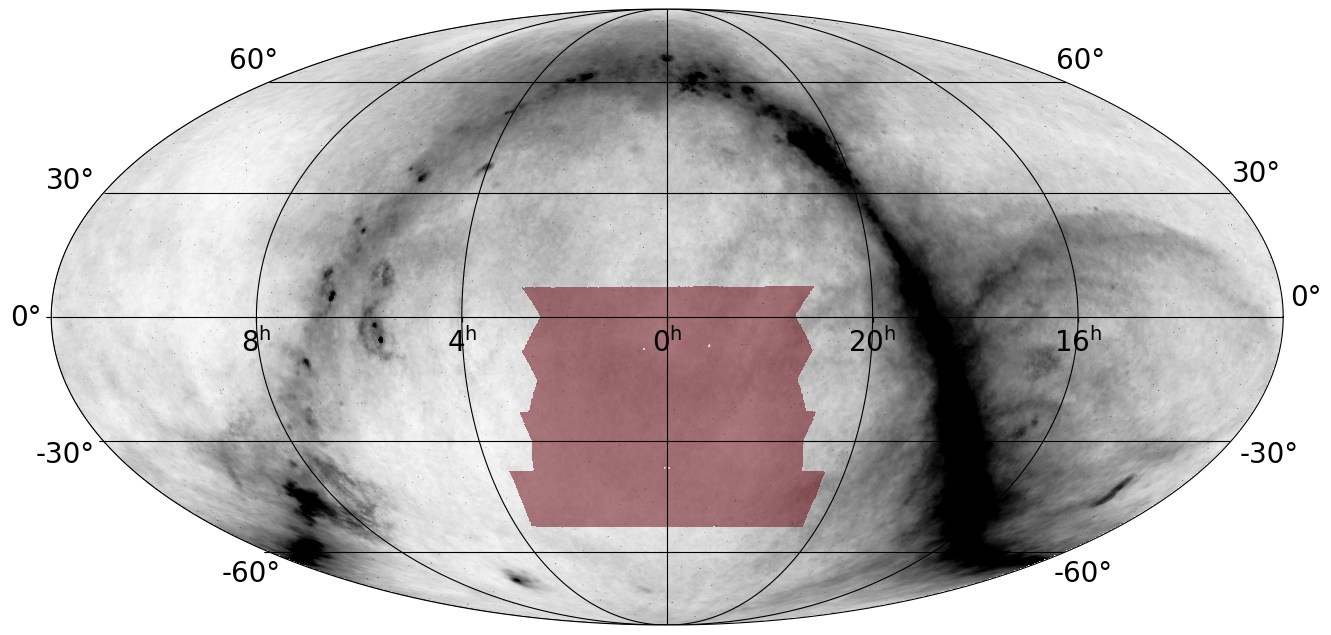} 
  \caption{
  An illustration of the footprint considered in this work (maroon shaded region) and its position with respect to the Galactic plane (in equatorial coordinates). The foreground emission model used for this plot is described in \secref{sec:PSM}
  and a more accurate description of the shape of the footprint can be found in \secref{sec:telescope}.}
  \label{fig:fgs512}
\end{figure}

\begin{table*}
\caption{A brief description of the components of the sky simulations. The cosmic neutral hydrogen signal (\HI) is presented in more detail in \secref{sec:HI}, while the foreground components are described in \secref{sec:sky_models}. We consider two  different sets of foreground simulations: the MS$_{05}$ and the PSM model (see main text for details).}\label{table:ingredients}
\centering
\begin{tabular}{|c|c|c|c|} 

\hline
Sky component & \multicolumn{3}{c}{Description} \\
 \hline\hline
  & \multicolumn{3}{c}{PINOCCHIO LPT light-cone halos painted with a $M_{\rm HI}-M_{\rm halo}$ relation} \\
\hinospace &  \multicolumn{3}{c}{extrapolated from the GAEA semi-analytical model}\\
& \multicolumn{3}{c}{\citep{Spinelli2020}} \\
\hline
\hline
  & & MS$_{05}$ Model (as in \citealt{MS05}), & PSM Model\\
Foregrounds &  & parameters of \autoref{eq:MS05}& (as in \citealt{Carucci:2020enz})\\
& & $\{A [{\rm mK}^2];\, \beta;\,  \alpha  \}$ &  \\
\hline
&  & & Haslam 408 MHz, \citep{has16}  \\
& Galactic Synchrotron & $\{ 700;\,2.4;\,2.80 \}$ &with spatially varying synchrotron spectral index  \\
& & & \citep{specind} \\
\hline 
& Free-Free & $\{ 0.088;\,3.0;\,2.15 \}$ & ${\rm{H}}\alpha$ template \citep{fink} \\
\hline
& Extragalactic Free-Free & $\{ 0.014;\,1.0;\,2.10 \}$ & $-$ \\
\hline
& Point Sources & $\{ 57;\,1.1;\,2.07 \}$ & Source count model with flux cut at $0.1$ Jy \citep{Olivari2018} \\
\hline
\hline
\end{tabular}
\end{table*}

\section{Simulations}\label{sec:sim}
In this section, we describe the various ingredients 
of our mock data. The sky simulations of the signal and the foregrounds are presented in \secref{sec:HI} and \secref{sec:sky_models}, respectively. We consider two different foreground models: a simplistic one based on \citet{MS05} (MS$_{05}$, \secref{sec:MS}) and a more realistic and physically motivated one based on available data and the Planck Sky Model \citep{psm} (PSM, \secref{sec:PSM}). All components of the sky simulation are summarised in \autoref{table:ingredients}. In \secref{sec:telescope} we describe the instrumental simulations, detailing the assumed beam model (\secref{sec:beam}) and the observing strategy and noise (\secref{sec:noise}). We focus on both a SKAO-MID telescope-like and a MeerKAT-like IM survey, considering for the former case a smaller beam and a lower noise level. 
We also explore two different beam models, a standard Gaussian beam and a more realistic beam model that includes side-lobes based on the apertures of the MeerKAT/SKAO-MID dishes, to which we will be referring to as the Airy beam. 
The different telescope and survey specifications are reported in \autoref{table:survey}. 
We focus on a frequency range covering 950 to 1400\,MHz binned into 512 observational channels, similar to the MeerKAT's L-band.
Our sky maps are created using  the HEALPix format \citep{healpix}, at $\rm{N_{side}} = 512$, providing $\sim 7$ arcmin resolution. 
\autoref{fig:fgs512} illustrates the footprint considered in this work.

We model the observed sky temperature, $T_{\rm obs}$, in the direction $\hat{\textbf{n}}$ and as a function of frequency $\nu$ as 
\begin{align}\label{eq:tmodel}
    T_{\rm obs}(\nu,\hat{\textbf{n}})=&\int {\rm d}{\bf \Omega}\, B(\nu,\hat{\textbf{n}},{\bf \Omega})\left[T_{\rm fg}(\nu,\hat{\textbf{n}})+T_{\rm HI}(\nu,\hat{\textbf{n}})\right]\nonumber\\
    & + T_\text{noise}(\nu,\hat{\textbf{n}}) \, ,
\end{align}
where $T_{\rm fg}(\nu,\hat{\textbf{n}})$ is the astrophysical foreground emission and $T_{\rm HI}(\nu,\hat{\textbf{n}})$ is the 21-cm signal from cosmic \hinospace. Both are convolved with the telescope beam  $B(\nu,\hat{\textbf{n}},{\bf \Omega})$, pointing in the direction $\hat{\textbf{n}}$ and covering the solid angle ${\bf \Omega}$. The response of the telescope also adds a thermal noise component $T_\text{noise}(\nu,\hat{\textbf{n}})$ that varies with frequency and also with direction since we take into account a scanning strategy.

Our simulations could be made more complex adding other systematics such as missing channels due to RFI, $1/f$ noise, or satellites contamination. In this work, we focus on the inclusion of realistic beam modelling and non-homogeneous noise in order to first establish their impact on the cleaning methods, leaving further systematics to future studies.

\subsection{Cosmological Simulation}\label{sec:HI}

Since the quality of the foreground cleaning procedure for IM experiments will inevitably depend on the properties of the \HI signal, having a realistic description of its large-scale distribution and evolution with redshift is crucial. At low redshifts, 
neutral hydrogen is expected to be hosted only in high density regions where, shielded from UV radiation, has survived the reionization process. Given the relatively poor spatial resolution of single-dish experiments, each voxel in the sky is expected to host a large number of galaxies. This implies that it is possible to simulate the \HI clustering without describing the single galaxies but by considering the total amount of neutral hydrogen mass $M_{\rm HI}$ hosted by a halo with mass $M_{\rm halo}$, i.e., the $M_{\rm HI}-M_{\rm halo}$ relation \citep[e.g.,][]{Bagla2010,Carucci2015,2017JCAP...12..018C,Modi2019,Asorey2020,Zhang2021_bingo}. 
In this work, we use the \HI Probe Populator (\texttt{HIP-POP}\footnote{Spinelli et al. \textit{in prep}})  that combines a full-sky halo light-cone with information on the baryonic content extrapolated from a semi-analytical model of galaxy formation and evolution. \texttt{HIP-POP} uses the PINOCCHIO code \citep{Monaco2002,Taffoni2002,Monaco2013,Munari2017} to generate catalogues of cosmological dark matter halos with a known mass, position, velocity, and merger history. PINOCCHIO is based on the Lagrangian Perturbation Theory (LPT) and is able to reproduce, with very good accuracy, the hierarchical formation of dark matter halos. We produce 1 Gpc $h^{-1}$ boxes using $2048^3$ particles to reach a minimum halo mass of $\lesssim 10^{11} {\rm M}_{\odot}h^{-1}$ and construct a full-sky light-cone. On the largest scale, there will be repetitions due to the limited size of the box that we replicate to fill the light-cone. 
This is not a problem in our case since we will select a relatively small patch at low redshift. 

We populate each halo following \citet{Spinelli2020}, who used the outputs of the semi-analytical model GAEA \citep{Delucia2004, Delucia2014,Hirschmann2016,Zoldan2017}. Specifically, we use the version of the code described in \citet{Xie2017}, run on the merger trees of the Millennium II simulation (MII, \citealt{Boylan-kolchin2009}). With $2160^3$ particles in a $100$~Mpc $h^{-1}$ box, it can describe galaxies down to \HI masses of $10^7~{\rm M}_{\odot} h^{-1}$. MII is based on a WMAP1 cosmological model \citep{Spergel2003} with $\Omega_{\rm m}=0.25$, $\Omega_{\rm b}=0.045$, $h=0.73$ and $\sigma_8=0.9$.
For consistency, our PINOCCHIO light-cone assumes the same cosmology.

For each available GAEA snapshot relevant for our purposes, we measure the 
$M_{\rm HI}$ as a function of $M_{\rm halo}$ and
model it using the $M_{\rm HI}-M_{\rm halo}$ relation:
\begin{equation}
    M_{\mathrm{HI}}(M_{\rm halo})=M_{\rm halo} \left[a_1 \left(\frac{M_{\rm halo}}{10^{10}}\right)^{\beta} e^{-\left(\frac{M_{\rm halo}}{M_{\mathrm{break}}}\right)^\alpha}+a_2\right] e^{-\left(\frac{M_{\mathrm{min}}}{M_{\rm halo}}\right)^{0.5}},\label{eq:M_HI}
\end{equation}
where $a_1$, $\beta$, $\alpha$, $M_{\rm break}$, $a_2$, and $M_{\rm min}$ are free parameters \citep{Spinelli2020}. 
We construct a Gaussian likelihood for these parameters and, assuming large flat priors, we reconstruct their posteriors through the \textsc{multinest} sampler \citep{Feroz2008,Feroz2009} using an MPI-enabled python wrapper \citep{Zwart2016}.  
We thus obtain a trend in redshift for each of the $M_{\mathrm{HI}}(M_{\rm halo})$ parameters that we interpolate with a spline. A similar procedure is followed for the scatter of the  $M_{\rm HI}-M_{\rm halo}$ relation. In this way, we have a prescription to populate halos with \HI at each needed redshift that we use for the full light-cone.

Since the \HI signal will be measured in redshift space, we use the plane-parallel approximation to displace the real-space halos positions using their peculiar velocities. 

We construct a HEALPix map with $N_{\rm side}=512$ for each of the 512 frequencies of interest binning the redshift space positions of the halo centres in slices of $\Delta \nu$ (see \autoref{table:survey}). Given the volume of each such defined portion of the light-cone and its total $M_{\rm HI}$ mass, one can compute the
\HI density $\rho_{\rm HI}$ and estimate the brightness temperature fluctuation in each pixel $\hat{\textbf{n}}$ \citep{Mao2012}:
\begin{equation}
    \delta T_{\rm HI}(\nu,\hat{\textbf{n}})=\overline{\delta T_{\rm HI}}(z) \left[ \frac{\rho_{\rm HI}(\hat{\textbf{n}})}{\overline{\rho_{\rm HI}}(z)}\right].
\end{equation}
The mean \HI brightness temperature at a given redshift $z$ can be computed following \citet{Furlanetto2006}
\begin{equation}
\overline{\delta T_{\rm HI}}(z)=23.88 x_{\rm HI} \left( \frac{\Omega_b h^2}{0.02}\right) \sqrt{\frac{0.15}{\Omega_m h^2}\frac{(1+z)}{10}}\; \mathrm{mK} \, ,
\end{equation}
where $x_{\rm HI} \equiv \Omega_{\rm HI}/ \Omega_{\rm H}$ is the fraction of neutral atomic hydrogen and $\Omega_{\rm HI}(z)=8 \pi G \overline{\rho_{\rm HI}}(z)/(3H_0^2)$.
The highest frequencies considered correspond to a very local universe and, in this case, the virial radius of the most massive halos can be comparable to the size of a voxel. To avoid such spurious over-densities, when in this regime, we do not assign all the \HI mass to the halo centre, but we distribute the \HI mass according to a NFW profile \citep{NFW1996}, thus spreading the \HI to neighbouring voxels.

\begin{figure}
  \centering
  \includegraphics[width=\columnwidth]{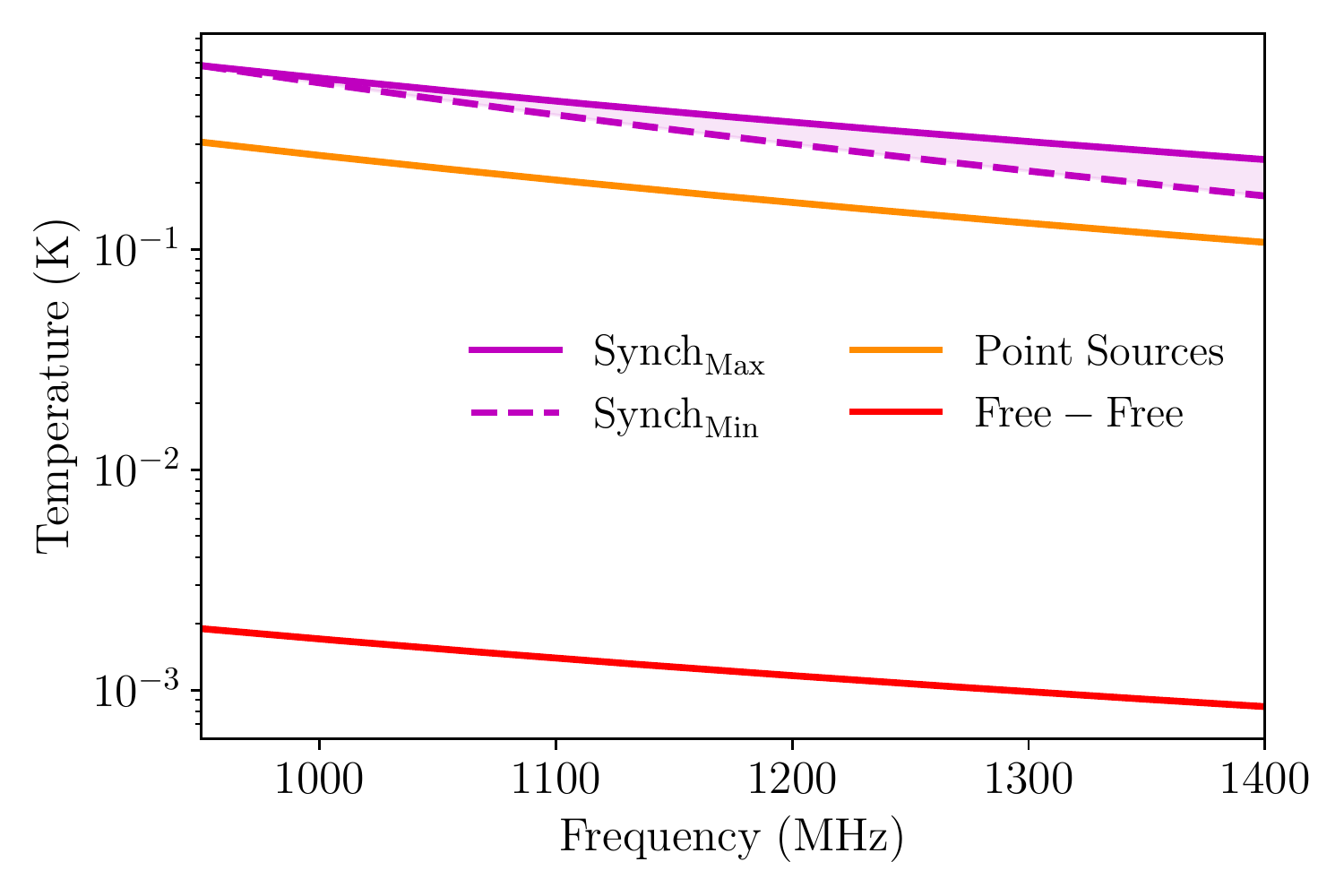} 
  \caption{The spectral forms of each PSM emission component normalised using the mean emission temperature within the sky region investigated by this work; synchrotron emission displayed as a shaded area between the steepest and shallowest spectral index present in the simulation.}
  \label{fig:sforms}
\end{figure} 

\subsection{Foreground Models}\label{sec:sky_models}

The dominant foregrounds present between 950 and 1400\,MHz are diffuse synchrotron emission, diffuse free-free emission, and extragalactic point sources. In this work, we explore two established models of IM foregrounds: a Gaussian realisation of the components based on \citet{MS05}, referred to as $\rm{MS_{05}}$ in this work, and a Planck Sky Model based simulation \citep{psm}, referred to as PSM.

\subsubsection{$\rm{MS_{05}}$}\label{sec:MS}

\citet{MS05} constructed Gaussian realisations of extragalactic and diffuse Galactic emissions to investigate their effect on the extraction of cosmological information from 21-cm IM data. The angular power spectrum for each foreground component takes the form:
\begin{equation} \label{eq:MS05}
C_{\ell}(\nu_i,\nu_j) = A \left( \frac{1000}{\ell} \right)^\beta \left( \frac{\nu_{\rm ref}^2}{\nu_i\nu_j}\right)^\alpha I_\ell^{ij} 
\, ,
\end{equation}
where the reference frequency used is $\nu_{\rm ref}=130$\,MHz, $A$ is the power spectrum amplitude, $\beta$ controls the angular scaling and $\alpha$ is the spectral index across the frequency range. 
Each of the foreground components is parametrised with a different set of $\{A, \beta,\alpha\}$ values, reported in \autoref{table:ingredients}.
The term $I_\ell^{ij}$ encodes the frequency coherency of the foreground and is expected to be unity for complete correlation.
\citet{MS05} considered departures from the complete correlation adding a decorrelation term. However, for simplicity, and to keep this as the most idealised of foreground models, we choose to omit this and assume $I_\ell^{ij}=1$.

\subsubsection{PSM}\label{sec:PSM}

With the aim of testing cleaning on realistic foreground contamination, we take advantage of the 
FFP10 (Full Focal Plane) all-sky simulations from the {\it Planck legacy archive}\footnote{\url{https://wiki.cosmos.esa.int/planck-legacy-archive/}}. We use the high frequency versions of the FFP10 simulations as these are available at ${\rm{N_{side}}}= 2048$ allowing us to downgrade the maps to our desired ${\rm{N_{side}}} = 512$. The FFP10 simulations take their foreground contributions from the Planck Sky Model (PSM, \citealt{psm}), which in turn uses empirical data sets to inform its estimates. Whilst these models only hold true under specific assumptions, which will be discussed, it is worthwhile to include foreground simulations which are 1) not Gaussian in nature and 2) have the possibility of being correlated with each other. We outline the main features of this set of foregrounds; for more details, we refer the reader to \citet{Carucci:2020enz}, where this model was first assembled for the frequencies of interest.

\paragraph*{Galactic synchrotron emission} We use the FFP10 synchrotron simulation at 217\,GHz, based on the source-subtracted and destriped version of the Haslam 408\,MHz map \citep{has16}, and scale it across frequencies using the synchrotron spectral index map of \citet{specind}. The Haslam 408\,MHz map is assumed to contain negligible amounts of Galactic free-free emission at high Galactic latitudes. The synchrotron spectral index map used has been formed from 408\,MHz and 23\,GHz data and so may in fact be slightly steeper than the true synchrotron spectral indices at MHz frequencies. For our study however, we only require
spatially varying spectral indices within the physically expected range for synchrotron emission. The spectral index map is at a lower resolution than our intended simulation resolution. We add in detail below the resolution threshold of $5\,\deg$ of the spectral index map using the following Gaussian realisation:
\begin{equation}
    C_{\ell} = A \left( \frac{1000}{\ell} \right)^{2.4} \, ,
\end{equation}
where the amplitude ($A$) is set using the angular power spectrum of the 5 degree spectral index map. 

\paragraph*{Galactic free-free emission} We scale the FFP10 free-free simulation at 217\,GHz, based on the all-sky ${\rm{H}}\alpha$ template of \cite{fink}, down to our MHz frequency range using a spatially constant spectral index of $-2.1$. It should be noted that free-free emission is the least dominant foreground component across our frequency and Galactic latitude range.

\paragraph*{Extragalactic point sources} This contribution is the only component not taken from the FFP10 simulations; for this we use the prescription outlined in \citet{Olivari2018}, which expands on the empirical 1.4\,GHz source count model of \citet{battyeps}. The model requires three selection criteria: 1) the cut-off flux, i.e., the value above which we assume point sources are bright enough to be identified and removed \citep[e.g.,][]{wang2010,Matshawule2020} 2) the average point source spectral index and 3) the distribution of this spectral index across the map. We use a flux cut-off of 0.1\,Jy and a Gaussian distribution for the source spectral index centred at $-2.7$ with $ \sigma = 0.2$. 

\medskip
The spectral forms of all components of our PSM-based foreground simulations are shown in
\autoref{fig:sforms}.

\subsection{Telescope Simulation}\label{sec:telescope}

\subsubsection{Beam Models}\label{sec:beam}
We aim to test how well component separation methods work in the presence of a beam model that includes not just the main lobe but also a side-lobe structure that changes with frequency. For these simulations we are considering the dishes used in the MeerKAT array and the SKAO-MID array. For the MeerKAT dishes, we assume unobstructed 13.5\,m apertures, and 15\,m unobstructed apertures with under illuminated primaries to reduce the side-lobe amplitude for the SKAO-MID
dishes. We do not model the final SKAO-MID survey which will include observations from both 13.5\,m and 15\,m dishes (integrating the MeerKAT dishes); instead, we focus on two separate surveys with different dish properties to analyse the effect of these characteristics distinctly.  

We generate the beam models for both dish types using modified Airy beam functions that allow for Gaussian tapered aperture distributions defined as \citep{wilson2009tools}
\begin{equation}\label{eqn:taper}
   E(\rho_\nu) = e^{-0.5(\rho_\nu/\sigma_\rho)^2} \, ,
\end{equation}
where $\sigma_\rho$ defines the width of a Gaussian taper, and $\rho_\nu$ is the number wavelengths across the dish at a given frequency defined as 
\begin{equation}
  \rho_\nu = \frac{D\nu}{2c} \, ,
\end{equation}
where $D$ is the dish diameter (either 13.5\,m or 15\,m), $\nu$ is the observing frequency, and $c$ is the speed of light. For the MeerKAT dishes we find that side-lobe structure is best represented by setting the Gaussian taper width in \autoref{eqn:taper} to be $\sigma_\rho = \infty$, which describes a dish that is being uniformly illuminated (i.e., the MeerKAT beam model is represented by an Airy beam). For the SKAO-MID dishes we expect the larger dishes to be under-illuminated to improve the side-lobe response, and we adopt a value of $\sigma_\rho = 20$.

\begin{figure}
\centering
\includegraphics[width=0.45\textwidth]{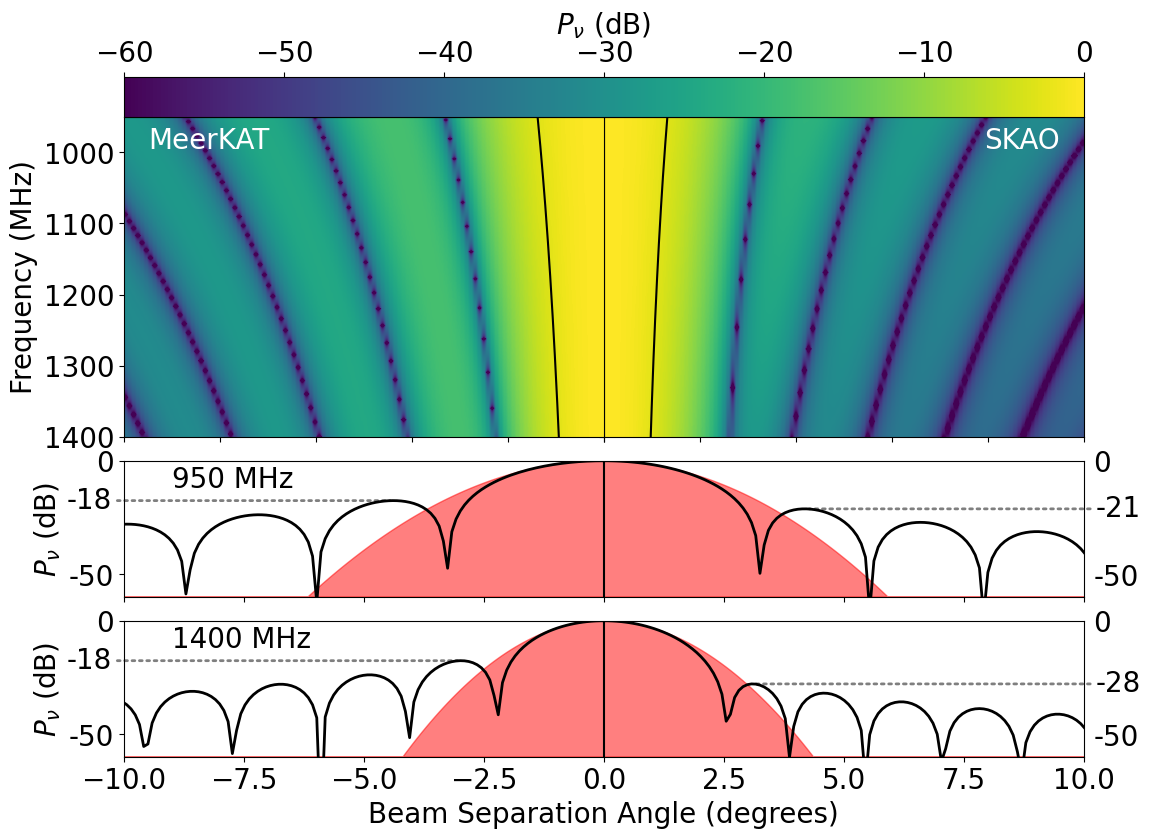}
\caption{Simulated beam models for MeerKAT (left) and SKAO-MID (right) dishes. The top panels show the change in the beam pattern with frequency out to a beam separation angle of $10\,\deg$. The black lines in the top panel show the FWHM of each beam model. The \textit{centre} and \textit{bottom} panels show cuts through the MeerKAT (\textit{left}) and SKA-MID (\textit{right}) beam patterns at 950 and 1400\,MHz (i.e. the top and bottom of the simulated band). The red shaded region shows the equivalent Gaussian beam of each instrument. The \textit{dashed-grey} lines mark the amplitude of the first sidelobe for the Airy beam model used for each instrument.}\label{fig:beams}
\end{figure}

\begin{table}
\caption{Simulated telescope and survey parameters.}
\centering
\begin{tabular}{|c|c c|} 
 \hline
Parameter & SKAO & MeerKAT\\
\hline
$N_{\rm dish}$ & 133 & 64 \\
$T_{rx}$      & 7.5\,K  & 9.8\,K \\
$T_\mathrm{spill}$ & 4\,K & 4\,K \\
$\Delta \nu$ & 1\,MHz & 1\,MHz \\
$\rho_\nu$    & 20 & $\infty$ \\ \hline
Strip Declinations & \multicolumn{2}{c}{-45, -30, -15, 0} \\
Strip Width & \multicolumn{2}{c}{15\,$\deg$} \\
Scan Speed  & \multicolumn{2}{c}{1$\,{\deg}$/s} \\
$\Omega_\mathrm{sky}$  & \multicolumn{2}{c}{$\approx 5000$\,$\deg^{2}$} \\
 \hline\hline

\end{tabular}
\label{table:survey}
\end{table}

To generate the beam pattern at each frequency we integrate over the aperture distribution frequency for each beam separation angle ($\theta$) as
\begin{equation}
  B(\nu,\theta) = \left| \frac{\int E(\rho_\nu)j_0(\rho_\nu\sin(\theta))\rho_\nu \mathrm{d}\rho_\nu}{\int E(\rho_\nu)\rho_\nu \mathrm{d}\rho_\nu} \right| \, ,
\end{equation}
where $j_0$ is the zeroth-order Bessel function. The resulting beam patterns for the MeerKAT and SKAO-MID dishes are shown side-by-side in \autoref{fig:beams} for the frequency range 950-1400\,MHz, and beam separation angle out to $40\,\deg$ from the main lobe. The upper panel in \autoref{fig:beams} shows how the beam models evolve with frequency. The marked black lines in the upper panel show how the FWHM of each beam model changes with frequency, changing by just 0.44 and $0.35\,\deg$ full width at half maximum (FWHM) for the MeerKAT and SKAO-MID dish models, respectively. The lower panel shows a slice of the beam model at 1175\,MHz. Here we can see that the first side-lobe in the MeerKAT model (-18\,dB) is 4\,times larger than the first SKAO-MID side-lobe at the same frequency (-24\,dB). For the MeerKAT model, the first side-lobe response is close to constant, while the SKAO-MID first side-lobe changes by a factor of 5 from $-21$\,dB to $-28$\,dB across the band. In the rest of the text and figures, we will refer to these beam models as Airy beam models.

We also produce a Gaussian beam model for each dish model that is representative of the main beam response of the telescope. We define these more approximate beam models as
\begin{equation}
  G_\nu(\theta) = e^{-4\mathrm{log}(2)(\theta/\theta_\mathrm{FWHM})^2} \, ,
\end{equation}
where the $\theta_\mathrm{FWHM}$ evolves with frequency as it is forced to match the measured FWHM of the Airy beam models.

Finally, we convolve the sky models described in \secref{sec:sky_models} with each beam model. The convolution is performed by transforming the map and the beam model into the spherical harmonic domain. For a radially symmetric beam model, the spherical harmonic transform of the beam pattern is defined as
\begin{equation}
  B_\ell(\nu) = 2 \pi \int B(\nu,\theta) P_\ell(\cos(\theta)) \sin(\theta) \mathrm{d}\theta \, ,
\end{equation}
where $P_\ell$ are Legendre polynomials. 

\subsubsection{Observing strategy}\label{sec:observing}

\begin{figure}
   \centering
   \includegraphics[width=\columnwidth]{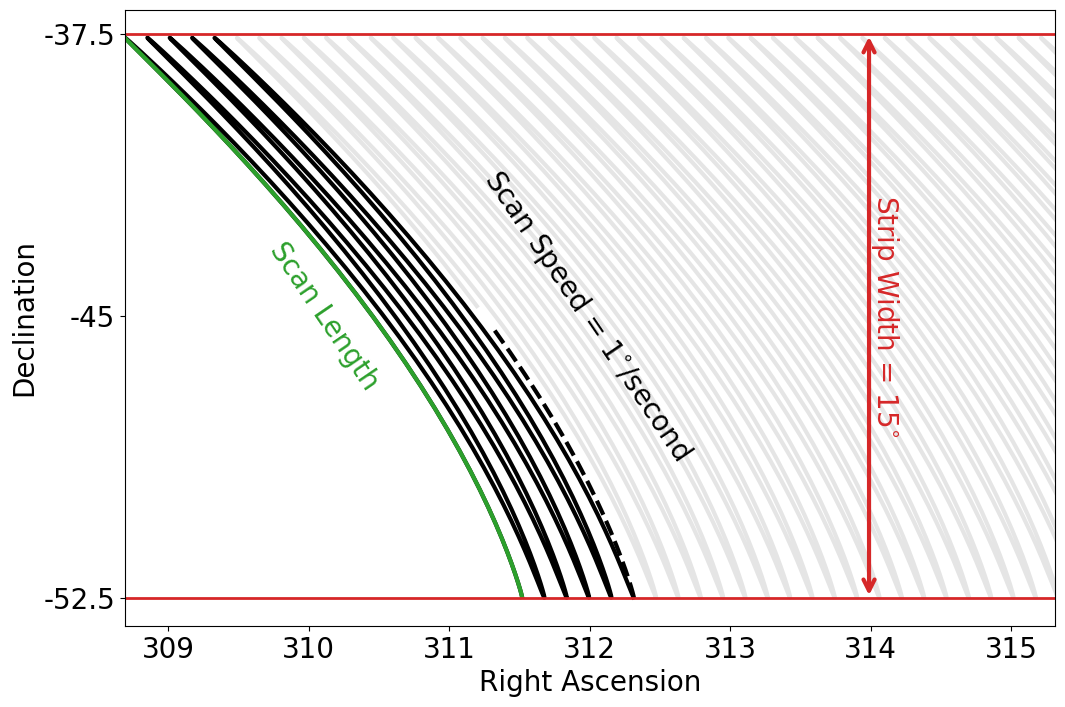} 
   \caption{Example of the constant elevation scanning strategy used to map out the strip centred on $-45\,\deg$ declination. For visualisation purposes we highlight the first scan in this example in \textit{green}, which defines a single scan length, the \textit{black} lines represent the area already scanned, while the \textit{faint-black} lines represent the upcoming scans. The \textit{red} lines mark the declination boundaries of the strip, which has a width of $15^{\circ}$. Each strip has the same width and scanning speed.}
   \label{fig:ObservingStrategy}   
\end{figure}

\begin{figure*}
   \centering
   \includegraphics[width=1.9\columnwidth]{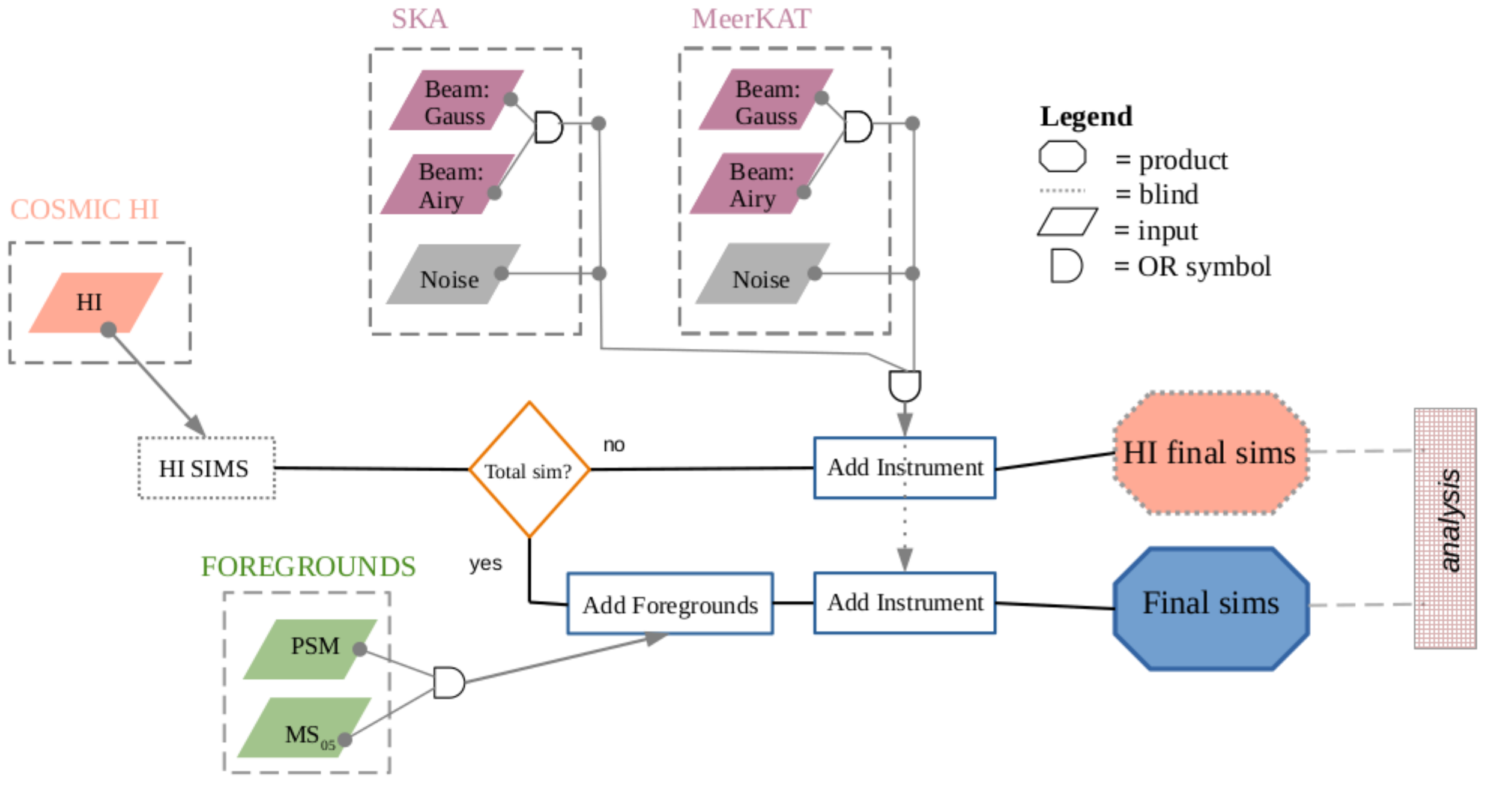} 
   \caption{Flowchart of the construction of the simulations described in \secref{sec:sim}, highlighting the various options for foreground modelling and instrumental effects. We recall that the HI level is not known to the participants of the challenge (dotted frames indicate the blind components).}
   \label{fig:flowchart_1}   
\end{figure*}

For our simulations we use a simulated observatory to create a mock IM data set that is closely 
representative of both the proposed SKAO-MID Band 2 survey set out in \citet{RedBook2018} and the ongoing MeerKLASS survey \citep{Santos:2017qgq, Pourtsidou:2017era, Wang2021}. The purpose of the simulations is to create inhomogeneities in the noise distribution around the map and create a patch shape that is representative of realistic observations. Both the realistic noise distribution and patch shape of the simulations will enable us to test the component separation methods on a \textit{quasi}-realistic data set.

The simulated survey scanning strategy used constant elevation azimuth scans to map out four strips in declination. \autoref{fig:ObservingStrategy} shows an example of how the strip centred on $-45\,\deg$ declination was mapped out. We observed each strip at half of the maximum elevation as seen from the centre of the SKAO-MID array, observing each strip both when it was rising and setting. The observation time for the simulated survey is approximately 40\,hours per dish, equating to a total observing time of approximately 5000\,hours for the SKAO-MID and 2500\,hours for the MeerKAT array. The total sky area mapped is approximately 5000\,$\deg^{2}$ spanning between $-52.5 < \delta < 7.5\,\deg$ in declination, and each strip is 70\,$\deg$ long centred at 0\,$\deg$ in right ascension. The choice of patch location was made to match preliminary \HI IM observations from MeerKAT \citep{Wang2021}, and also to have minimal galactic foreground contributions. However, we are aware that due to strong satellite RFI any real ground based survey would not choose to observer near $\delta \sim 0\,\deg$ \citep[e.g.,][]{Harper2018}, however this is not an issue here as we are not including RFI within the simulation and the exact declination of the patch will not significantly change the results.

The fixed elevation azimuth scanning strategy was simulated using a simple sine-wave model of the telescope motion described as
\begin{equation}
  A = \frac{\Delta A}{2} \sin(2\pi t/T) + A_0 \, ,
\end{equation}
where $A$ is the telescope azimuth, $A_0$ is the central azimuth corresponding to the declination of each strip, $T$ is the time to complete a single scan defined as $T = \Delta A/v_\mathrm{scan}$ where $v_\mathrm{scan}$ is the scan speed of the telescope, and $\Delta A$ is the scan length which is dependent on the strip width, the scanning speed, and the elevation which we calculate numerically for each scan. The choice of a sine function to model the telescope azimuth motion as opposed to a triangular waveform was to also include the effect of the telescope turnaround time. The elevation is modelled as a constant value for each strip and ranges between $25$ and $30\,\deg$.  For a summary of the simulation parameters see \autoref{table:survey}.

\subsubsection{Noise Model}\label{sec:noise}

For both the SKAO-MID and MeerKAT receiver noise models we assume the noise to be Gaussian and white. The noise per pixel is calculated by
\begin{equation}\label{eq:radiometer}
  \sigma = \frac{T_\mathrm{sys}}{\sqrt{N_\mathrm{dish} \tau \Delta \nu}} \, ,
\end{equation}
where $\tau$ is the integration time in seconds per pixel which is defined by the observing strategy described in \secref{sec:observing}, $\Delta \nu$ is the bandwidth of each frequency channel, $N_\mathrm{dish}$ is the number of dishes in the array, and $T_\mathrm{sys}$ is the system temperature\footnote{\autoref{eq:radiometer} is strictly for a single polarisation receiver. In principle, two polarisations would be available but the resulting factor two in the equation is within our uncertainty in the total system temperature budget. We have thus ignored it.}.

We define the system temperature for both receiver types as
\begin{equation}\label{eq:Tsys}
    T_\mathrm{sys}\left(\hat{\textbf{n}}\right)  =T_\mathrm{rx} + T_\mathrm{CMB} + T_\mathrm{spill} + T_\mathrm{sky}\left(\hat{\textbf{n}}\right) \, ,
\end{equation}
where $T_\mathrm{CMB} = 2.73$\,K is the CMB monopole contribution, $T_\mathrm{spill} = 3$\,K is the approximate contribution to spill-over, $T_\mathrm{sky}\left(\hat{\textbf{n}}\right)$ is the brightness of the sky along line-of-sight $\hat{\textbf{n}}$, and $T_\mathrm{rx}$ is the receiver temperature\footnote{For these simulations we do not include any atmospheric contribution but it is expected to be only a few K at 1\,GHz  \citep{Bigot2015}.}. For the receiver temperature of the SKAO-MID dishes we used the band 2 receiver temperature prediction given in \citet{RedBook2018} which gives $T^\mathrm{SKAO}_{\rm rx} = 7.5$\,K. For MeerKAT we use the mean of the measured receiver temperature response defined as \citep{Braun2019arXiv}
\begin{equation}
      T^\mathrm{MeerKAT}_\mathrm{rx} = \left< 7.5 + 6.8 \left|\nu_\mathrm{GHz}-1.65\right|^{1.5}\right>\:\mathrm{K}\,,
\end{equation}
which gives $T^\mathrm{MeerKAT}_\mathrm{rx} = 9.8$\,K.

\begin{figure*}
   \centering
   \includegraphics[width=2.1\columnwidth]{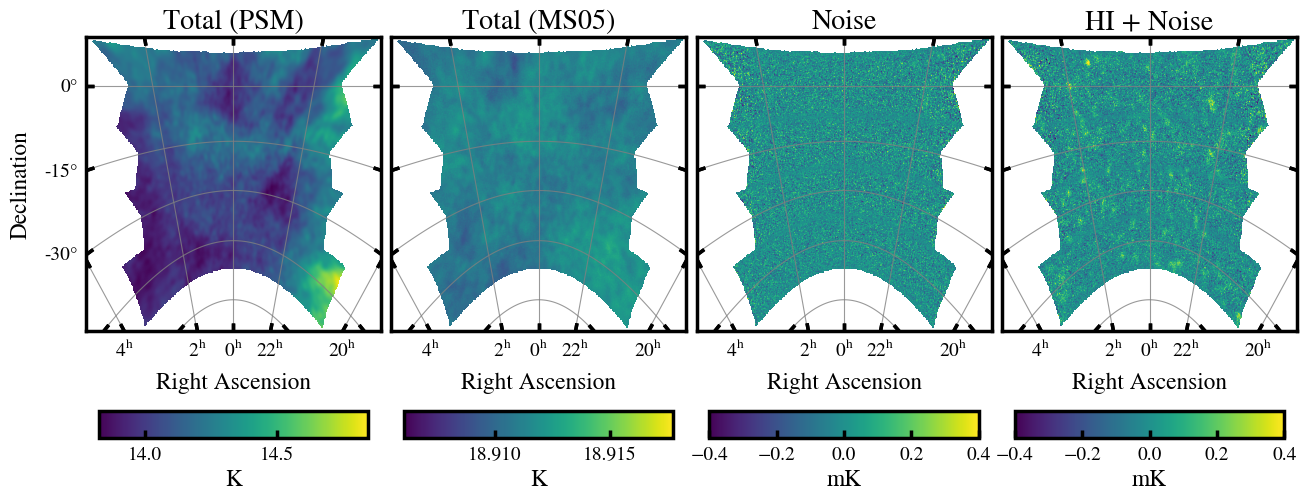} 
   \caption{Maps of the contributing components to the final simulated signal at $1000\,\text{MHz}$ for the SKAO-MID Airy beam case. The first two panels show the two different foreground simulations we implement; the Planck Sky Model (PSM) \citep{psm} and the Gaussian realisations presented in \citet{MS05} (MS$_{05}$). The last two panels show the simulated thermal noise alone and with the cosmological \HI signal  that we aim to recover, i.e., $\boldsymbol{\mathsf{R}}$ in \autoref{eq:systosolve}. The noise in these examples is generated using the PSM as the input for $T_\text{sky}$ in \autoref{eq:Tsys}. 
   }
   \label{fig:componentmaps}   
\end{figure*} 

\subsection{Final Combined Data Product}\label{sec:data_product}

Our final data sets for the Blind Challenge are composed of the cosmological \HI (\secref{sec:HI}) added to the foreground model (either MS$_{05}$ or PSM - \secref{sec:sky_models}). The maps are then processed by the telescope simulation outlined in \secref{sec:telescope}, which emulates the effects from the particular beam. We add in some instrumental noise specified by the type of telescope and the scanning strategy. The procedure is schematically summarised in  \autoref{fig:flowchart_1}.

Since our PSM model is based on empirical data, it inherently includes a zero-point (or monopole) signal. On the other hand, for the MS$_{05}$ model we have Gaussian realisations with mean zero amplitude. 
To balance this effect, we add an artificial monopole to the MS$_{05}$ model, which is given by the $T_\text{CMB}$ and $T_\text{sky}$ components in \autoref{eq:Tsys}. For the latter, the offset is derived by roughly scaling the mean value of the sky at 408\,MHz \citep{Wehus2017} with the expected synchrotron spectral index at low frequencies \citep[e.g.,][]{Platania1998,Platania2003}.
Whilst this results in some differences between the MS$_{05}$ and PSM models for the monopole amplitude, it is not overly important for our investigation since the monopole is used mostly to fix the total system temperature at each frequency and most of the foreground cleaning methods are not concerned with the monopole level.  

\autoref{fig:componentmaps} shows the final simulated maps of the different components in our combined data product for a frequency of $1000\,\text{MHz}$. The first two panels show the two different foreground models, PSM and MS$_{05}$, respectively. There is more spatial structure in the PSM model as expected, whereas the MS$_{05}$ is uniformly Gaussian distributed. The third panel shows the noise, in this example for the SKAO-MID instrument. Close inspection reveals subtle horizontal stripes of lower noise due to our scanning strategy (i.e., regions observed more often have lower noise). The final panel shows the addition of the cosmological \HI signal: the noise floor is quite high and dominates the small scales, yet we can notice by eye the large scale features of the \hinospace.

\section{Foreground subtraction methods}\label{sec:methods}

The observed temperature maps defined in \autoref{eq:tmodel} can be represented by two-dimensional (frequency and pixel) data-cubes, $\boldsymbol{\mathsf{X}} \equiv  T_{\rm obs}(\nu,\hat{\textbf{n}})$. Most of the cleaning algorithms we use assume that we can linearly decompose the matrix $\boldsymbol{\mathsf{X}}$ in a set of \Nfg sources in pixel space $\boldsymbol{\mathsf{S}}$ modulated in frequency through a mixing matrix $\boldsymbol{\mathsf{A}}$ plus some residuals $\boldsymbol{\mathsf{R}}$ that should in principle contain most of the cosmological signal that we aim to recover together with the white instrumental noise:
\begin{equation}\label{eq:systosolve}
    \boldsymbol{\mathsf{X}}=\boldsymbol{\mathsf{A}}\boldsymbol{\mathsf{S}} + \boldsymbol{\mathsf{R}} \,.
\end{equation}
In practice, we do not expect the above decomposition to hold perfectly for a data-cube, as leakage between the frequency-correlated and uncorrelated parts is unavoidable. The foreground cleaning process boils down to solving \autoref{eq:systosolve} to find $\boldsymbol{\mathsf{R}}$, once the number of sources (foregrounds) is set to \Nfg. More explicitly, using \autoref{eq:tmodel}, for each frequency channel $i$ and pixel $n_p$, we relate the cleaned residual $\boldsymbol{\mathsf{R}}$ to the input signal through:
\begin{equation}
    R_{ip}=\int {\rm d}{\bf \Omega}\left[ B(\nu,n_p,{\bf \Omega})  T_{\rm HI}(\nu_i,n_p)\right]+ T_\text{noise}(\nu_i,n_p)\,.
\end{equation}
The assumptions to be made in order to find the matrix $\boldsymbol{\mathsf{A}}$ and components $\boldsymbol{\mathsf{S}}$ that satisfy \autoref{eq:systosolve}, vary from method to method.

The methods used in this challenge are summarised in 
\autoref{table:methods}; we describe them in more detail in the following sections.

\begin{table*}
\caption{Summary of the nine foreground cleaning pipelines used in this work. See \secref{sec:methods} for details.}
\centering
\begin{tabular}{|c | c| c| c|} 
 \hline
Method & Assumption on & Pipeline &  Brief Description and References\\
& foreground components & & \\
  \hline
  \hline
 Principal Component & Statistically uncorrelated & PCA(a) & Classical PCA with no weighting (see \citet{Cunnington:2020njn})\\ 
Analysis & & PCA(b) & \emph{fg\_rm} code \citep{Alonso:2014dhk}, with inverse \emph{rms} weighting\\ 
 &  & PCAwls & Classical PCA applied on the wavelet-transformed data\\ 
   \hline
Independent Component & Non-Gaussian & \fastica(a) & Based on \texttt{Scikit-learn} package \\  Analysis & & \fastica(b) & \emph{fg\_rm} code \citep{Alonso:2014dhk} \\ 
\hline 
Generalised Morphological  & Sparse in a given domain & GMCA & Sparsity enforced in the wavelet domain (see \citet{Carucci:2020enz})\\ 
Component Analysis & and morphologically diverse & mixGMCA & PCA on the coarse scale $+$ GMCA on small scales \\
\hline 
Polynomial Fitting & Smooth in frequency & poLOG  & In log-log space \citep[\emph{fg\_rm} code]{Alonso:2014dhk} \\
\hline 
Parametric Fitting & Assumptions on spectral indices & LSQ & Fit to individual foregrounds\\
\hline 
\hline 
\end{tabular}
\label{table:methods}
\end{table*}

\subsection{PCA} \label{sec:PCA}

Principal Component Analysis (PCA) can be used to identify an estimate for the mixing matrix $\boldsymbol{\mathsf{A}}$, the columns of which will be given by the first $N_\text{fg}$ principal components. The principal components are essentially the eigenvectors of the mean-centred data $\nu\nu'$ covariance matrix $\boldsymbol{\mathsf{C}}$, given by
\begin{equation}\label{eq:InverseNoiseCov}
    C_{ij}=\frac1{N_{\hat n}}\sum^{N_{\hat n}}_{p=1} w_i \Delta T(\nu_i,n_p)~ w_j \Delta T(\nu_j,n_p)\,,
\end{equation}

where $\Delta T(\nu_i,n_p) = T(\nu_i,n_p)-\bar T(\nu_i)$ and the summation is over all $N_{\hat n}$ pixels . The $w$ factors provide an optional map weighting.
The eigendecomposition is given by $\boldsymbol{\mathsf{C}}\boldsymbol{\mathsf{V}} = \boldsymbol{\mathsf{V}}\mathbf{\Lambda}$, where $\mathbf{\Lambda}$ is the diagonal matrix of $N_\nu$ eigenvalues. The first $N_\text{fg}$ columns from the eigenvector matrix $\boldsymbol{\mathsf{V}}$ represent the entries for the mixing matrix. 
Computing the covariance matrix is useful since the magnitudes of its eigenvalues $\mathbf{\Lambda}$ offer some guidance on how many principal components to include, i.e., the choice of $N_\text{fg}$ (see for example \autoref{fig:eigen} that we will comment in the discussion in \secref{sec:discussion}). In brief, since we know that foregrounds have undoubtedly 
higher amplitude and higher variance than the cosmological signal, we expect them to be well characterised by the first few principal eigenvalues and eigenvectors.

As summarised in \autoref{table:methods}, in this Challenge we use three PCA  implementations: PCA(a), PCA(b) and PCAwls. PCA(a) uses a straightforward implementation of the process described in this section, with no weighting  ($w_i=w_j=1$), replicating the pipeline used in \citet{Cunnington:2020njn}. PCA(b) uses the publicly available code \textit{fg\_rm}\footnote{\href{https://github.com/damonge/fg_rm}{https://github.com/damonge/fg\_rm}} (see \citet{Alonso:2014dhk}) with the implemented inverse noise weighting, i.e., we use the root mean square (rms) of the map at each frequency:
\begin{equation}\label{eq:weights}
    w_i=\frac1{\sigma_i}\,,\quad \sigma_i=\sqrt{\frac1{N_{\hat n}} \sum^{N_{\hat n}}_{p=1}\Delta T(\nu_i, n_p)^2}\,,
\end{equation}
designed to minimise the influence of noise on the identification of dominant foreground modes.
Lastly, PCAwls is an implementation of PCA on wavelet-transformed data with no weights and will be described in \secref{sec:gmca}.

\subsection{\fastica}
\label{sec:fastica}

Fast Independent Component Analysis (\fastica) is a widely used method developed in \citet{Hyvrinen1999FastAR} and employed for foreground cleaning on simulated \HI data \citep{Chapman:2012yj,Wolz:2013wna, Cunnington:2019lvb,Carucci:2020enz} as well as real data \citep{Wolz:2015lwa,Wolz:2021ofa, Hothi2021}. 

\fastica\ estimates the mixing matrix $\boldsymbol{\mathsf{A}}$ by assuming the sources are statistically independent of each other. The method, therefore, aims to maximise statistical independence that can be assessed using the central limit theorem, which states that the greater the number of independent variables in a distribution, the more Gaussian that distribution will be (that is, the probability density function of several independent variables is always more Gaussian than that of a single variable). Hence, by maximising any statistical quantity that measures non-Gaussianity, we can identify statistical independence.

Before assessing non-Gaussianity, \fastica\ begins by mean-centring the data, then carries out a \textit{whitening} step that aims to achieve a covariance matrix equal to the identity matrix for this whitened data (i.e., the components will be uncorrelated and their variances normalised to unity). Since this whitening step can be achieved with a PCA analysis, \fastica\ is essentially an extension of PCA, and hence in most cases in the context of foreground cleaning, will provide very similar results.

For maximising non-Gaussianity, an approximation of the negentropy is used. In the context of 21-cm foreground cleaning, the approximation of negentropy uses a set of optimally chosen non-quadratic functions which are applied to the data and averaged over for all available pixels. The maximisation of negentropy by averaging over angular pixels means that for purely Gaussian sources, \fastica\ will be unable to improve upon the initial PCA step carried out in the whitening step due to Gaussian sources having an equivalent zero negentropy. This explains the similarity in results often found between PCA and \fastica\ when most of the simulated components are Gaussian fields \citep{Alonso:2014dhk,Cunnington:2020njn}. 

As summarised in \autoref{table:methods}, in this Challenge we use two \fastica\,  implementations: \fastica(a) and \fastica(b). The \fastica(a) pipeline uses the \fastica\ module in \texttt{Scikit-learn}\footnote{\href{https://scikit-learn.org/stable/modules/generated/sklearn.decomposition.FastICA.html?highlight=fastica}{https://scikit-learn.org/}} \citep{Pedregosa:2012toh}. \fastica(b) uses the public  \emph{fg\_rm} code \citep{Alonso:2014dhk}. Despite the fact that the two implementations use different codes to apply the same \fastica\ methodology, their differences lie on pre-processing choices of input data and the choice of number of modes to remove (see \autoref{fig:hist_Nfg}).

\subsection{GMCA and Wavelet Decomposition}
\label{sec:gmca}

Generalised Morphological Component Analysis (GMCA) is a blind component separation method based on sparsity \citep{gmca1}. It assumes that the $N_\text{fg}$ foreground components verify two hypotheses: they are sparse in a given transformed domain (i.e., most samples are zero-valued) and their supports are disjoint; in other words, the foreground components are {\it morphologically} diverse (i.e., their non-zero samples appear at different locations). GMCA has been successfully applied in various astrophysical contexts (e.g., Cosmic Microwave Background data \citep{bobin2013,bobin2014}, high-redshift 21-cm interferometry \citep{2013MNRAS.429..165C,Patil2017}, X-ray images of Supernova remnants \citep{Picquenot2019}, gravitational waves \citep{Blelly2020}).

\citet{Carucci:2020enz} showed the wavelet domain to be optimal to sparsely describe foregrounds and contaminants in the low-$z$ \HI IM context. Firstly, we project the data $\boldsymbol{\mathsf{X}}$ onto wavelet space. The GMCA algorithm aims at minimising the following cost function:
\begin{equation} \label{eq:GMCAmaster}
   \min_{\boldsymbol{\mathsf{A}}, \boldsymbol{\mathsf{S}}}  \sum_{i=1} ^{N_\text{fg}}  \lambda_i \left\lvert \left\lvert S_{i} \right\rvert \right\rvert_1+ \left\lvert \left\lvert \boldsymbol{\mathsf{X}} - \boldsymbol{\mathsf{A}}\boldsymbol{\mathsf{S}} \right\rvert \right\rvert_{2}, 
\end{equation}
where the first term is the $\ell_1$ norm, i.e. $\sum_{j,k} \left\lvert  S_{j,k} \right\rvert$: this constitutes a constraint for sparsity, mediated by the regularisation coefficients $\lambda_i$. The second term is an usual data-fidelity $\ell_2$ norm term. We find solutions for $\boldsymbol{\mathsf{A}}$ and $\boldsymbol{\mathsf{S}}$ by iterating a projected alternate least-squares procedure: we fix $\boldsymbol{\mathsf{A}}$ and perform a least-squares update to determine $\boldsymbol{\mathsf{S}}$, we compute the thresholds $\lambda_i$ via mean absolute deviation of $S_i$, we update $\boldsymbol{\mathsf{A}}$ with $\boldsymbol{\mathsf{S}}$ fixed and so on. The key point is the thresholding: it allows us to keep the samples with the highest amplitudes, which are the most informative to retrieve the mixing matrix $\boldsymbol{\mathsf{A}}$ (i.e., they most likely belong to the foreground components and are the least likely to be contaminated by the cosmological signal and noise), and it provides robustness in terms of convergence since the thresholds decrease with the progressive iterations.

In the Challenge described in this work, we decided to test three different cleaning methods based on wavelet decomposition and GMCA.
\begin{enumerate}
    \item {\bf PCAwls}. We perform a PCA decomposition as described in \secref{sec:PCA} on the wavelet-transformed data. We expect it to be equivalent to PCA in standard pixel-space as the PCA algorithm does not depend on the domain in which data is described. The purpose of using PCAwls has been to add an extra set of solutions with the PCA method, i.e., a different participant using a different pipeline and choosing a different number of components $N_\text{fg}$ to remove (see later \autoref{fig:hist_Nfg} for a summary of the $N_\text{fg}$ choices).
    
    \item {\bf GMCA}. We apply GMCA as it is described above and by \citet{Carucci:2020enz}.
    
    \item {\bf mixGMCA}. We apply PCA on the largest scale of the wavelet-transformed data and GMCA on the remaining scales. By largest scale, we mean the coarse approximation of the maps resulting from the initial low-pass filtering of the wavelet decomposition \citep{JLbook}. We assemble the two solutions back together before re-transforming the maps into pixel-space. This allows to have two different mixing matrices $\boldsymbol{\mathsf{A}}$ and two different numbers $N_\text{fg}$ of components for the small and the large spatial scales of maps.
\end{enumerate}
Ongoing work on optimising the GMCA method in the \HI IM context resulted in the development of mixGMCA, which we use here for the first time in the literature. \citet{Carucci:2020enz} highlighted the need of having a different number of components for different spatial scales, and \citet{Cunnington:2020njn} highlighted how, in the IM context, the sparse assumption might not suit the largest scales, yet it holds well in the small ones. Analysis of LOFAR observations also supports the idea of having \Nfg dependent on scale \citep{Hothi2021}. The wavelet decomposition offers a straightforward framework for analysing multi-scale data. With mixGMCA, we further developed this idea by allowing different mixing matrices to describe the data-cube at different scales.

\subsection{Logarithmic Polynomial Fitting}
One of the first approaches to foreground subtraction methods is to come up with a base of smooth functions in frequency which we can then use to model the foregrounds. This has been extensively used \citep{2006ApJ...650..529W,Ghosh2011_oscillations,2012A&A...540A.129A,wang2013}, and here we follow the approach of \citet{Alonso:2014dhk}
and perform a power-law base expansion in log-log space. In particular, we will use polynomials of the logarithm of the frequency, i.e.,
\begin{equation}
    \log T_{\rm fg}(\nu,\hat{\textbf{n}})=\sum_{n=1}^{N_{\rm fg}}\alpha_n(\hat{\textbf{n}}) [\log(\nu)]^{n-1}\,.
\end{equation}
We then solve the log-log space equation equivalent to \autoref{eq:systosolve}. For this purpose we used the code \textit{fg\_rm} \citep{Alonso:2014dhk} with the frequency logarithm polynomials, and we also weight the data using the rms (see \autoref{eq:weights}) translated into logarithmic space, $\sigma_{i,\log T}\simeq\sigma_i/T$. In this Challenge, we set \Nfg$=6$. We refer to this pipeline as poLOG.

\subsection{Parametric Fitting}
\label{sec:paramfit}

Parametric methods, unlike blind component separation, assume that a considerable portion of the measured total signal is well-known due to prior empirical knowledge. Specifically,
we could make the following assumptions: 
\begin{itemize}
    \item Diffuse synchrotron and free-free emission are non-negligible at MHz frequencies, with synchrotron emission dominating at high Galactic latitudes, as indicated by the numerous ground-based surveys collated for use by the Global Sky Model \citep{gsm}.
    
    \item Diffuse free-free emission has a spatially constant spectral index which can also be considered spectrally constant over our 500\,MHz frequency range \citep{ben92}.
    
    \item Free-free emission, synchrotron emission, and the extragalactic point source contributions are heavily degenerate with each other due to their similar spectral forms (power laws with similar indices as in \autoref{fig:sforms}) \citep{plancklow}.
\end{itemize}

Here we fit for the diffuse emissions only: free-free and synchrotron. We attempt to use the foreground degeneracy to our advantage by trialling the assumption that the extragalactic temperature contribution will be absorbed into either our estimate of Galactic synchrotron emission or our estimate of Galactic free-free emission or both. 

For our parametric fit we require the zero-level at each frequency map to be set solely by the diffuse foreground emissions we intend to fit: no additional temperature contributions can be present. Zero-level contributions can include 1) the CMB monopole, which is both spatially constant and constant across frequency and hence easy to subtract; 2) the receiver temperature, which we subtract under the assumption that this component can be measured by each experiment e.g., \citet{Wang2021} and 3) the average temperature of all the unresolved extragalactic point sources. Regarding the latter contribution, values for these averages at various frequencies are available in the literature (e.g., \citet{gervasi,mdeep}); hence, we decide to subtract the true value for this average (i.e., the fiducial value used in our simulation) from the total temperatures at each frequency before beginning our fit. 

We aim to determine the true $\boldsymbol{\mathsf{A}}$ in \autoref{eq:systosolve} for the combination of free-free and synchrotron emission; for this we require both the synchrotron and free-free emission spectral index per pixel. For free-free emission we use the true (i.e., the fiducial value used in our simulation) value of -2.1 at each map pixel. 

We find that it is optimum to first obtain the synchrotron spectral index from the total temperature data assuming that the free-free contribution is negligible. This works in practice by performing a least-squares fit at each pixel using the \texttt{python} module \texttt{lmfit} with two free parameters: the amplitude and spectral index of synchrotron emission. The parameter space of the synchrotron spectral index is restricted to within $10$ per cent of the total temperature spectral index across the first three frequencies. We weight our fit using the FFP10 free-free emission map smoothed to $1.5\,\deg$ and scaled to each frequency as an estimate for noise. Having fitted for the synchrotron spectral index at each pixel $\beta_{\rm{sy}}(\hat{\textbf{n}})$ our mixing matrix estimate can then be expressed as: 
\begin{equation}
\tilde{\boldsymbol{\mathsf{A}}} = \left( \begin{array}{c}
(\nu/\nu_{0})^{\beta_{\rm{sy}}(\hat{\textbf{n}})} \\
(\nu/\nu_{0})^{-2.1} \end{array} \right) \, .
\end{equation}
The matrix of emission amplitudes ($\boldsymbol{\mathsf{S}}$) is computed by again minimising the standard least-squares problem: 
\begin{equation}
\label{eq:inv}
\boldsymbol{\mathsf{S}} = {(\tilde{\boldsymbol{\mathsf{A}}}^{T} \tilde{\boldsymbol{\mathsf{A}}})}^{-1} \tilde{\boldsymbol{\mathsf{A}}}^{T} \boldsymbol{\mathsf{X}} \, ,
\end{equation}
where $\boldsymbol{\mathsf{X}}$ are the total temperature data. Any components of the total data that can be characterised by a power law with spectral indices similar to the range of the indices within our mixing matrix estimate will be grouped together as foregrounds. We present the residual between the total data and our estimated combined foregrounds at each frequency as an estimate for \HI emission plus noise.
We refer to this pipeline as LSQ.

\begin{figure}
   \centering
   \includegraphics[width=\columnwidth]{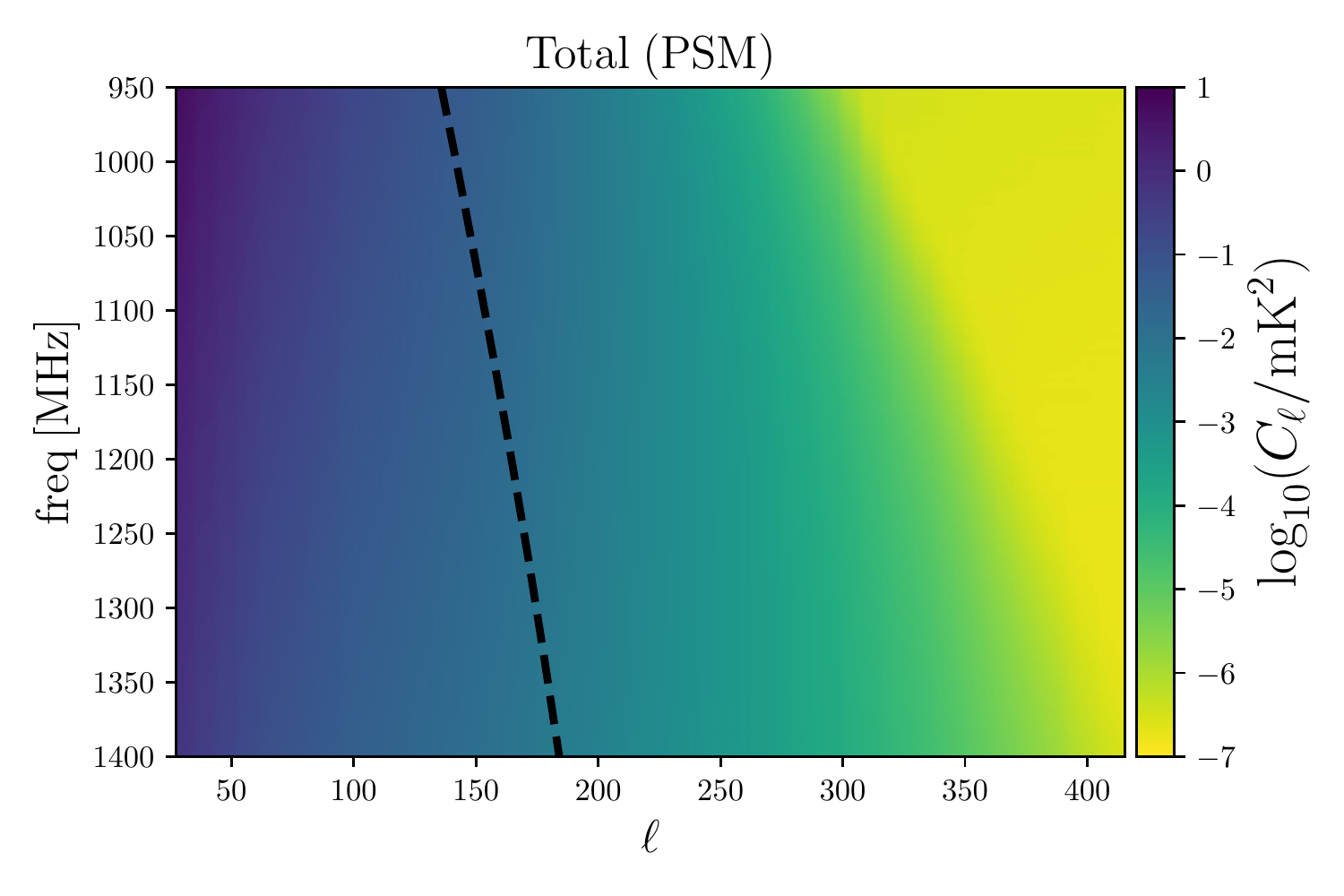}\\
   \includegraphics[width=\columnwidth]{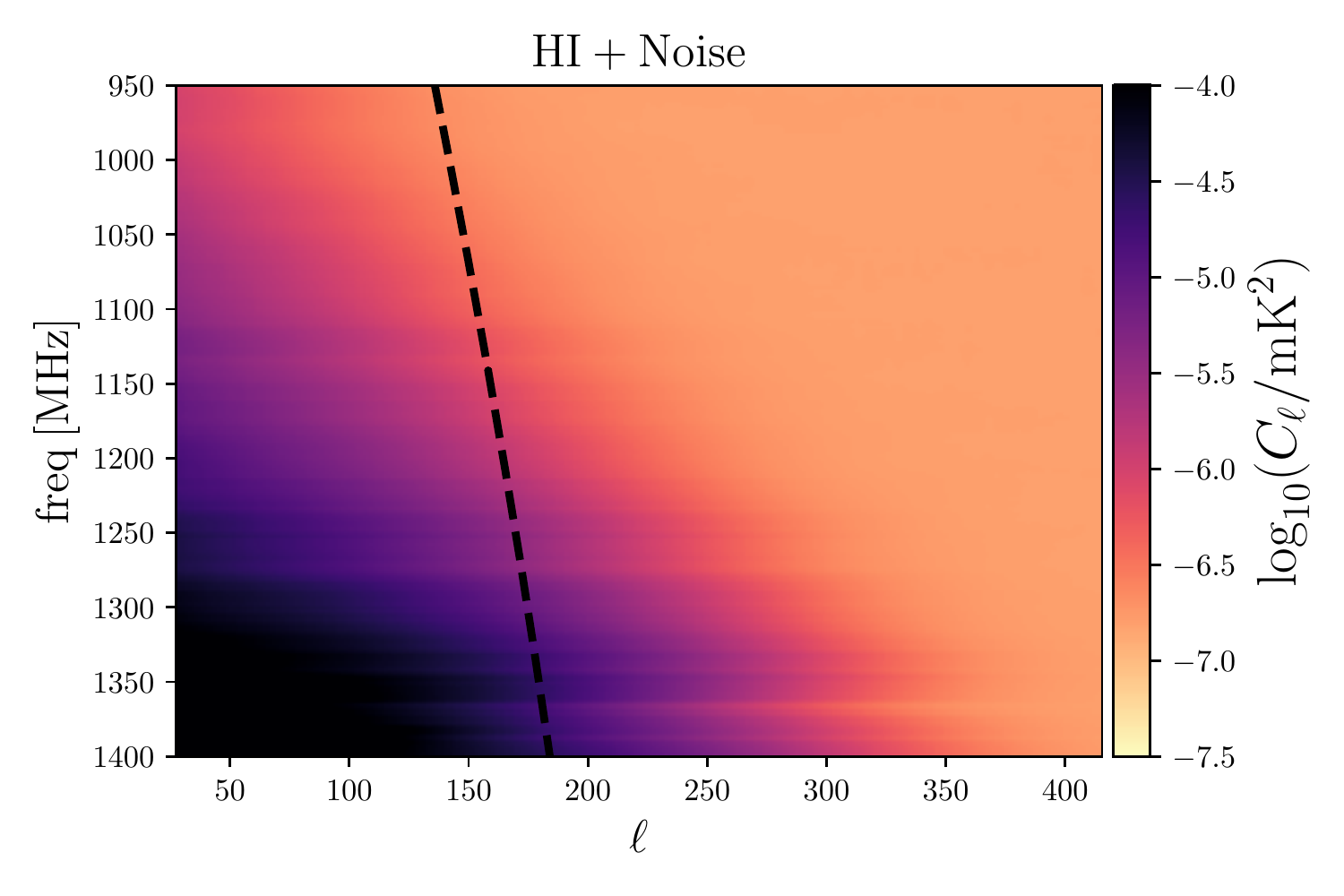} 
   \caption{The angular power spectrum $C_\ell$ of the total sky emission considering the PSM foregrounds (upper panel) and the \HI signal plus noise (lower panel) as a function of frequency, considering an SKAO-MID instrument and the Airy beam model. The black dashed line in both panels traces the evolution with frequency of the angular scale of the FWHM of the telescope beam.
   }
   \label{fig:cl_input}   
\end{figure}

\begin{figure*}
   \centering
   \includegraphics[width=17cm]{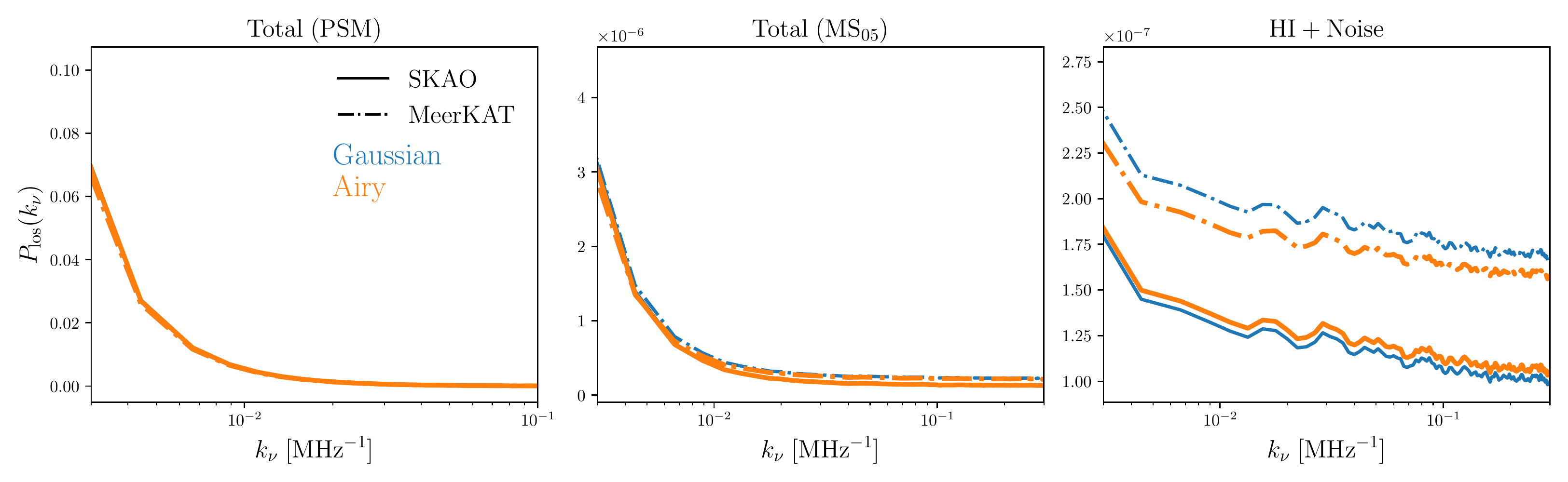}
   \caption{
   The line-of-sight power spectrum $P_{\rm los}$ for the total sky emission considering the PSM foregrounds (left) or the MS$_{05}$ model (centre), compared with the \HI signal plus the noise (right), for both an SKAO-MID-like survey (solid lines) and a MeerKAT-like survey (dot-dashed lines). We show results after convolution with a Gaussian beam model (in blue) and the Airy beam model (in orange).}
   \label{fig:ps_input}   
\end{figure*}

\section{Summary statistics}\label{sec:estimators}

The analysis of the simulations and the quality assessment for the residual maps after cleaning require estimators to compress the three-dimensional information contained in the data-cubes.
Although some studies have started to explore observational effects in higher-order statistics \citep{Cunnington:2021czb, 2021JCAP...06..039J}, in this work, we focus on $2$-point summary statistics, looking both at the angular and line-of-sight directions, keeping the two separated to distinguish features that could show up independently in each direction. In particular, we compute the angular power spectrum as a function of frequency $C_\ell(\nu)$ (\secref{sec:estcl}) and the one-dimensional line-of-sight power spectrum $P_{\rm los}(k_{\nu})$ (\secref{sec:estpk}). These choices relieve us from making extra assumptions (e.g., flat-sky approximation and thin-channel assumption to translate observed frequencies into distances).
The same summary statistics are computed for the residual maps and the input signal plus noise, allowing a straightforward quantitative estimation of the performance of the various methods.
We acknowledge that, by comparing the statistics, we can not properly discriminate between a true reconstructed signal or a contribution of leaked foregrounds with a resulting power spectrum similar to the input signal. To this end, other strategies --although with different caveats-- could be used,  such as the cross-correlation of the residuals maps with the input signal and, for cleaning methods that involve the construction of a mixing matrix, the estimation of the leakage through appropriate projections of such matrix \citep[e.g.,][]{Carucci:2020enz}.
The direct comparison of summary statistics offers a simple and efficient way to test all different cleaning methods; moreover, the auto-spectra of the recovered maps represent the final product of observations before the cosmological analysis.
Therefore, in this work, we rely on these statistics and their comparison with the input counterparts.

For future work, where we plan to assess the cosmological content of the cleaned maps,  
a proper error estimation of the reconstructed 2-point statistics and covariance analysis will be crucial. In this analysis, we have roughly estimated the uncertainties on these statistics both using jackknife and theoretical errors and found that prominent features of the various methods persist even considering these estimated uncertainties. This implies that enough meaningful comparison of the cleaning methods can be achieved even without the errors, and we thus postpone a detailed analysis of uncertainties to a follow-up project.

\subsection{Angular Power Spectrum}\label{sec:estcl}
At a given frequency, the simulated sky patch has been constructed as a HEALPix map and can be decomposed in spherical harmonics.
For the full sky case, the angular power spectrum can be estimated from the spherical harmonic coefficient $a_{\ell m}(\nu)$ of this decomposition,
\begin{equation}\label{eq:cl}
    \hat{C}_{\ell}(\nu) \equiv \frac{1}{2\ell+1}\sum_{m=-\ell}^{m=+\ell} |a_{\ell m}|^2 \,.
\end{equation}
This estimator is no longer valid for sky patches, but can be corrected, in first approximation, by dividing by the sky fraction covered by the patch. 

However, in the presence of sharp edges, such as the ones caused by the single-dish scanning strategy assumed here (see \autoref{fig:componentmaps}), the coupling induced by the mask can be important, and should be corrected for.
One efficient and commonly used solution is the 
Monte Carlo Apodized Spherical Transform Estimator \citep[MASTER,][]{hivon2002}. In this work, we compute this correction using  the \texttt{NaMaster} software\footnote{\url{https://github.com/LSSTDESC/NaMaster}} \citep{NaMaster}. 
Although \texttt{NaMaster} has been optimised to deal with partial sky coverage, a complete validation of the result would require further studies and possibly a refinement of the final patch footprint. For the purpose of this work, whose main intent is to compare performances of different foreground methods, there is no such concern.

We present in the upper panel of \autoref{fig:cl_input} the evolution of the angular power spectrum as a function of frequency for one of the final data-cubes, where the more realistic PSM foregrounds are considered. Dominated by the smooth foreground emission, the $C_\ell$ shows the effect of the beam suppressing signal at progressively larger scales going to lower frequencies (for reference, the black dashed line corresponds to the beam FWHM). As expected, the emission is stronger at lower frequencies. We show results for the SKAO-MID case, which we find similar to the MeerKAT case.

The lower panel of \autoref{fig:cl_input} shows the \HI cosmic signal plus noise we are aiming to recover, orders of magnitude fainter than the astrophysical foreground emission. In this case, it is not only the beam effect that dictates the amplitude of the power spectrum, but also the interplay between the structured \HI signal (whose intensity fades at lower frequencies) and the instrumental noise (that increases at lower frequencies). Indeed, at small scales, we can recognise the (quasi) scale-invariant noise floor at $\sim 10^{-7}\,\text{mK}^{-2}$ covering the structure of the signal (differently at different channels), confirming what shown in \autoref{fig:componentmaps}.

\subsection{Radial Power Spectrum}\label{sec:estpk}

We compute the one-dimensional power spectrum directly in frequency space, $P_{\rm los}(k_{\nu})$ with  $k_\nu = 1/ \nu$. It is the most straightforward choice to investigate how well the radial information is recovered \citep{Alonso:2014dhk,Villaescusa2017}.
Here, we follow the procedure described in \citet{Carucci:2020enz}. In short, for each pixel --i.e., line-of-sight-- we Fourier transform the temperature along the frequency direction $\tilde{T}=\mathcal{F}[T(\nu)]$, and we compute $P_{\rm los}(k_{\nu})$ by averaging over the power spectra from each pixel $n_p$:
\begin{equation}\label{eq:plos}
    P_{\rm los}(k_{\nu}) = \Delta \nu \langle |\tilde{T}(k_{\nu},n_p)|^2 \rangle_{N_{\hat n}}\:.
\end{equation}
We expect the smooth foregrounds, that are strongly correlated in frequency, to display more power at small $k_\nu$. We can see from \autoref{fig:ps_input} that this is indeed the case, for both the more realistic PSM foregrounds and for the MS$_{05}$ model. The effect of the different instrumental response is very small when looking at the total sky signal in the first two panels, whereas we can clearly see the offset between the SKAO-MID and MeerKAT cases for the cosmic signal plus noise $P_{\rm los}$, caused by the different noise levels and beam models in the right panel.
The amplitude of $P_{\rm los}$ for the MS$_{05}$ model is lower than the PSM one which is intrinsic to the MS$_{05}$ model construction that simply adds a small mean-centred, Gaussian oscillation on top of $T_{\rm sys}$ (see \secref{sec:MS}).

\section{The Blind Challenge}\label{sec:blind}

\begin{figure*}
   \centering
   \includegraphics[width=\textwidth]{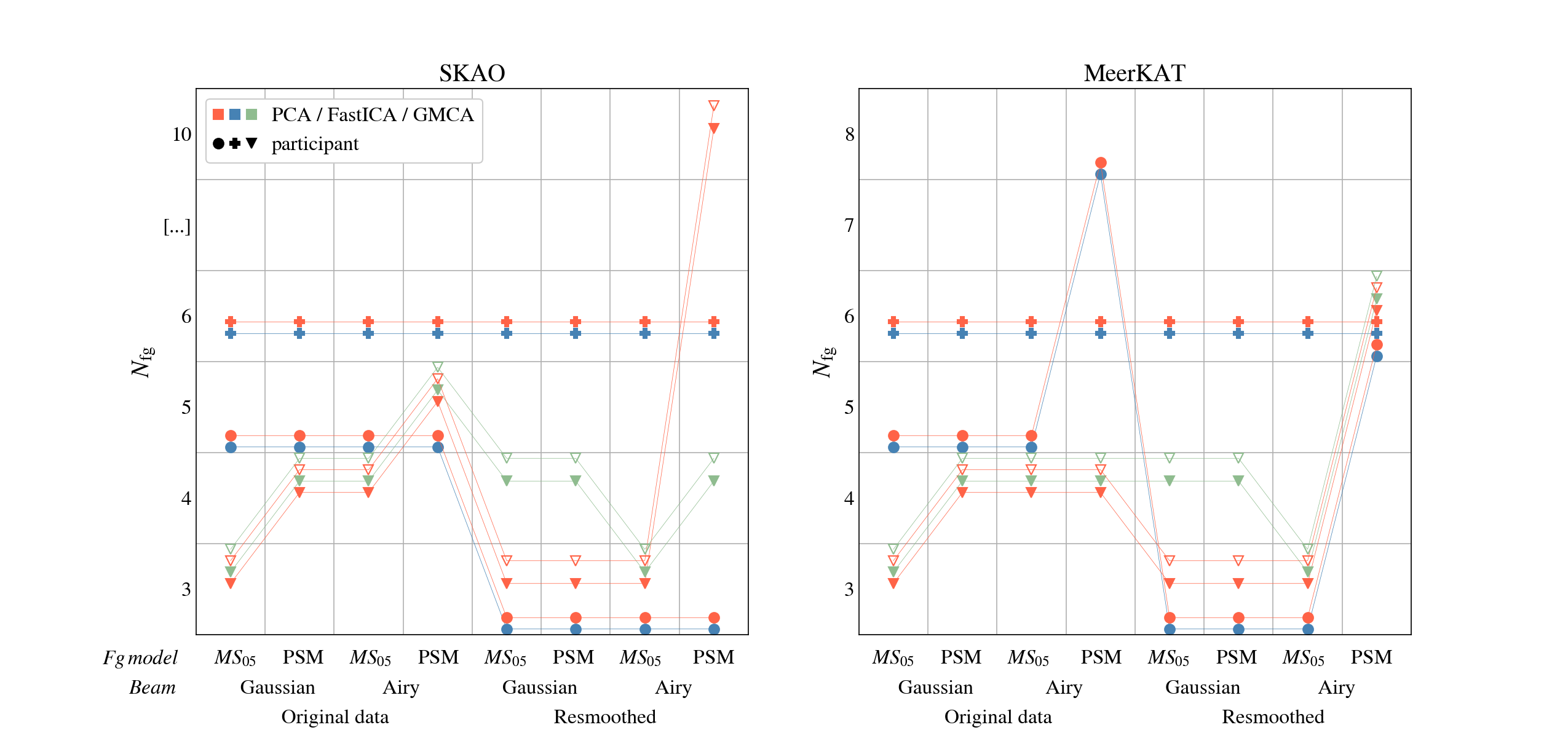}
   \caption{Graphical representation of the \Nfg used for the various cleaning methods as reported in \autoref{table:Nfg}, shown separately for the case of SKAO (left panel) and MeerKAT (right). Each row represents the integer number $N_\text{fg}$ used in each method, with the lines offset to facilitate reading. Each column refers to a specific data set as described by the legend below the $x$-axis. Different colours correspond to different cleaning methods, different symbols to the different participants who performed the cleaning (i.e., different pipeline too). In the case of mixGMCA, two values of \Nfg need to be set: for a PCA run at the large scale and a GMCA run at small scales; we decompose the mixGMCA information into two PCA/GMCA cases that we highlight using empty symbols.}
   \label{fig:hist_Nfg}
\end{figure*}

\begin{figure*}
   \centering
   \includegraphics[width=1.6\columnwidth]{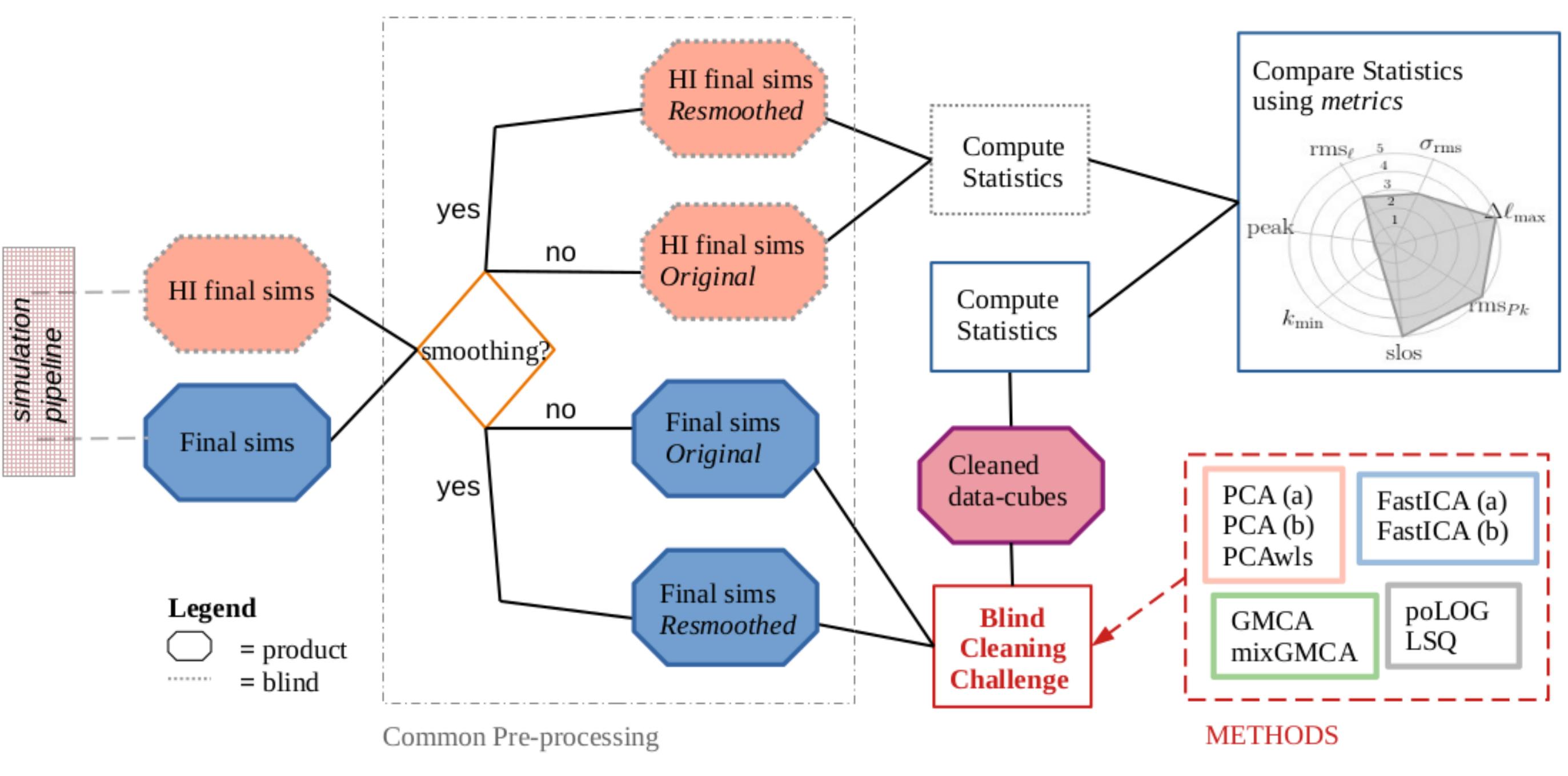} 
   \caption{Flowchart of the analysis procedure. The input simulations (see \autoref{fig:flowchart_1}) are pre-processed as described in \secref{sec:preproc}; we then compute the summary statistics described in \secref{sec:estimators} for all data-cubes and compare the maps recovered by different pipelines through the metrics defined in \secref{sec:spider}. Finally, results are compressed in the radar charts of \autoref{fig:spiderSKAO} and \autoref{fig:spiderMK}.}
   \label{fig:flowchart_2}   
\end{figure*}

In this section, we describe the procedure of the Blind Foreground Subtraction Challenge. 
This type of approach is increasingly adopted in cosmological studies \citep[e.g.,][]{Kitching_2013,Nishimichi_2020} and is a useful and transparent test for the maturity of analysis pipelines \citep{Prat:2021uzb}. 
In this work, both the simulation of the \HI signal and the details of the assembly of the components' maps (including beam convolution and addition of instrumental noise) have been kept \textit{blind} to the participants that attempted the foreground cleaning.

The final data-cubes, summarised in \secref{sec:data_product}, can thus be effectively treated as mock observations. A common pre-cleaning processing is described in \secref{sec:preproc}, while the details of the blind challenge procedure are presented in \secref{sec:blindcleaning}.

\subsection{Common Pre-Processing}\label{sec:preproc}

For a diffraction limited antenna, the FWHM of the beam pattern is proportional to the dish size and the observing frequency, resulting in a variable resolution in frequency across the data-cubes. Real data analyses 
have found it useful to counteract this effect by \textit{resmoothing} the maps \citep{Switzer:2015ria,Wolz:2021ofa}, i.e., by convolving them to a common FWHM (often $10-20$ per cent lower than the one of the lowest frequency). 
To test the advantage of the resmoothing, 
we opt for two approaches: 1) cleaning the data-cubes at the native channel-dependent resolutions; 
2) resmoothing all maps of the data-cube to a common resolution.
We thus created an extra set of resmoothed data-cubes where all maps have been deconvolved to a Gaussian beam with FWHM equal to $1.05$ times the FWHM of the lowest frequency channel. The resmoothing Gaussian kernel is defined as
\begin{equation}\label{eq:recv}
    a_{\ell m}^{\rm \textit{Res}}(\nu) = a_{\ell m}^{Or}(\nu) e^{-(\ell \left(\ell + 1\right))/(16\ln(2)) \left[\theta_R^2 - \theta^2(\nu)\right] },
\end{equation}
where $\theta_R$ is the FWHM to convolve the data to, and $\theta(\nu)$ is the FWHM at frequency $\nu$ (see \secref{sec:beam}).

Because of the border effects of the Gaussian smoothing, we had to define a new (smaller) footprint, going roughly from a coverage of $11$ per cent of the sky to $10$ per cent. Moreover, because the SKAO-MID and MeerKAT beams are different, these new footprints are also (slightly) different for the two instrumental setups. 
To avoid the inclusion of a different footprint in the comparison of the results, the final footprint created for the resmoothed case has been used on the original data-cubes too.

The combinations of two foreground models, two beam models, two instrumental setups, and frequency-dependent vs constant angular resolution, resulted in having a total of 16 different input data-cubes to analyse.

\subsection{Blind Cleaning}\label{sec:blindcleaning}

The cleaning of the various data-cubes has been performed with the nine pipelines summarised in \autoref{table:methods}. As discussed in section \secref{sec:preproc}, for each pipeline, sixteen residual data-cubes (expected to contain only the \HI signal and the noise) have been submitted. 
Most of the used cleaning methods are blind source separation techniques (PCA, \fastica, GMCA and mixGMCA), where the only assumption is related to a statistical property of foregrounds (e.g., non-Gaussianity, sparsity); poLOG explicitly assumes frequency smoothness of the foreground emission while LSQ tries to reconstruct known properties of their emission\footnote{Unlike the other methods, LSQ requires prior information on the map monopole and it 
could only be run on the PSM foreground model cases as it relies upon the foreground spectral forms each being known well enough to be parameterised.}.

For the blind methods, each participant was free to choose the number of components \Nfg to subtract. 
The variety of choices made for the different data-cubes by the various participants are presented in \autoref{fig:hist_Nfg} and reported further in \autoref{table:Nfg}. We separate the SKAO-MID and the MeerKAT cases, and the type of data-cubes, specifying the foreground and the beam models used and the original or resmoothed scenarios. \autoref{fig:hist_Nfg} highlights the difficulty 
and subjectivity in
choosing $N_{\rm fg}$, especially facing increasingly realistic sky mocks.

\medskip
A summary diagram of the procedure is reported in \autoref{fig:flowchart_2}, together with the subsequent steps for the analysis and comparison of the results, detailed in the next section.

\section{Results}\label{sec:results}

\begin{figure*}
  \centering
  \includegraphics[width=\columnwidth]{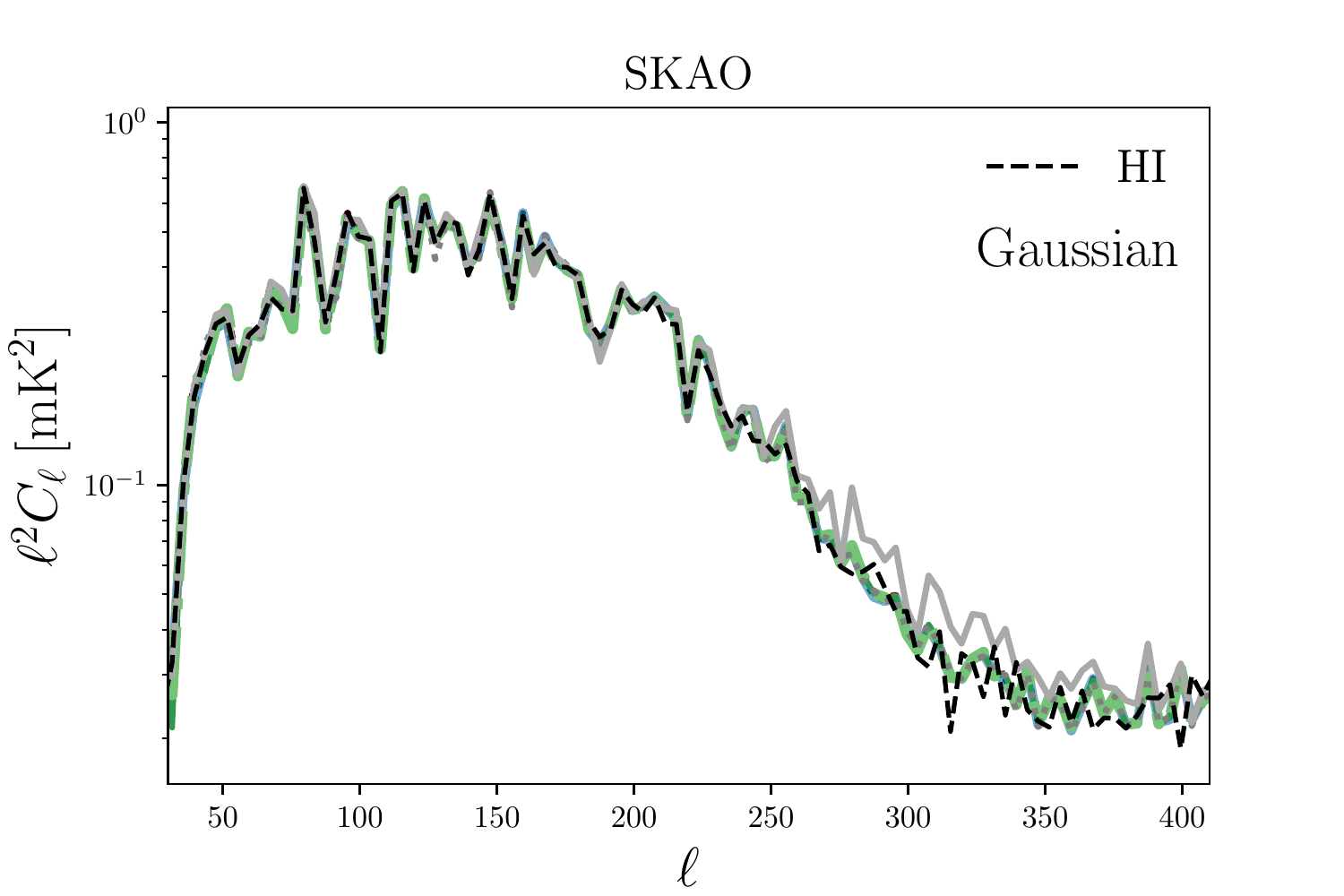}       \includegraphics[width=\columnwidth]{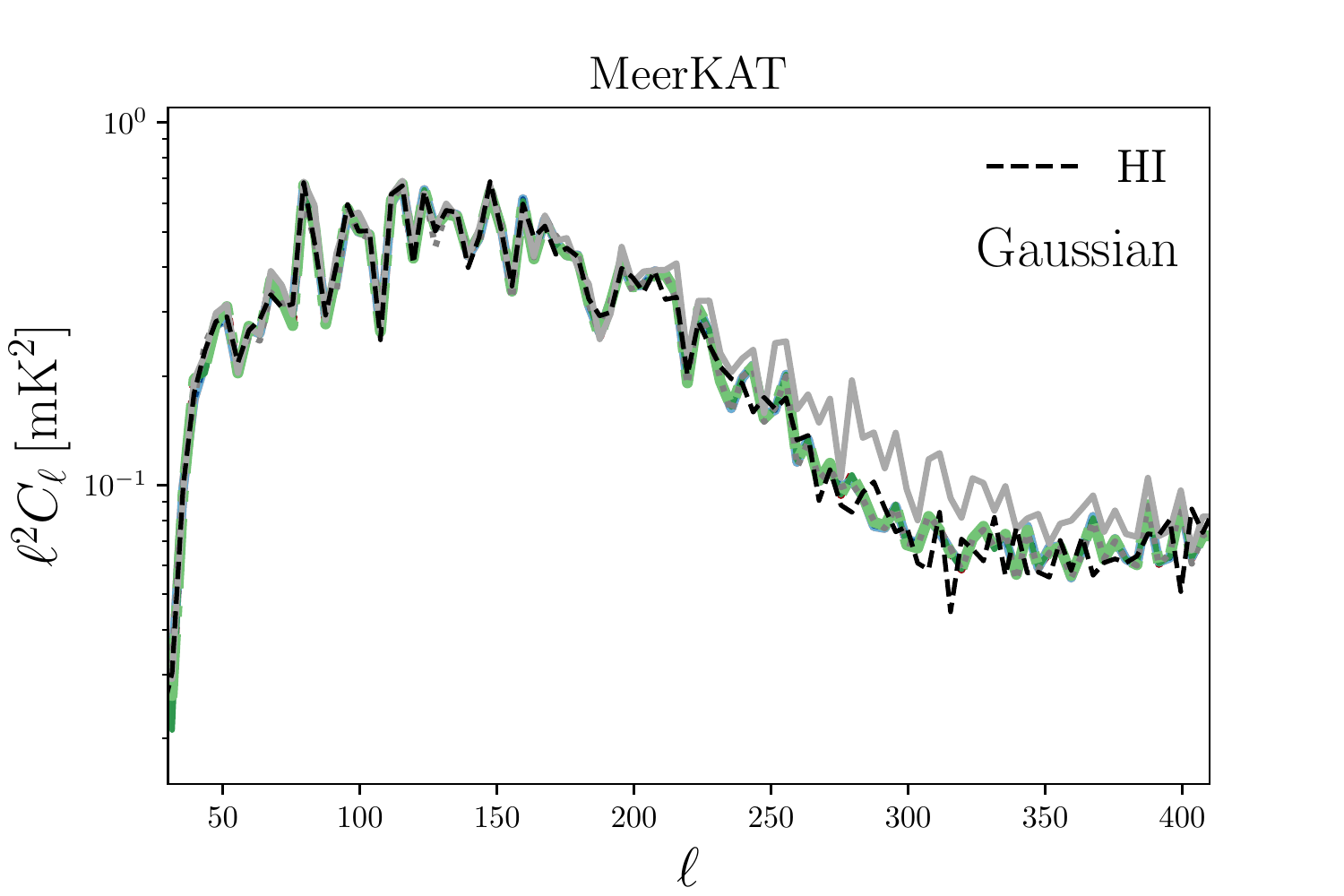}\\
  \includegraphics[width=\columnwidth]{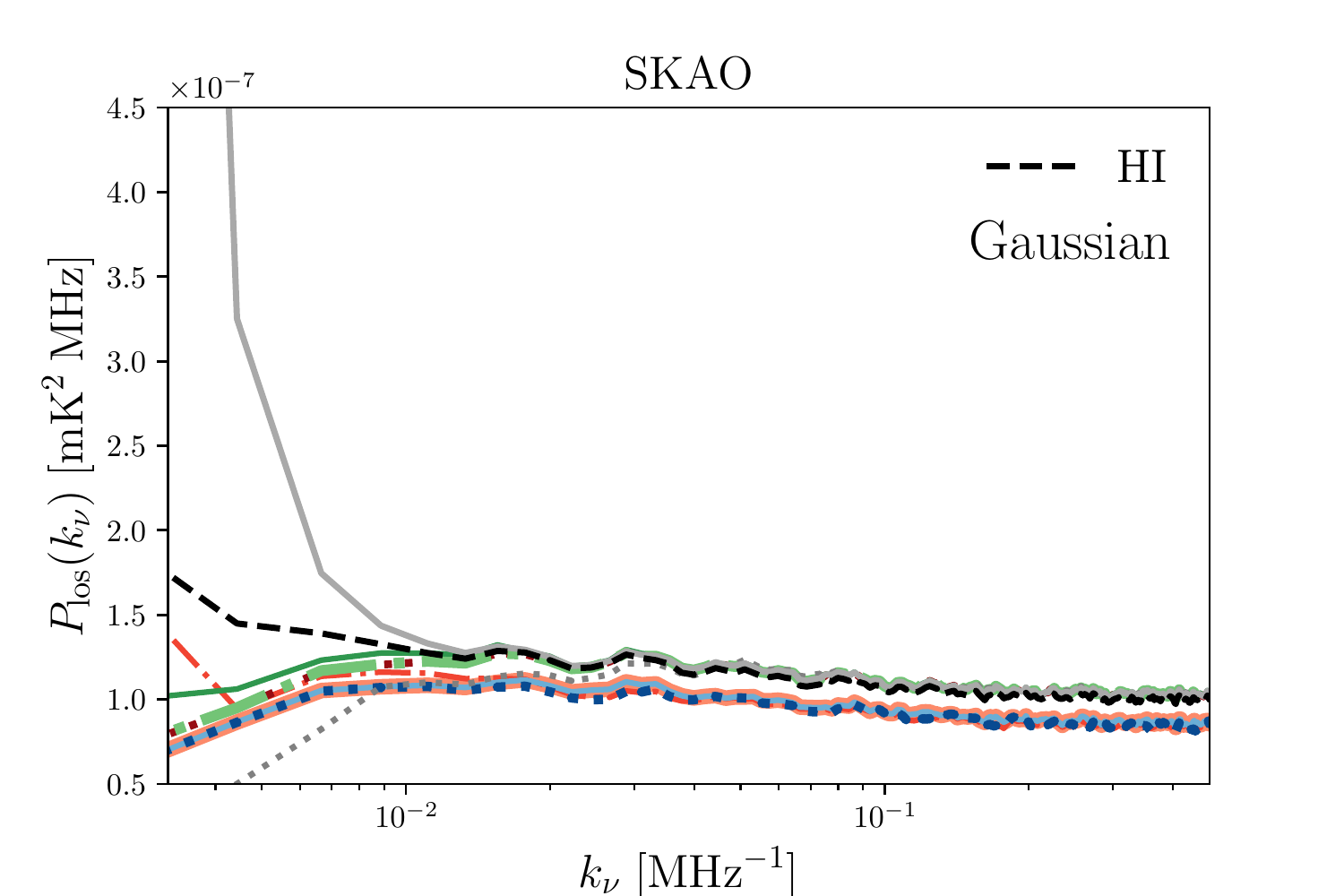} \includegraphics[width=\columnwidth]{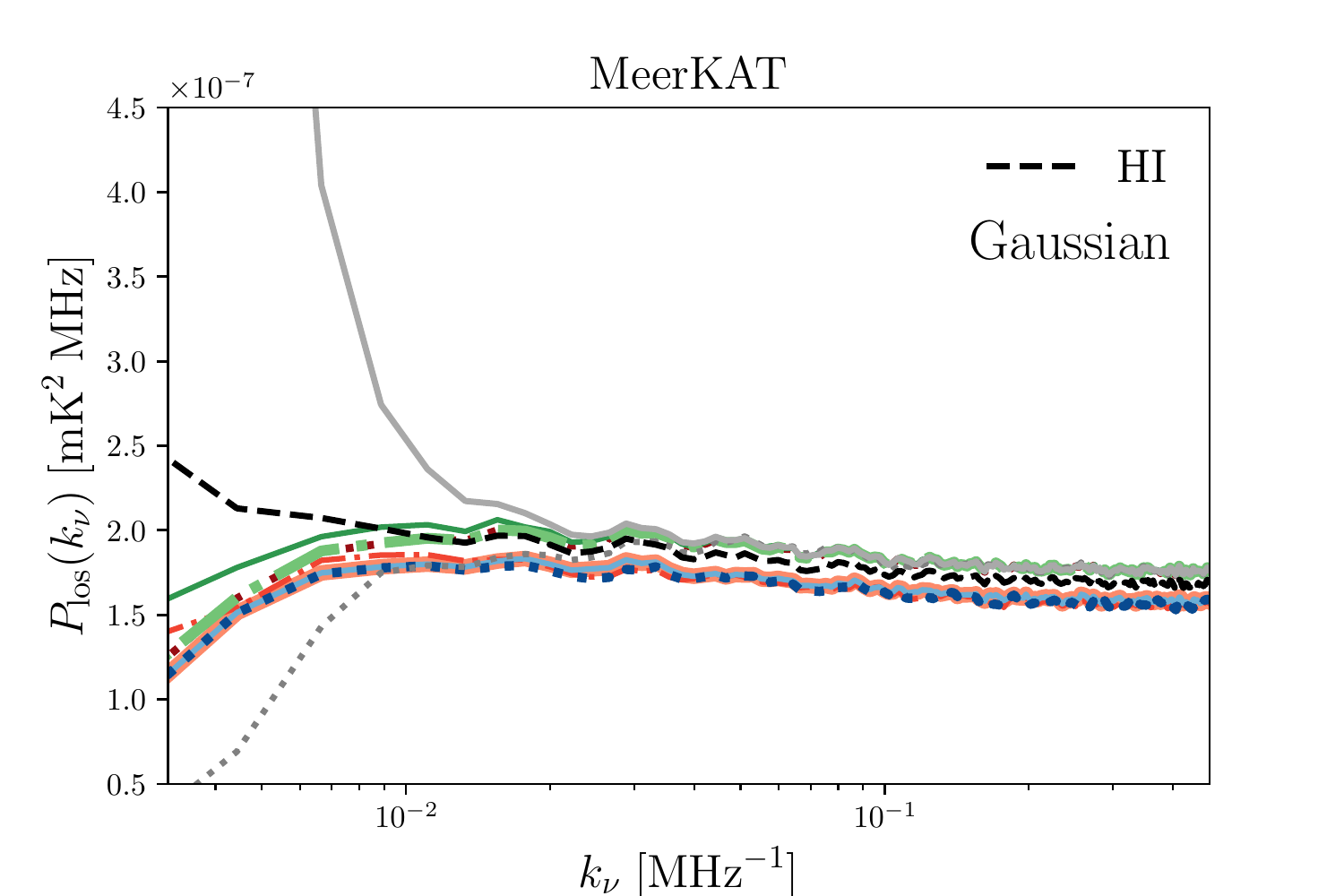}
      \includegraphics[width=1.5\columnwidth]{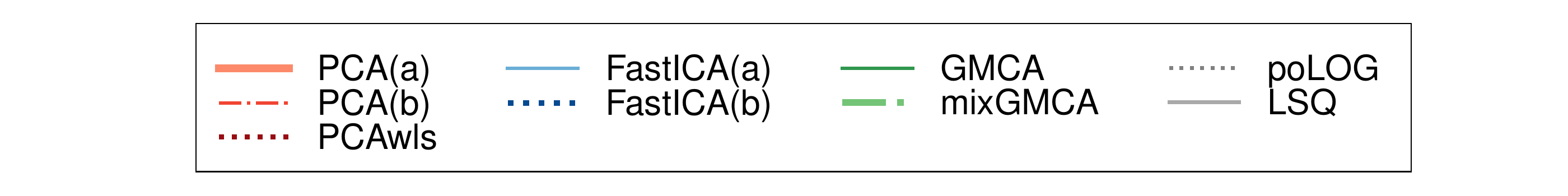}
  \caption{Angular power spectrum $C_\ell$ at 1225~MHz (top) and line-of-sight power spectrum $P_{\rm los}$ (bottom) of the residual maps for both an SKAO-like (left) or a MeerKAT-like (right) survey. We focus on the realistic foreground (PSM) case. The results are colour-coded to distinguish the various cleaning methods. Similar techniques and/or different implementations of the same algorithm are grouped together: PCA in reds, \fastica in blues, GMCA in greens and non-blind method in grays. All panels show results when the adopted beam model is Gaussian. The true \HI signal, convolved with the appropriate beam, is shown as a black dashed line.}
  \label{fig:various_methods_gauss}
\end{figure*}

\begin{figure*}
  \centering
  \includegraphics[width=\columnwidth]{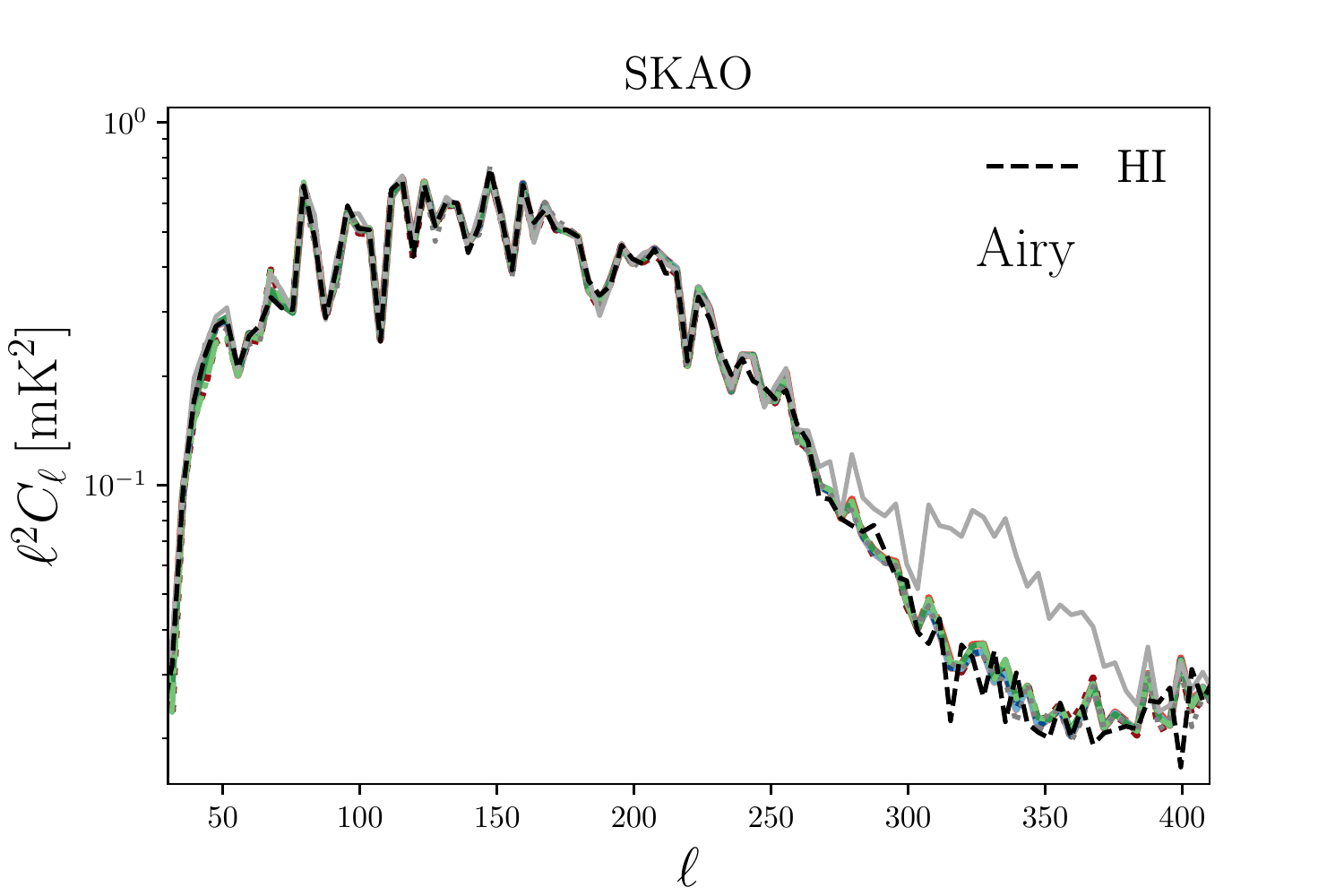}       \includegraphics[width=\columnwidth]{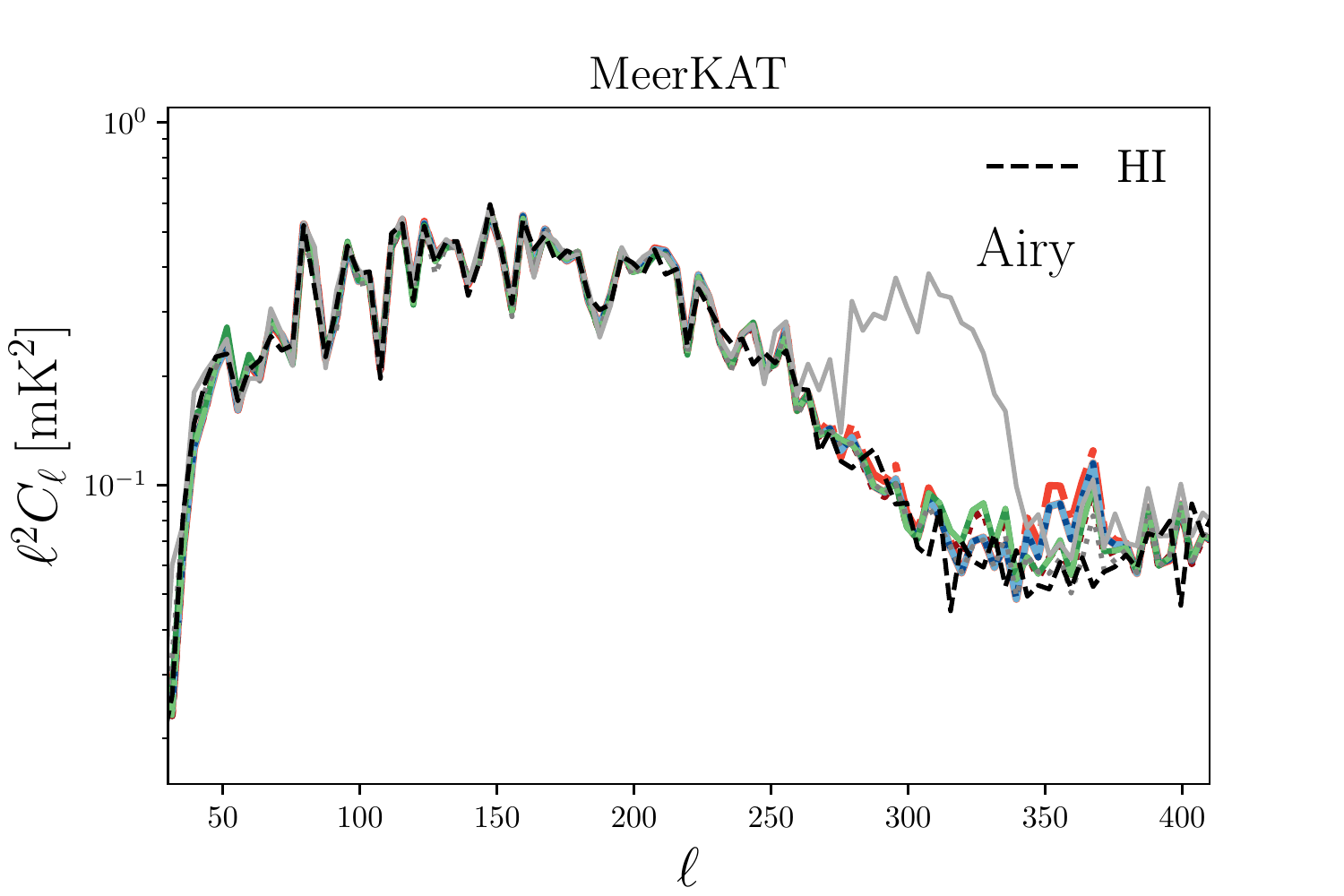}\\
  \includegraphics[width=\columnwidth]{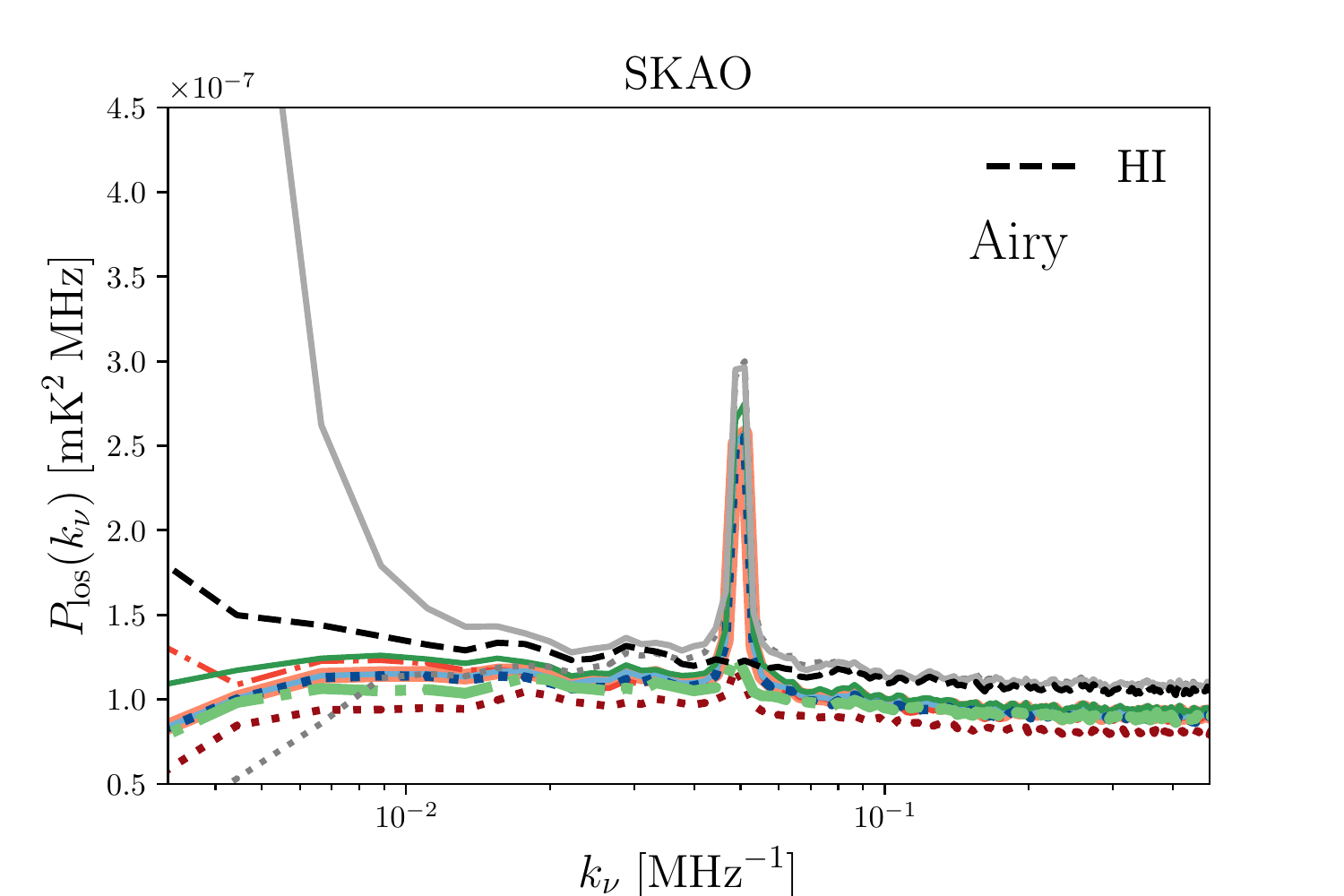} \includegraphics[width=\columnwidth]{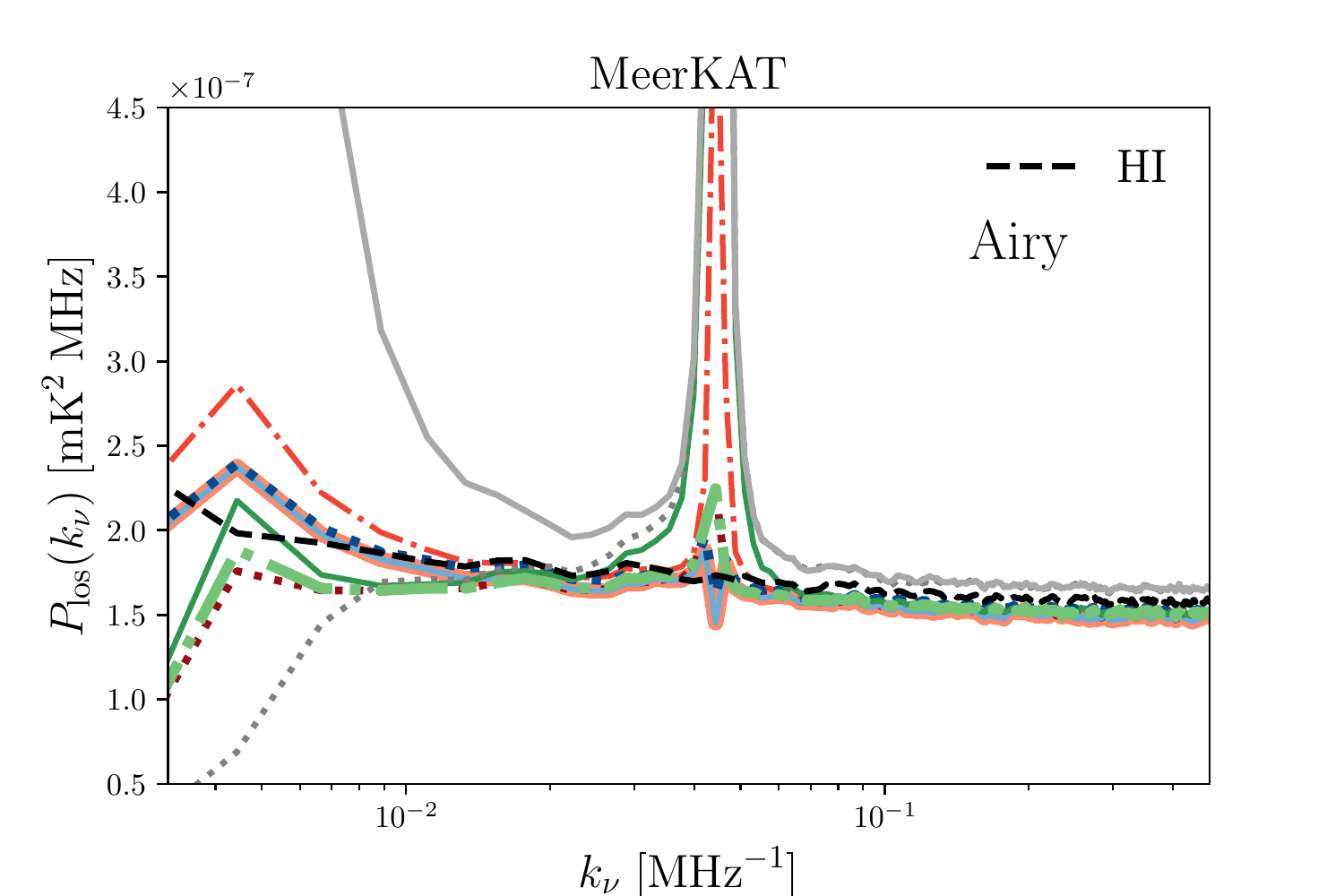}
      \includegraphics[width=1.5\columnwidth]{figures/legend.pdf}
  \caption{Same as \autoref{fig:various_methods_gauss} but for the Airy beam model. PCA(a), \fastica(a) and \fastica(b) often overlap.}
  \label{fig:various_methods_airy}
\end{figure*}

\begin{figure}
    \centering
    \includegraphics[width=\columnwidth]{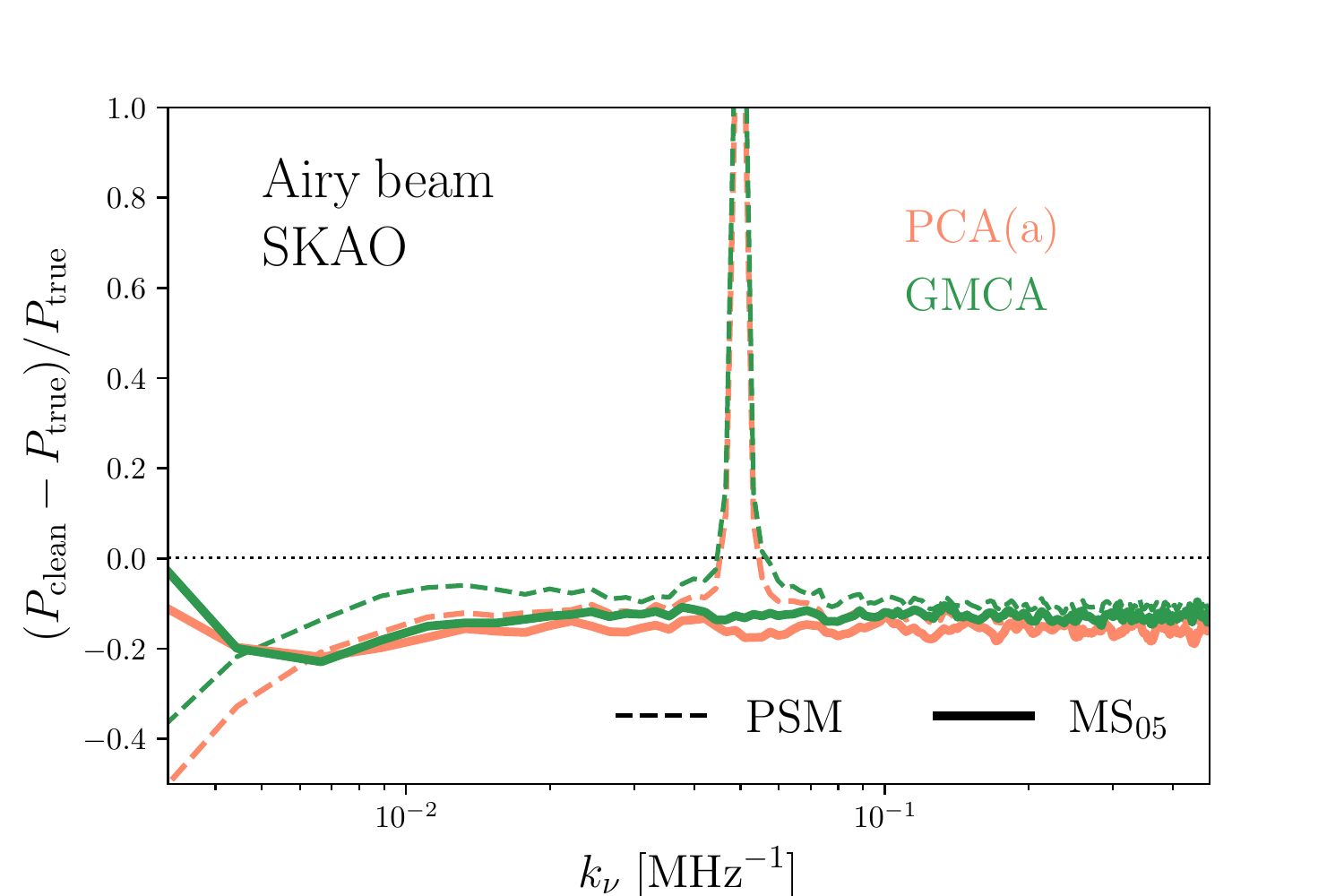}
    \caption{
    The different effect of the Airy beam model on the cleaning when the sky model is constructed with the MS$_{05}$ foreground (solid lines) or the PSM model (dashed lines).
    The line-of-sight power spectrum is shown for the SKAO case and for two different cleaning methods as an example: PCA(a) in orange and GMCA in green.}
    \label{fig:MS5vsPSM}
\end{figure}

\begin{figure*}
   \centering
   \includegraphics[width=\columnwidth]{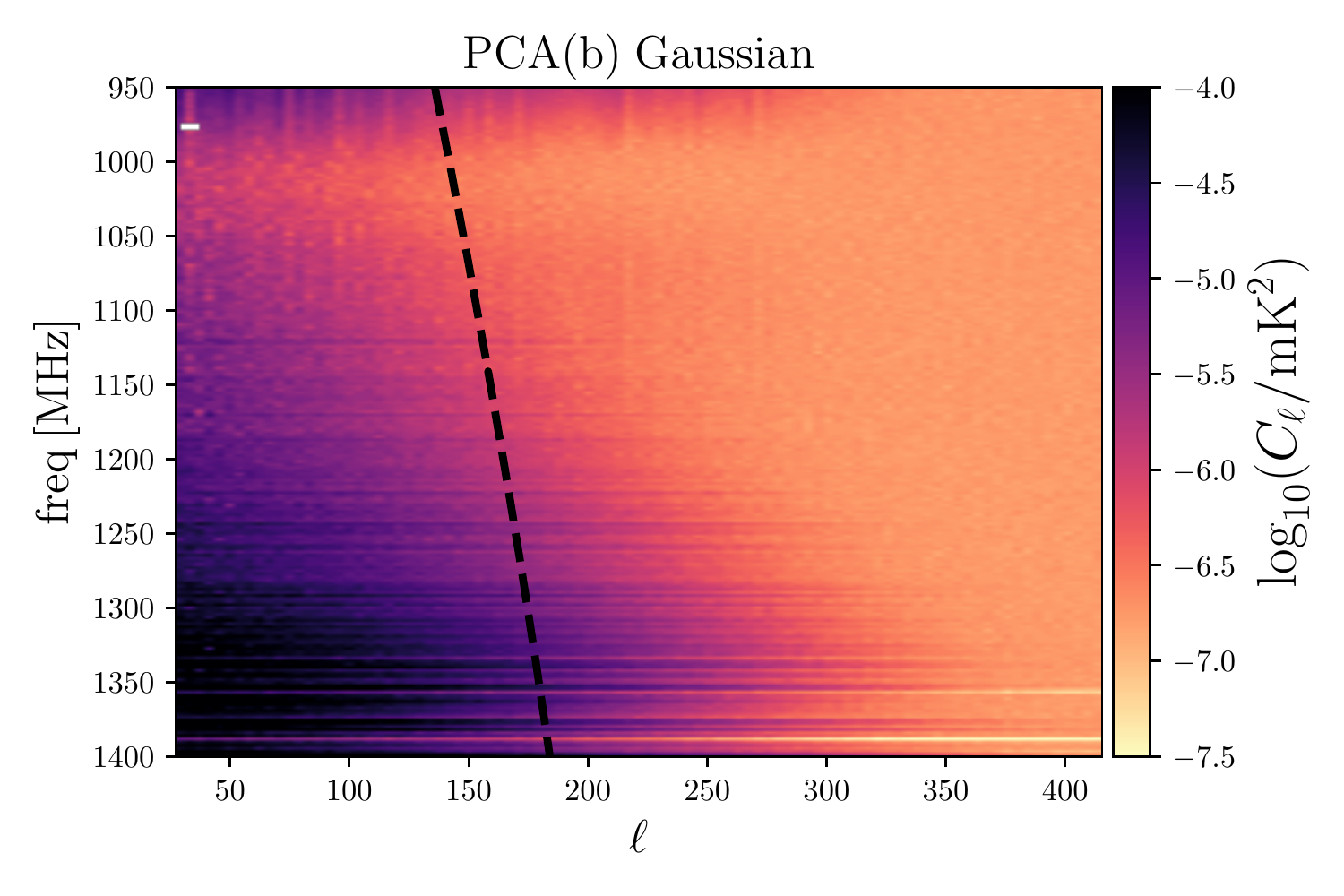} 
      \includegraphics[width=\columnwidth]{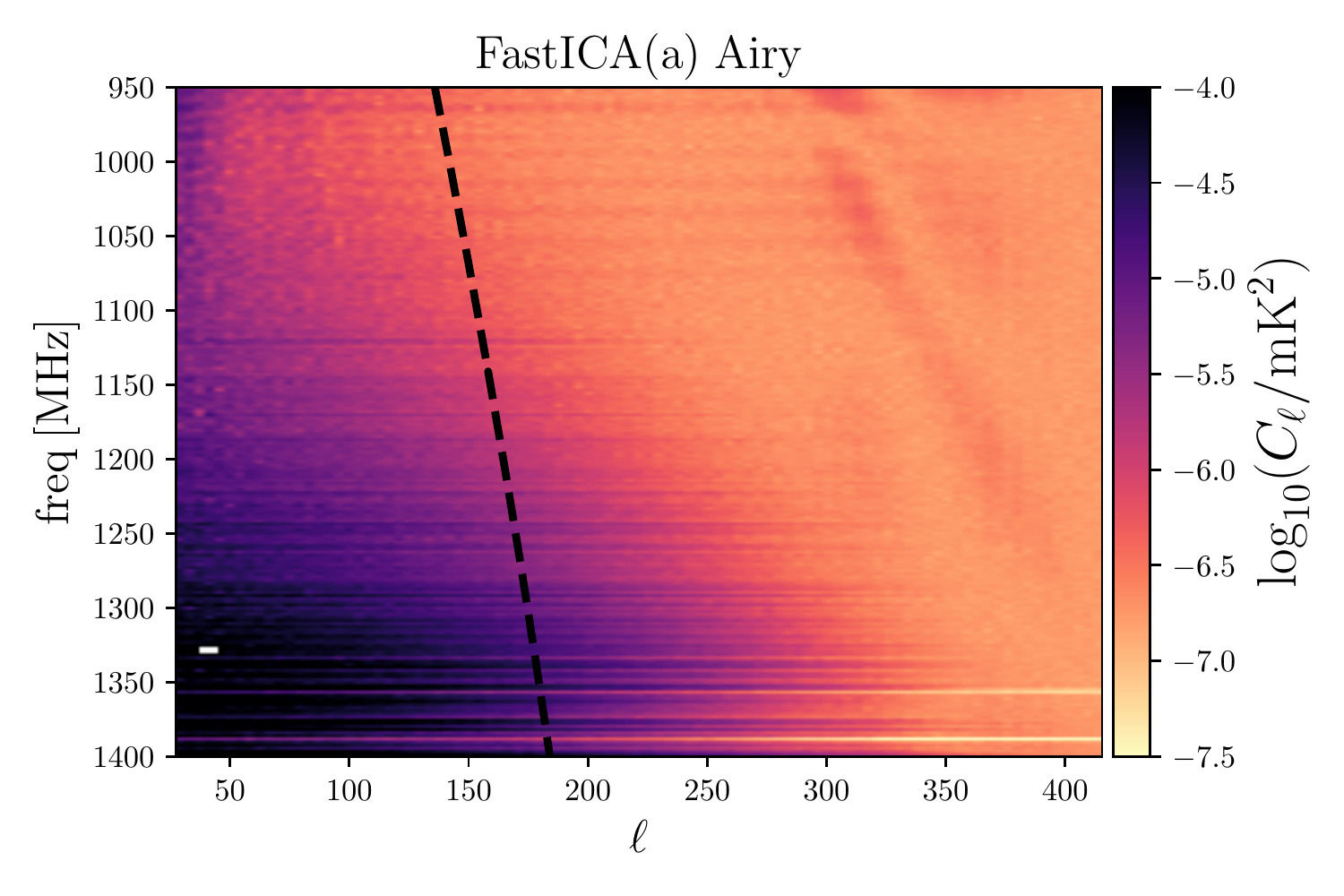}\\
    \includegraphics[width=\columnwidth]{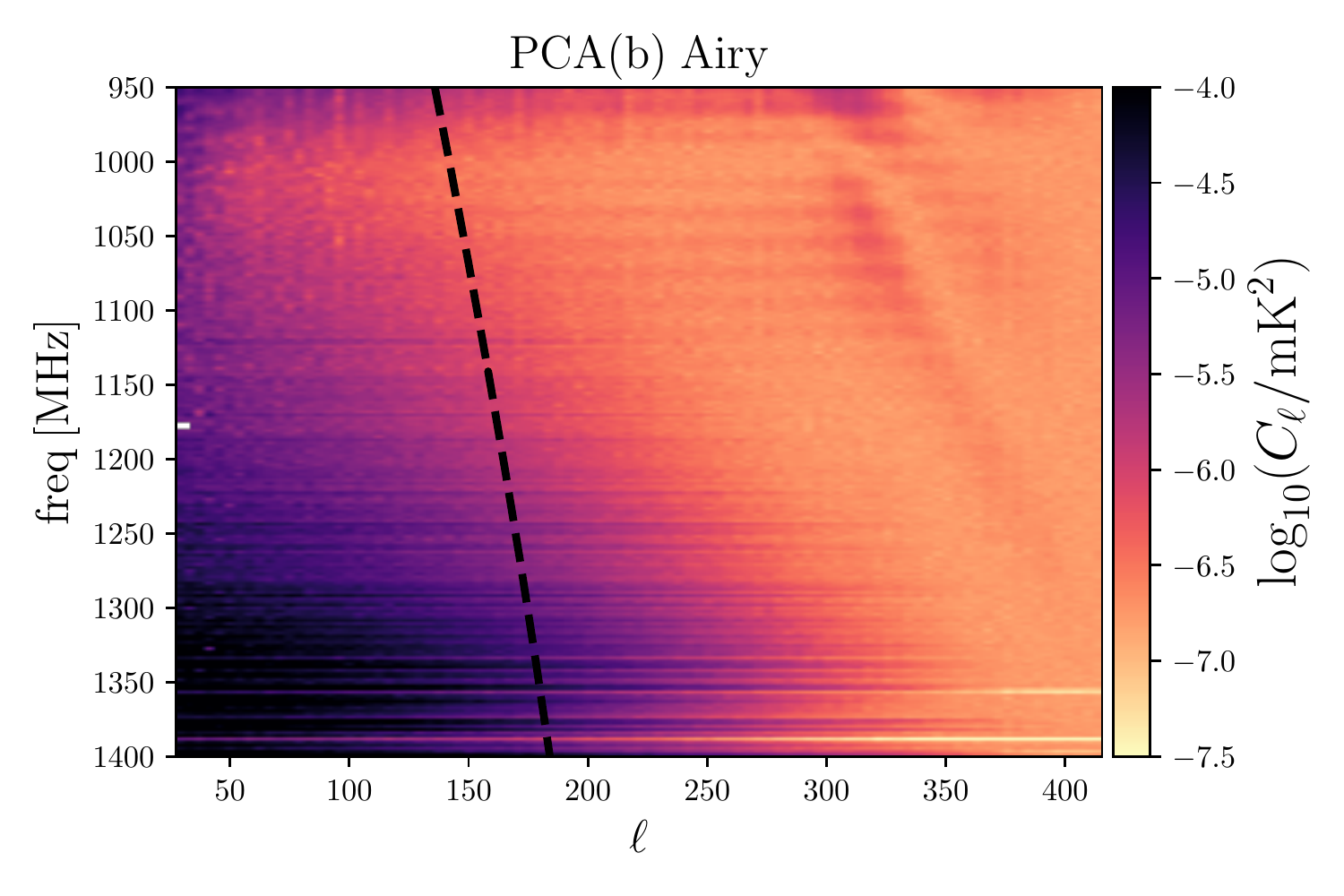}
    \includegraphics[width=\columnwidth]{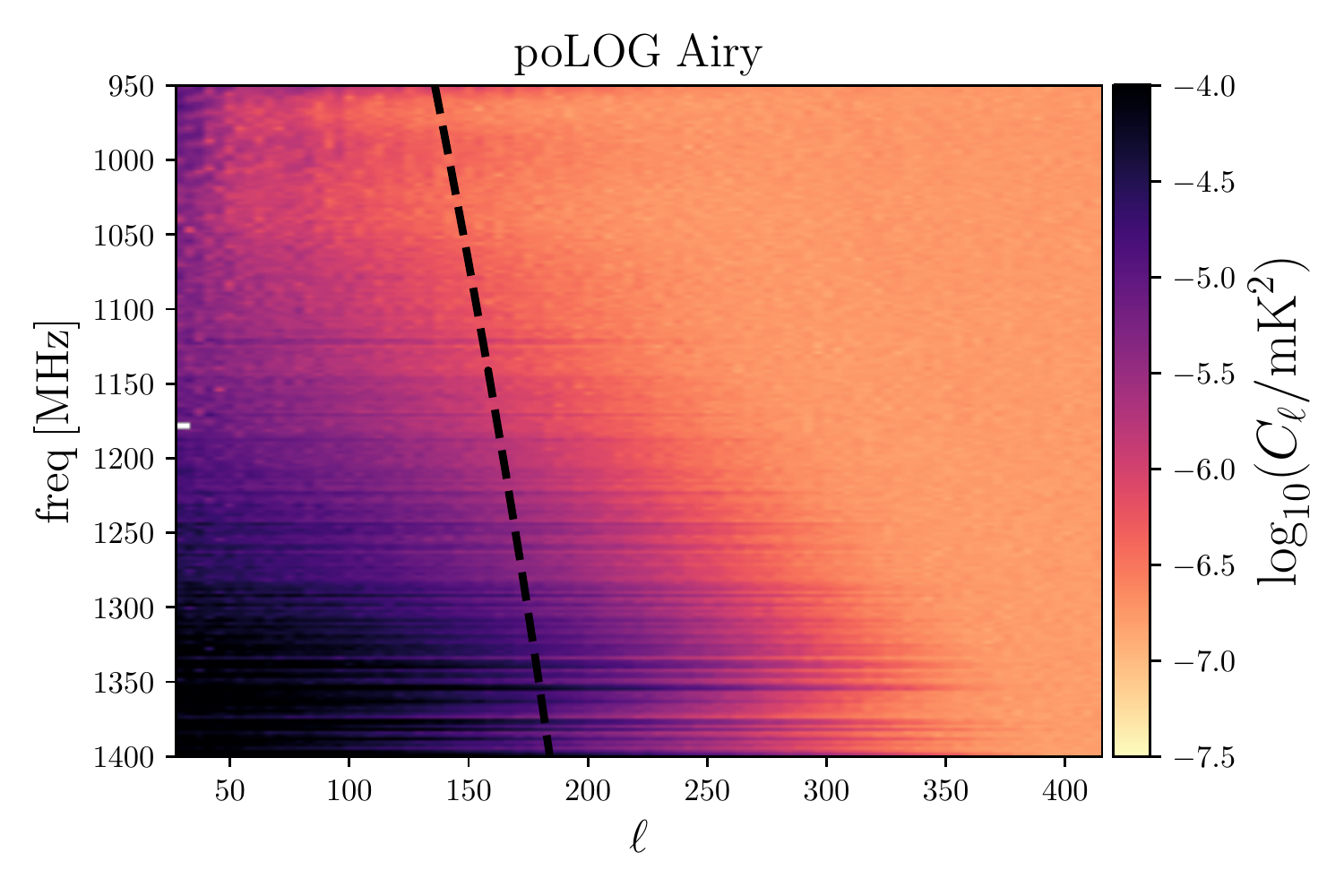}\\
   \caption{The angular power spectrum $C_\ell$ as a function of frequency of the residual maps (SKAO case) after cleaning with one of the implementation of the PCA method (left panels). The more realistic case where the sky data-cube has been convolved with the Airy beam model (lower left panel) presents a complex frequency behaviour that is not present in the Gaussian case (upper left panel). The right panels show how this frequency feature, induced in the cleaning by the presence of the Airy beam, affects the cleaning by other methods. We show  \fastica(a) (upper right panel) that has a similar shape as most of the other methods, and the poLOG (lower right panel) which, enforcing smoothness by construction, does not display any frequency features in the angular power spectra. The black dashed line in all panels traces the evolution with frequency of the angular scale of the FWHM of the telescope beam.}
   \label{fig:beam_effect}
\end{figure*} 

\begin{figure*}
    \centering
    \includegraphics[width=\columnwidth]{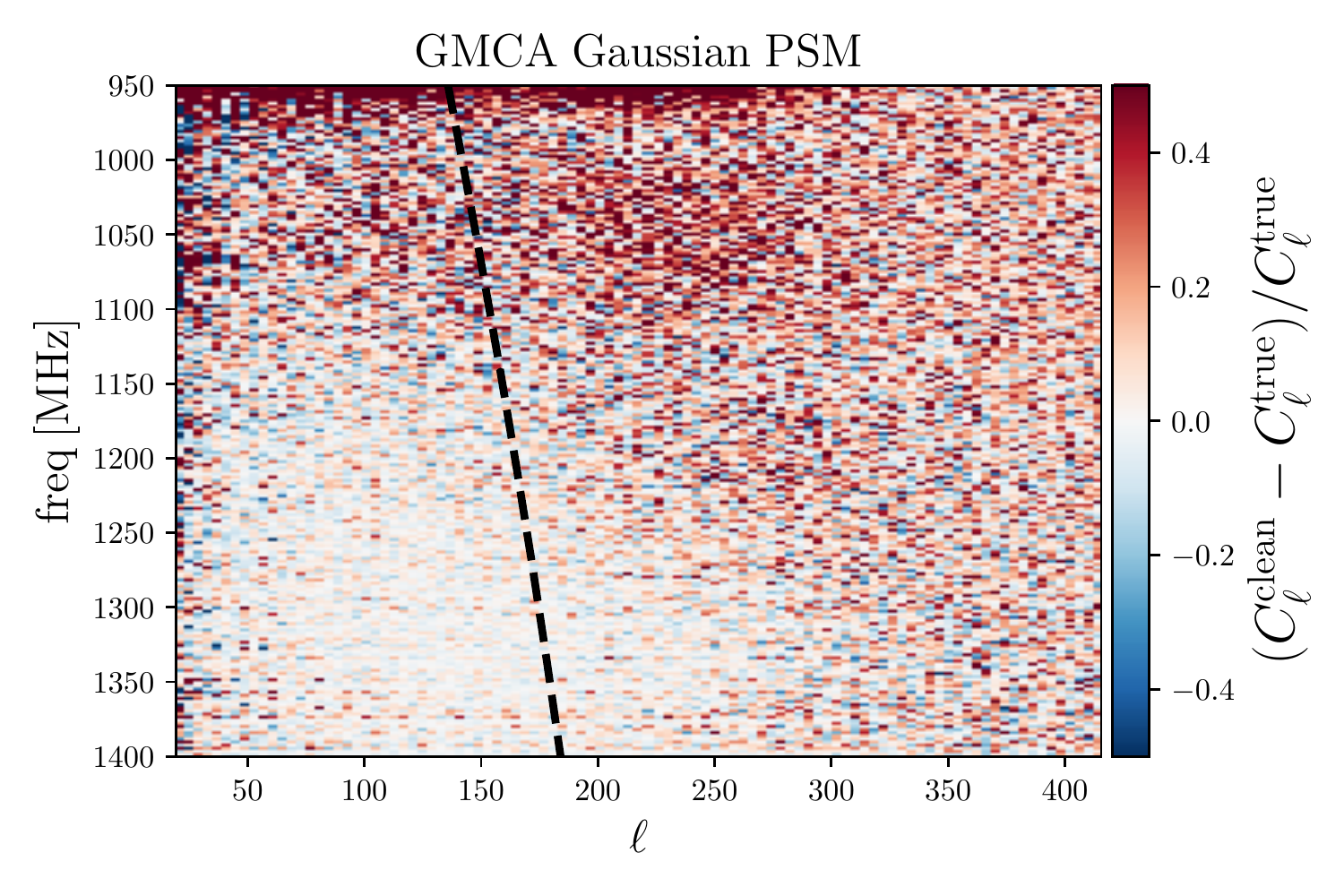}
    \includegraphics[width=\columnwidth]{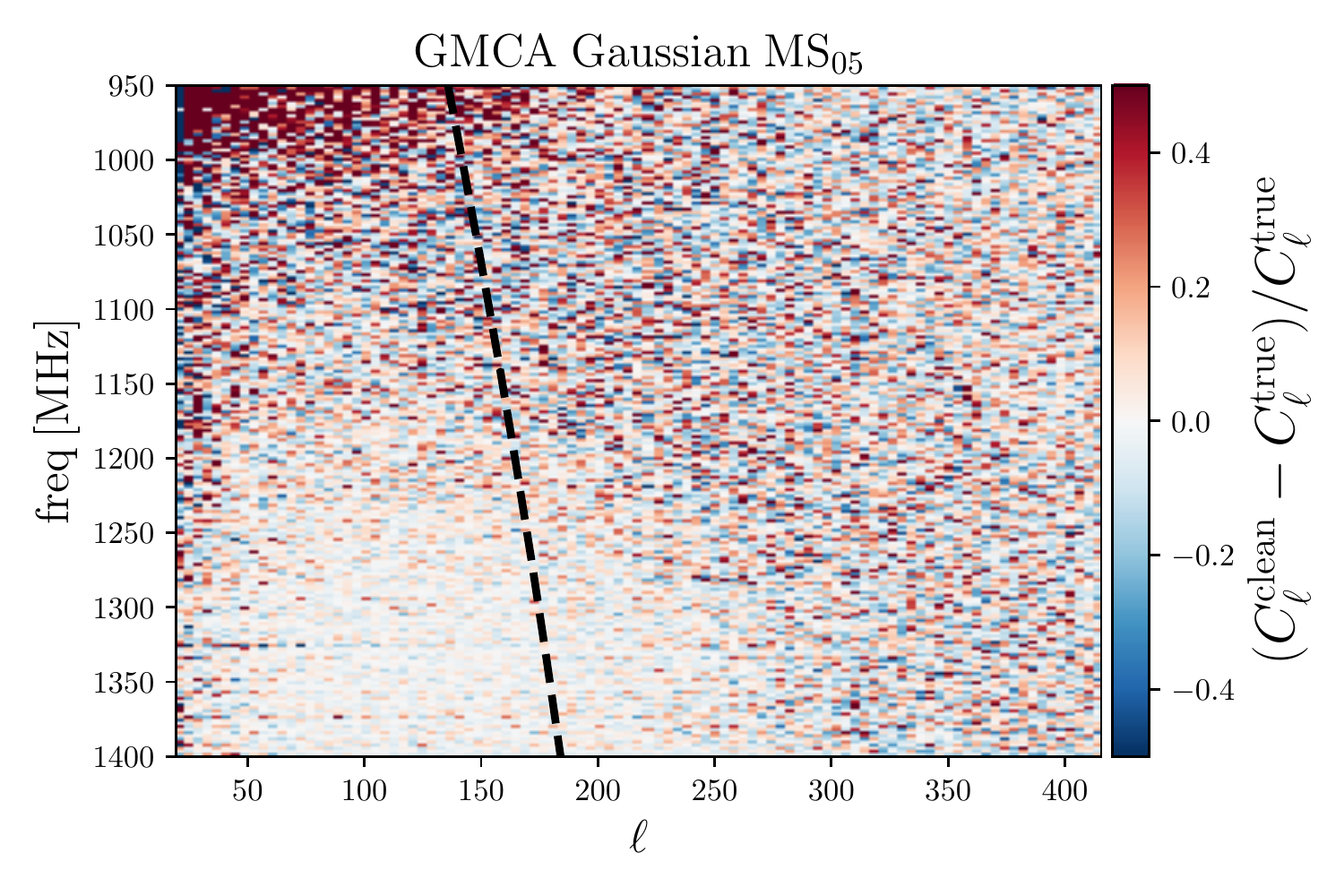}\\
        \includegraphics[width=\columnwidth]{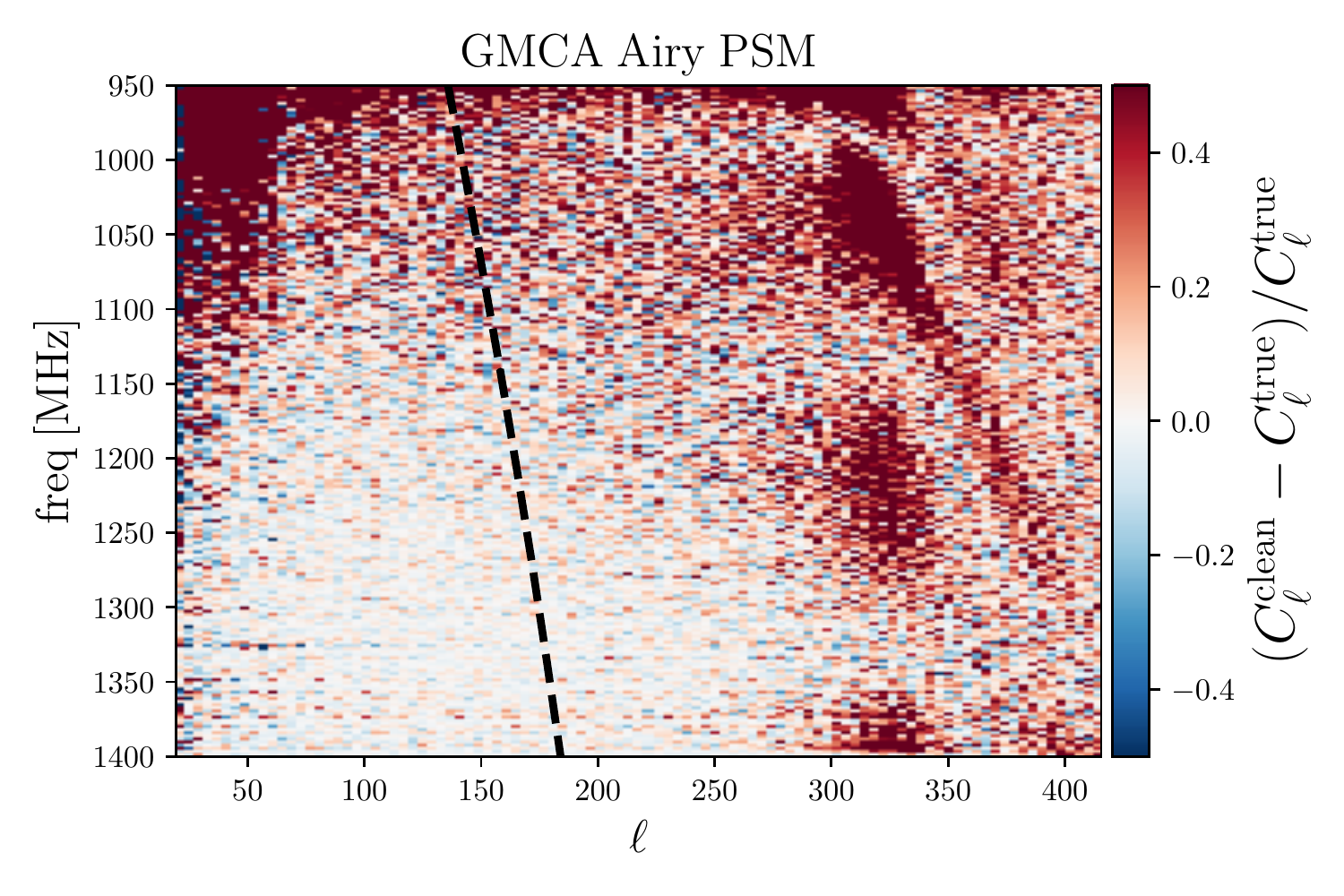}
    \includegraphics[width=\columnwidth]{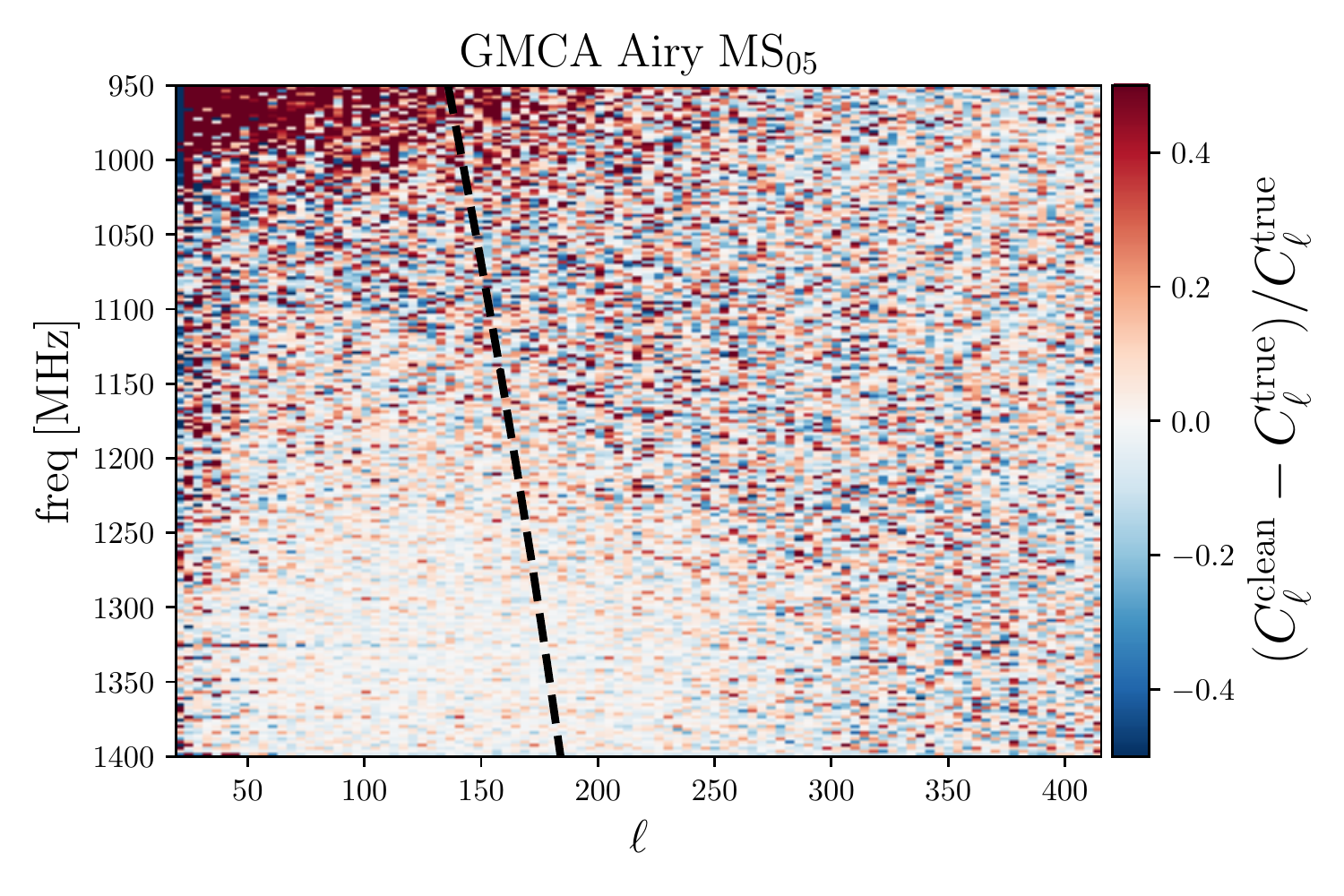}
    \caption{The estimator $(C_\ell^{\rm clean}-C_\ell^{\rm true})/C_\ell^{\rm true}$ for one of the cleaning methods (GMCA in this example), where $C_\ell^{\rm clean}$ is the angular power spectrum of the cleaned maps, while $C_\ell^{\rm true}$ is the angular power spectrum for the input signal and noise, as a function of frequency. We present results for the Original data-cubes. The more realistic PSM foreground model (left panels) is compared with the MS$_{05}$ foreground model (right panels). We consider both the Gaussian and the Airy beam and focus on the SKAO case. Results are qualitatively similar in all cases, showing that cleaning is easier in the MS$_{05}$ case and that in absence of foreground structure, there are no frequency features induced by the Airy beam. For reference, the black dashed line in all panels traces the evolution with frequency of the angular scale of the FWHM of the telescope beam.}
    \label{fig:cellMSvsPSM}
\end{figure*}

\begin{figure*}
   \includegraphics[width=\columnwidth]{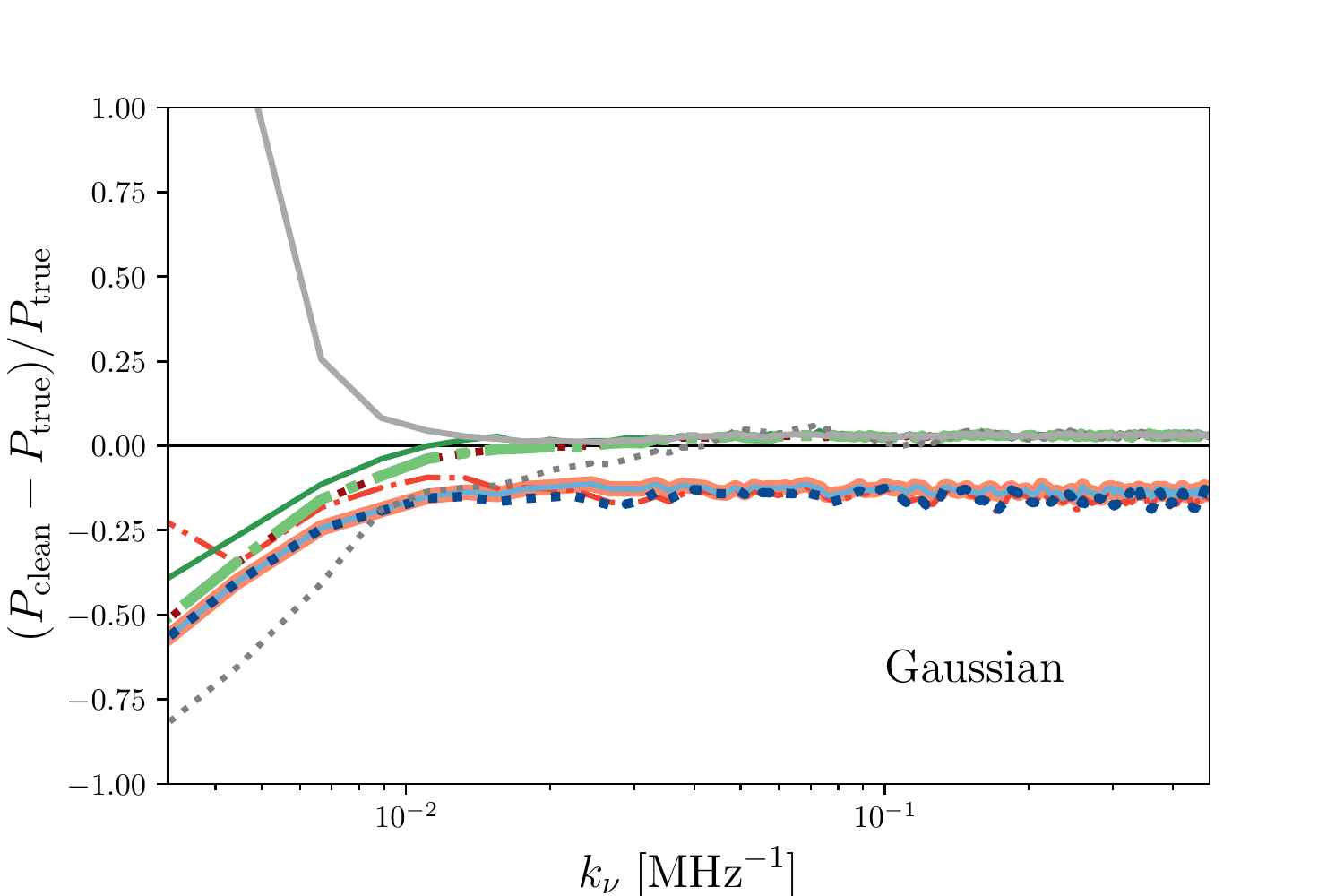}
\includegraphics[width=\columnwidth]{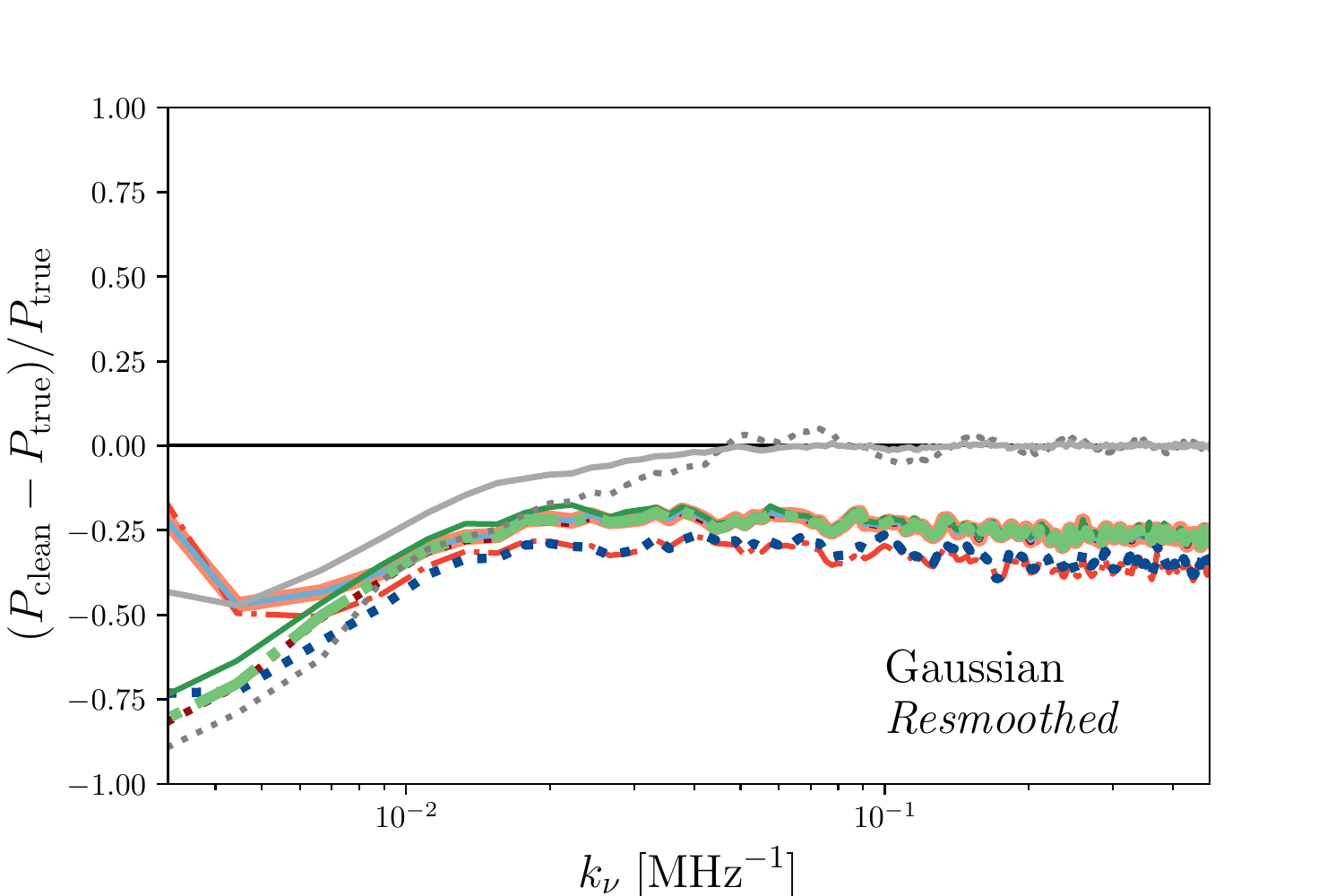}\\

   \includegraphics[width=\columnwidth]{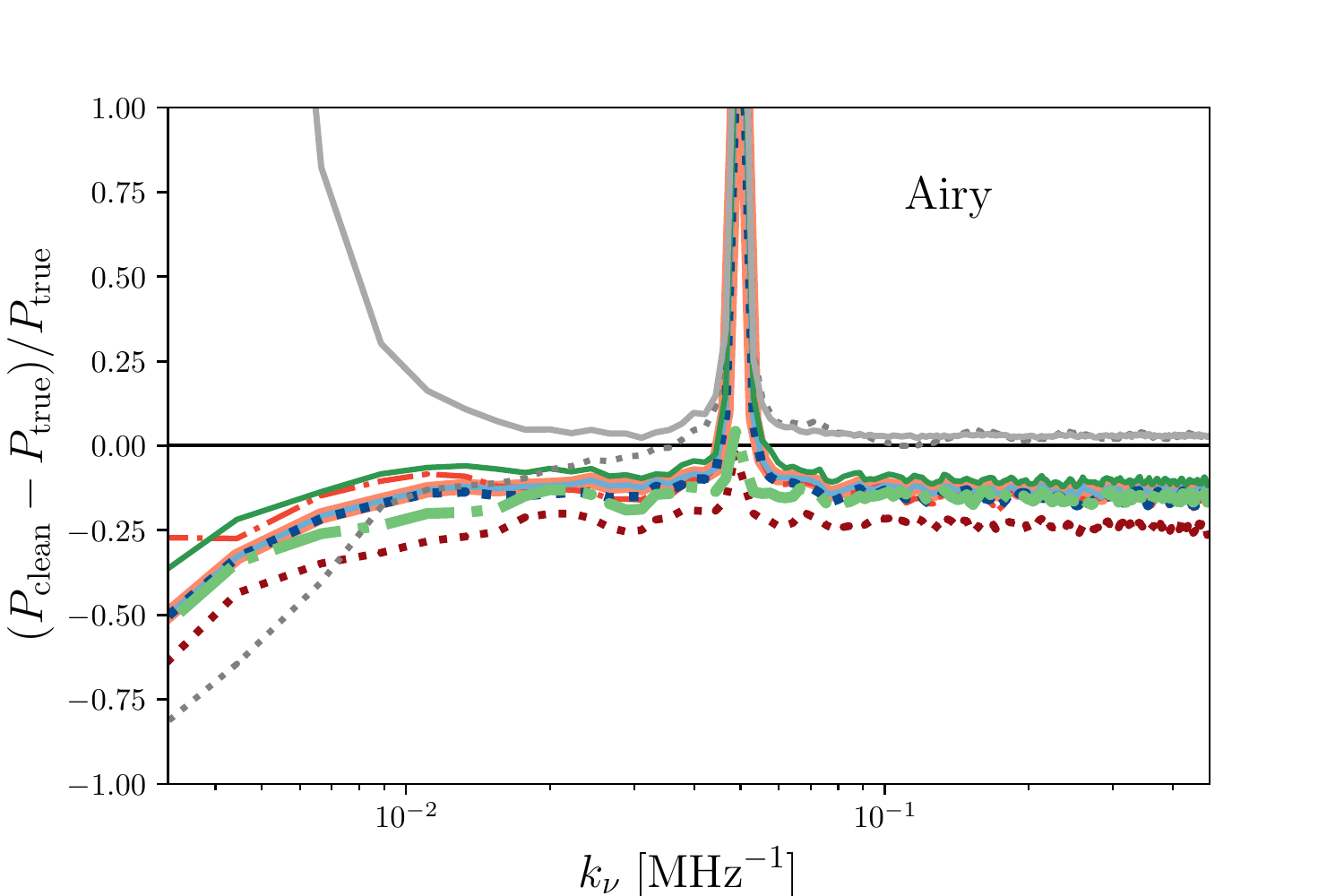}
\includegraphics[width=\columnwidth]{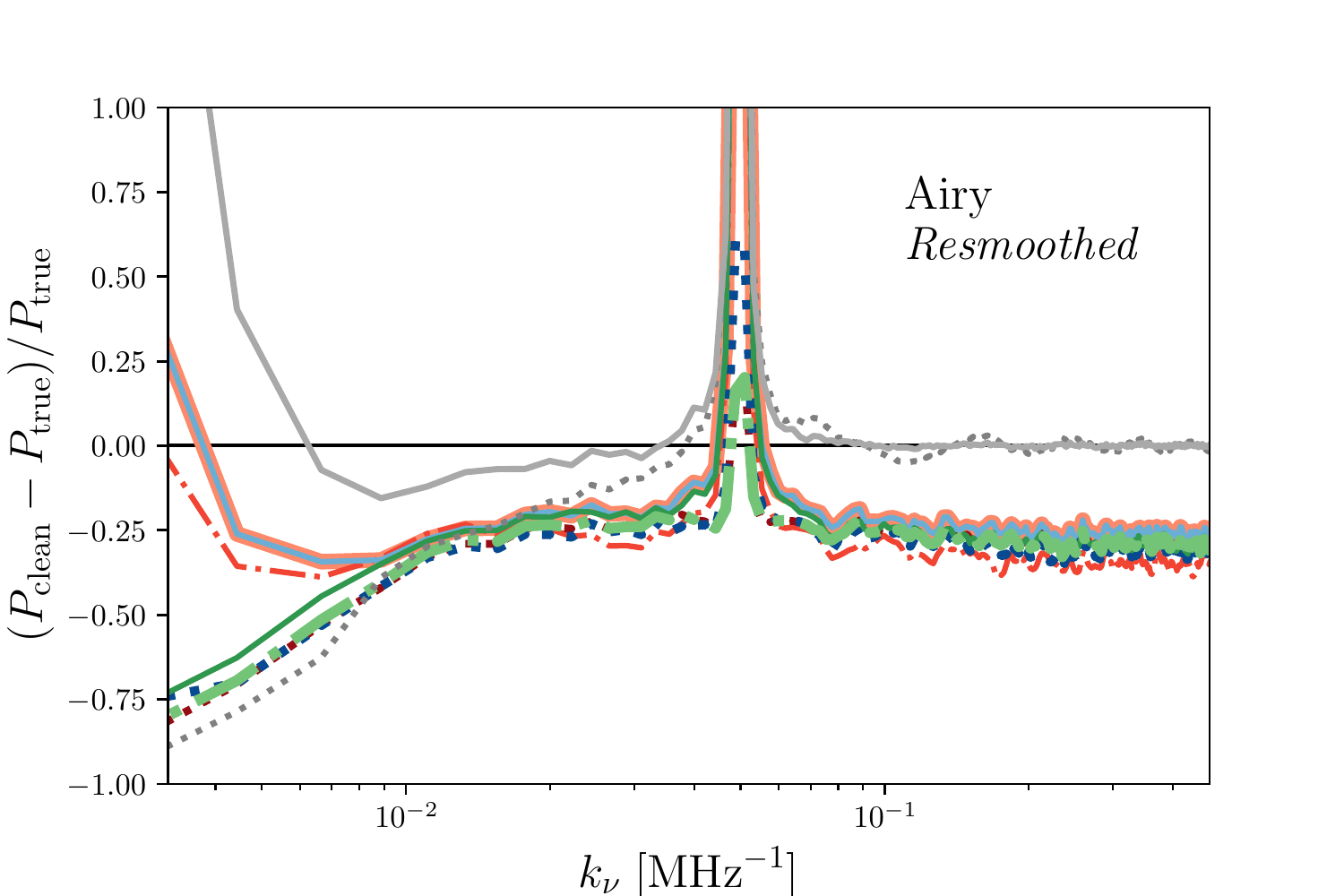}\\
     \includegraphics[width=1.5\columnwidth]{figures/legend.pdf}
   \caption{Comparison of the estimator $(P_{\rm clean}-P_{\rm true})/P_{\rm true}$, where $P_{\rm clean}$ is the power spectrum of the residual maps  for the various cleaning methods, while $P_{\rm true}$ is the input signal and noise original or resmoothed depending on the specific case. The left panels show the difference between the Gaussian and Airy beam model for the original data (as in the lower panels of \autoref{fig:various_methods_gauss} and \autoref{fig:various_methods_airy}).
   The right panels present the same cases but for the resmoothed data. The resmoothing procedure affects the line-of-sight power spectrum as the Airy beam {\it peak} broads and the reconstructed signal has an enhanced offset with respect to the true one. All panels refer to SKAO - PSM cases.}
   \label{fig:unblind_res_pk}
\end{figure*}

\begin{figure*}
   \includegraphics[width=0.9\columnwidth]{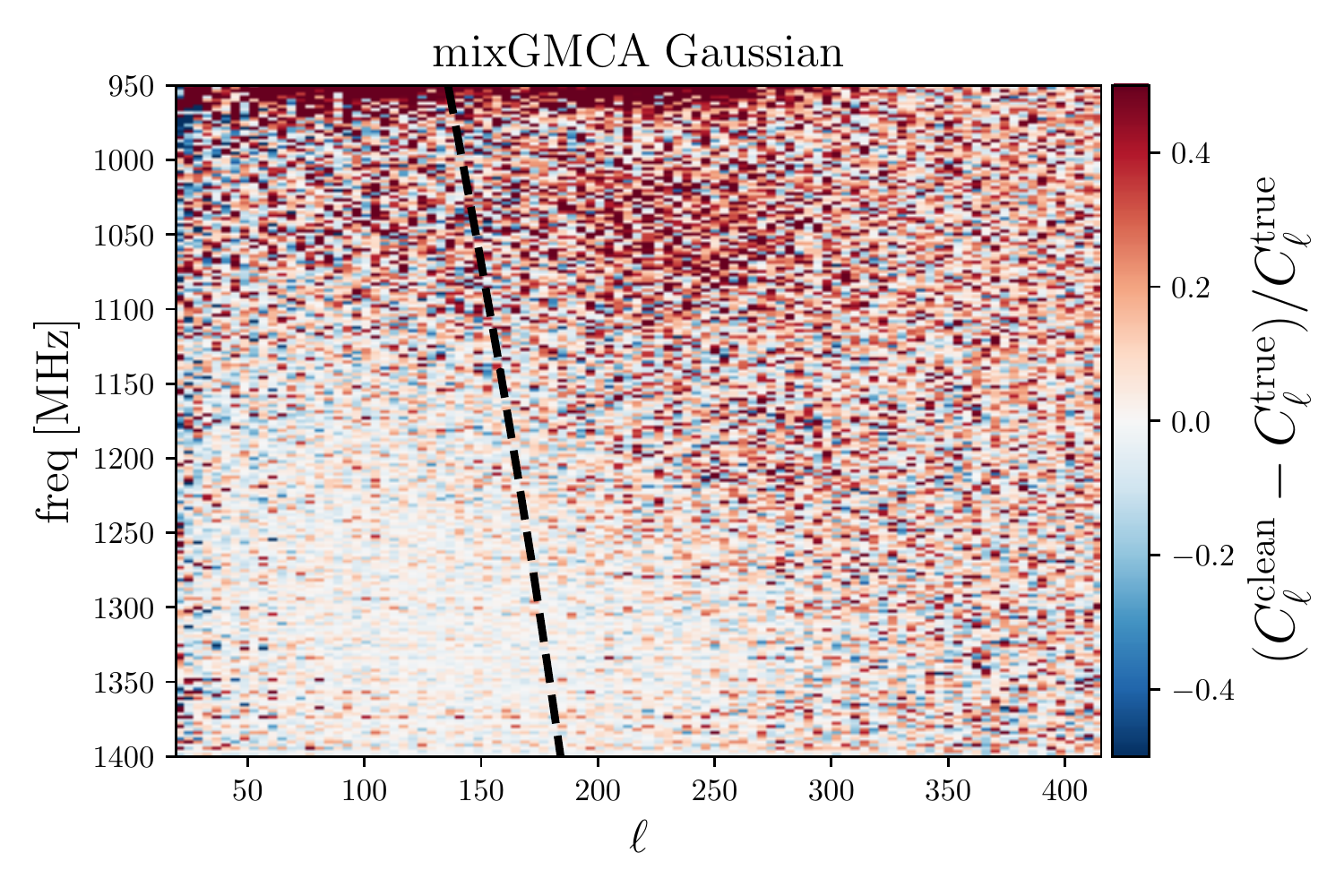}
   \includegraphics[width=0.9\columnwidth]{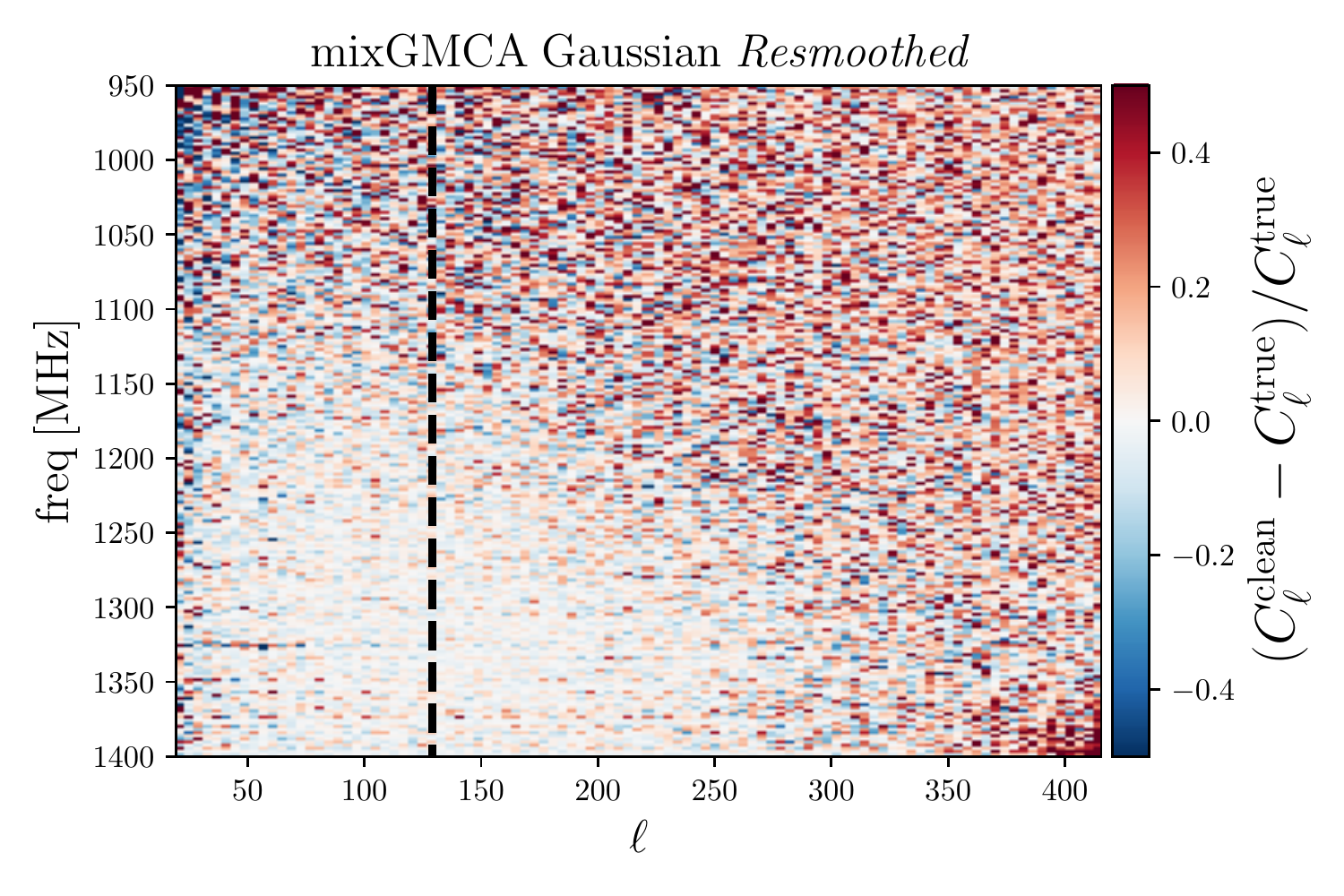}\\
      \includegraphics[width=0.9\columnwidth]{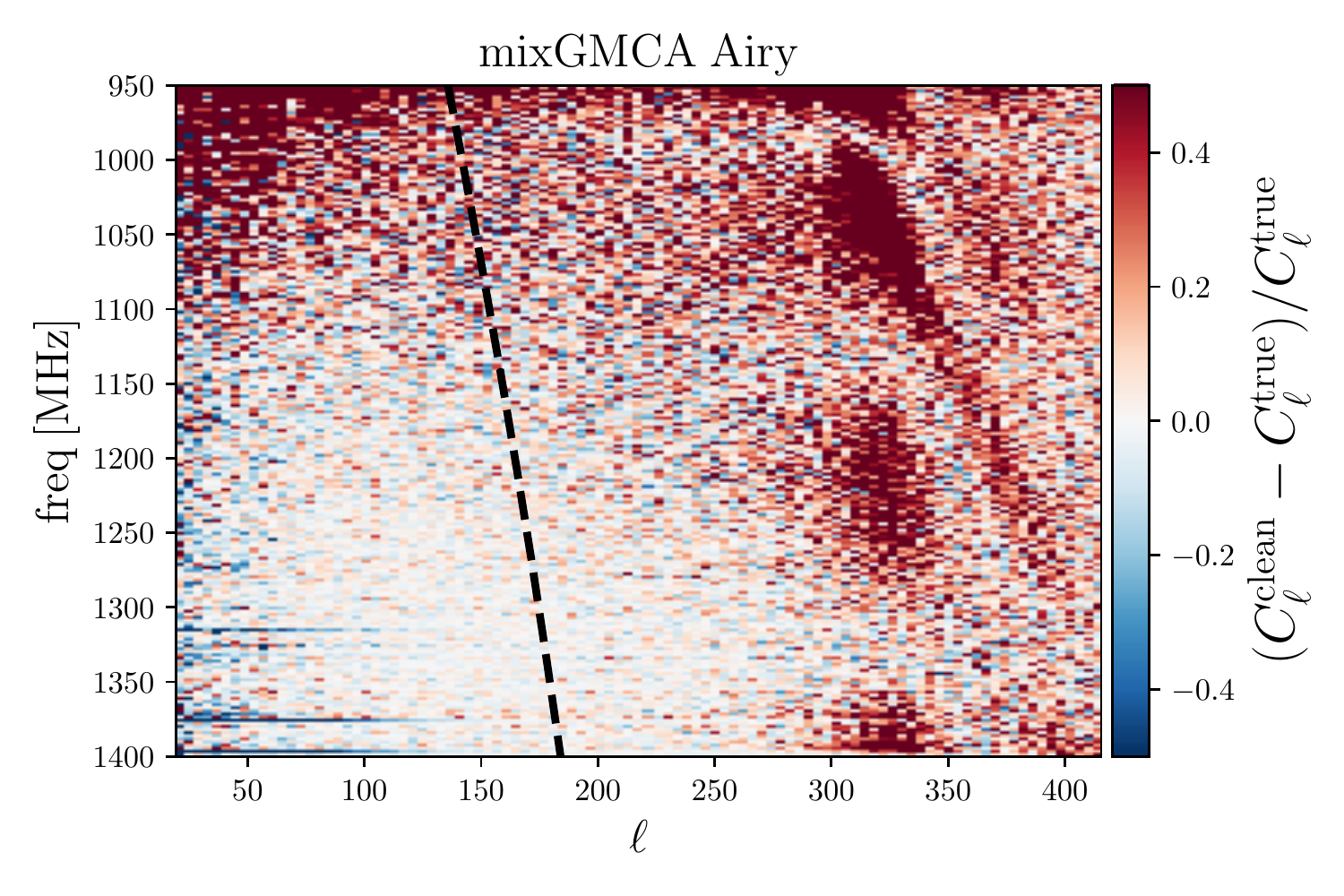}
   \includegraphics[width=0.9\columnwidth]{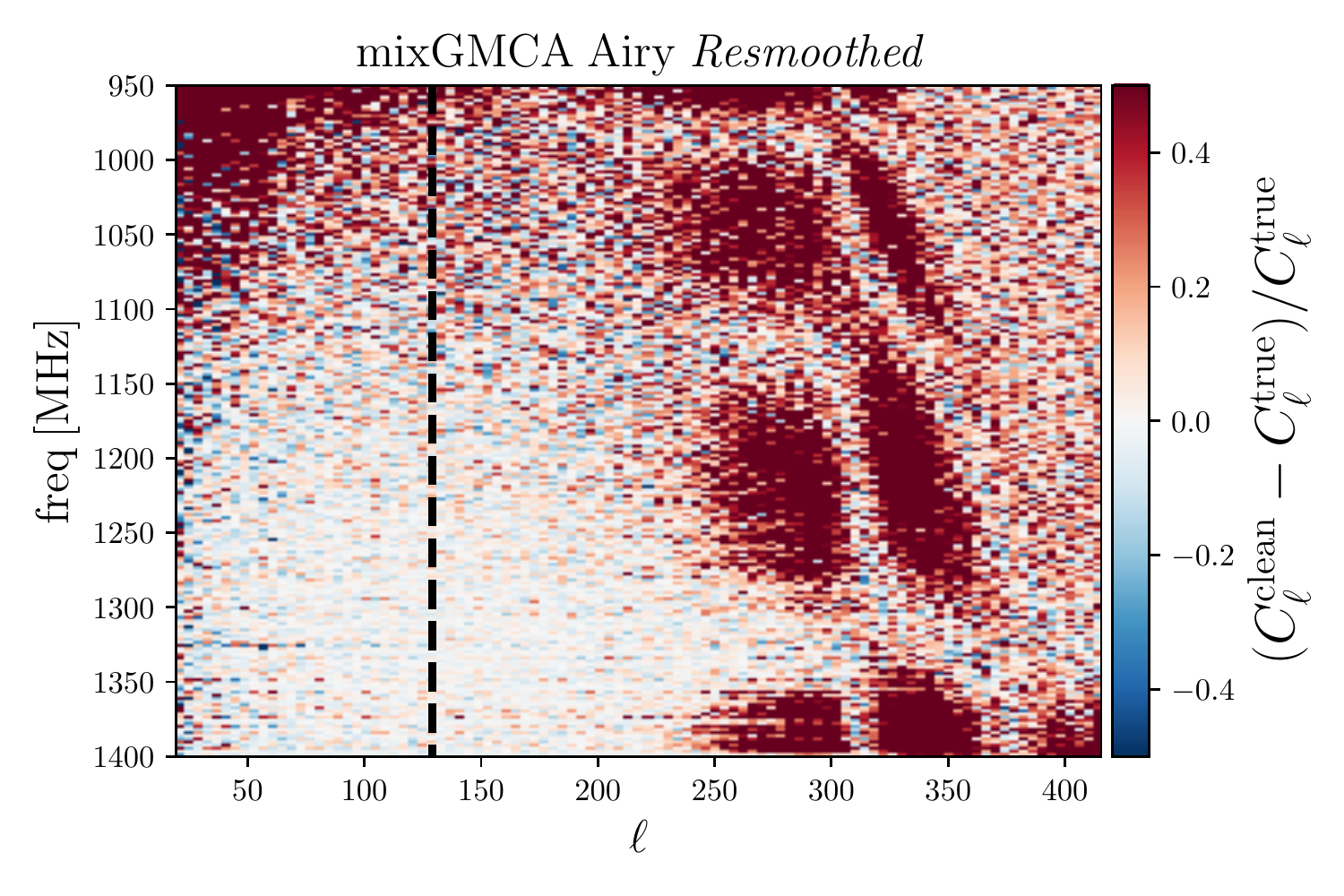}\\
   \caption{The estimator $(C_\ell^{\rm clean}-C_\ell^{\rm true})/C_\ell^{\rm true}$ for one of the cleaning method (mixGMCA in this example) in the PSM and SKAO scenarios, where $C_\ell^{\rm clean}$ is the angular power spectrum of the cleaned maps, while $C_\ell^{\rm true}$ is the angular power spectrum for the input signal and noise, as a function of frequency. From top to bottom the beam model changes from Gaussian to Airy, while from left to right is shown the effect of resmoothing. For reference, the black dashed line in all panels indicates the angular scale of the FWHM of the telescope beam. After resmoothing, this scale is constant with frequency.}\label{fig:unblind_res_cl}
   \end{figure*}
   
\begin{figure*}
   \includegraphics[width=0.9\columnwidth]{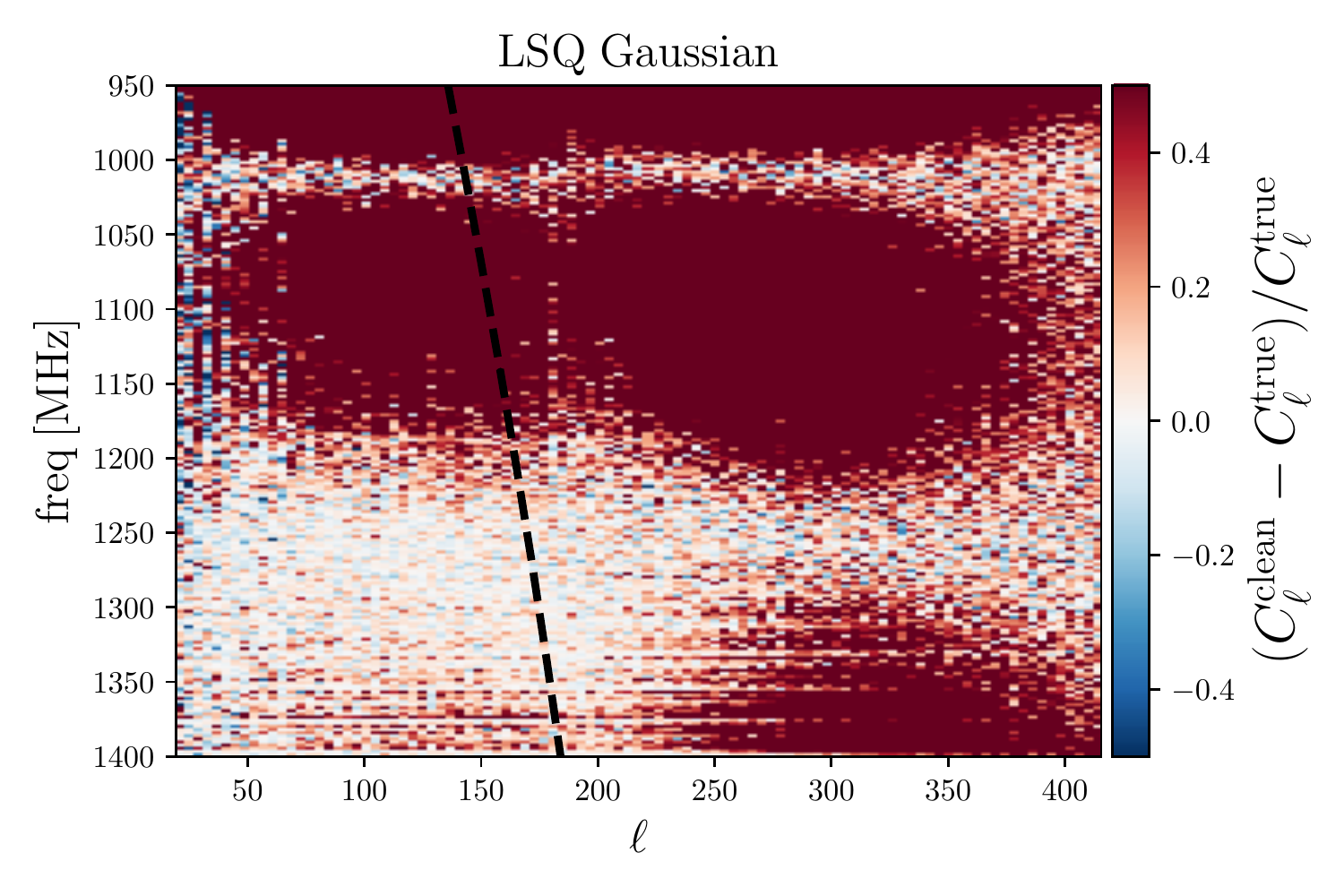}
    \includegraphics[width=0.9\columnwidth]{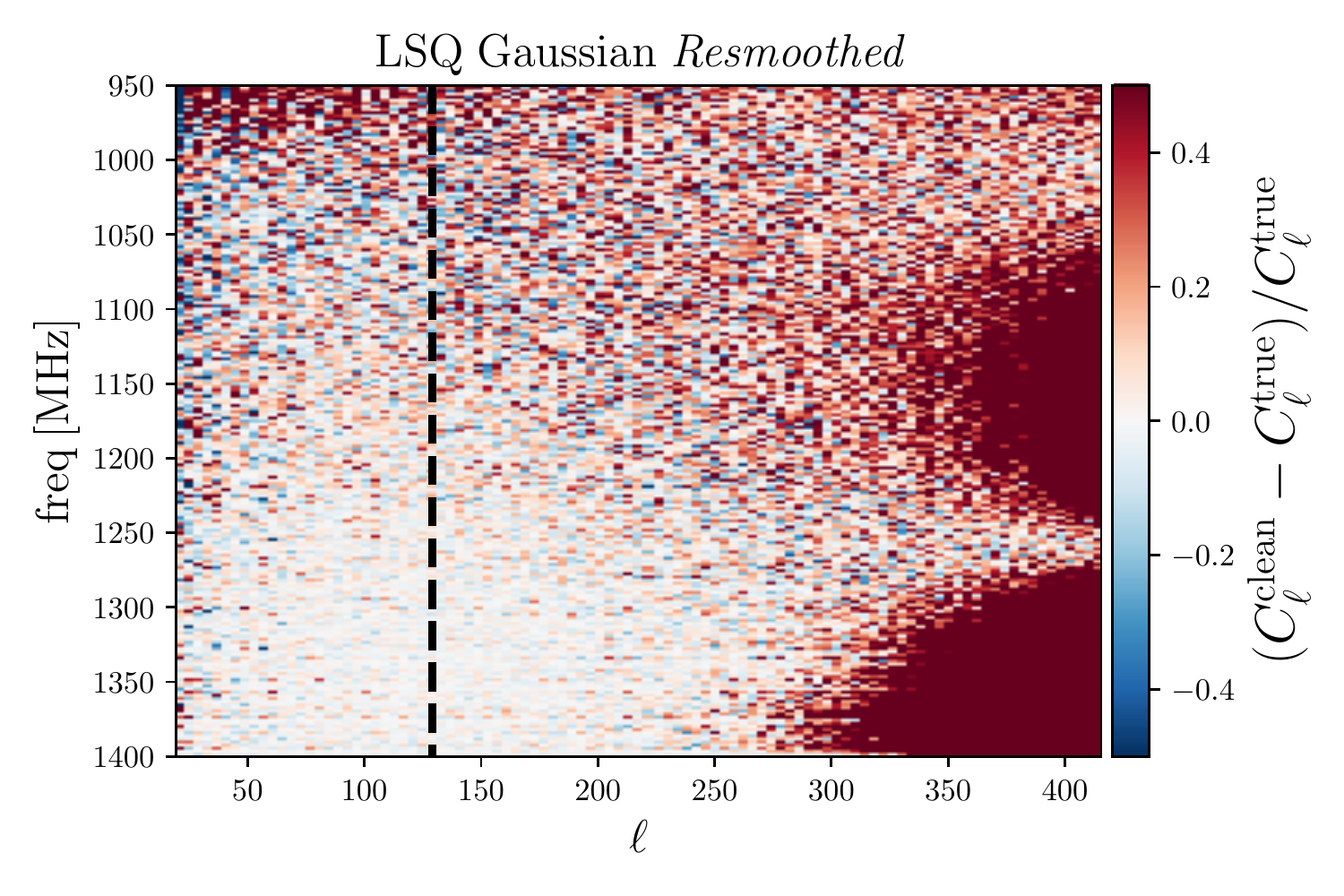}\\
   \includegraphics[width=0.9\columnwidth]{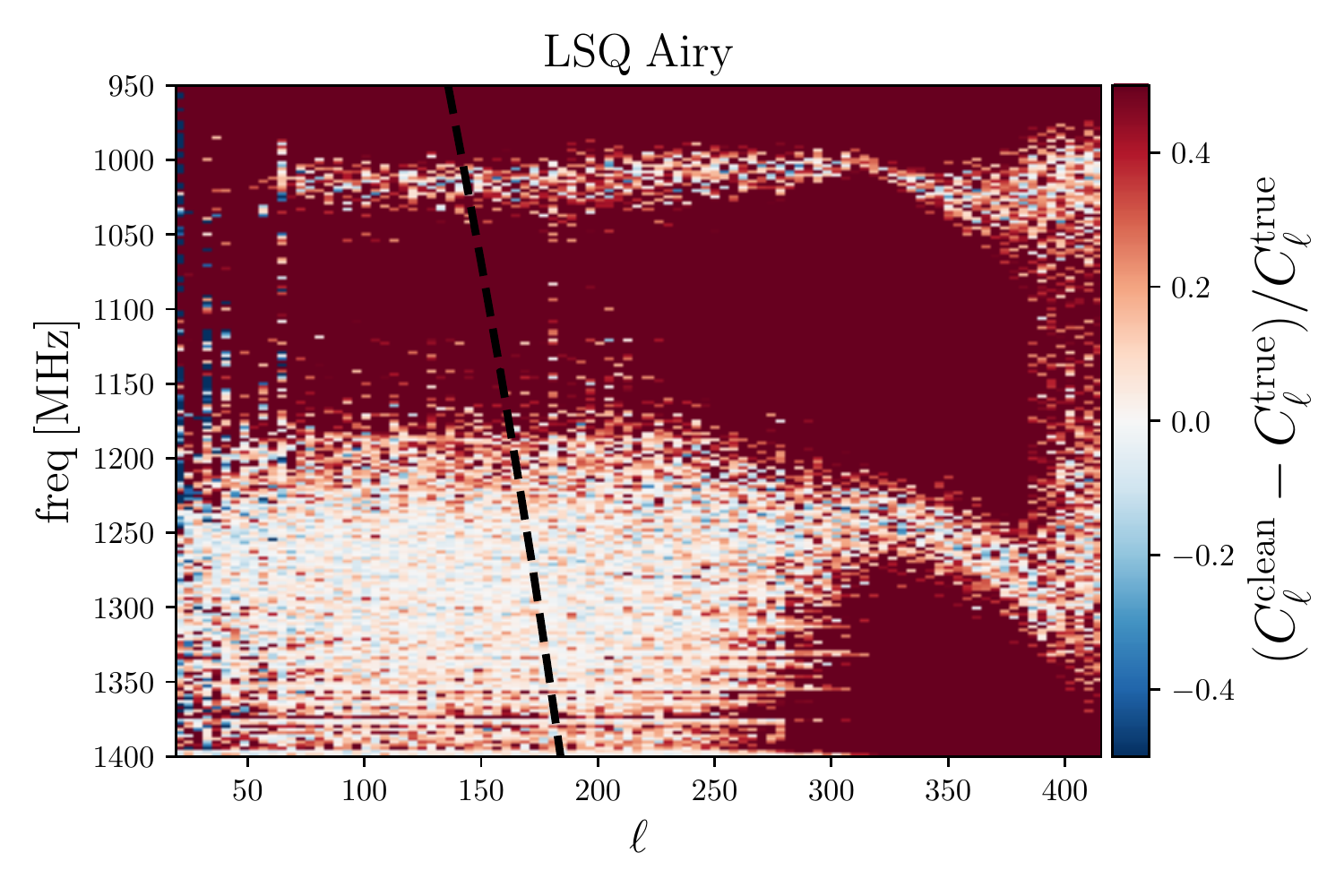}
    \includegraphics[width=0.9\columnwidth]{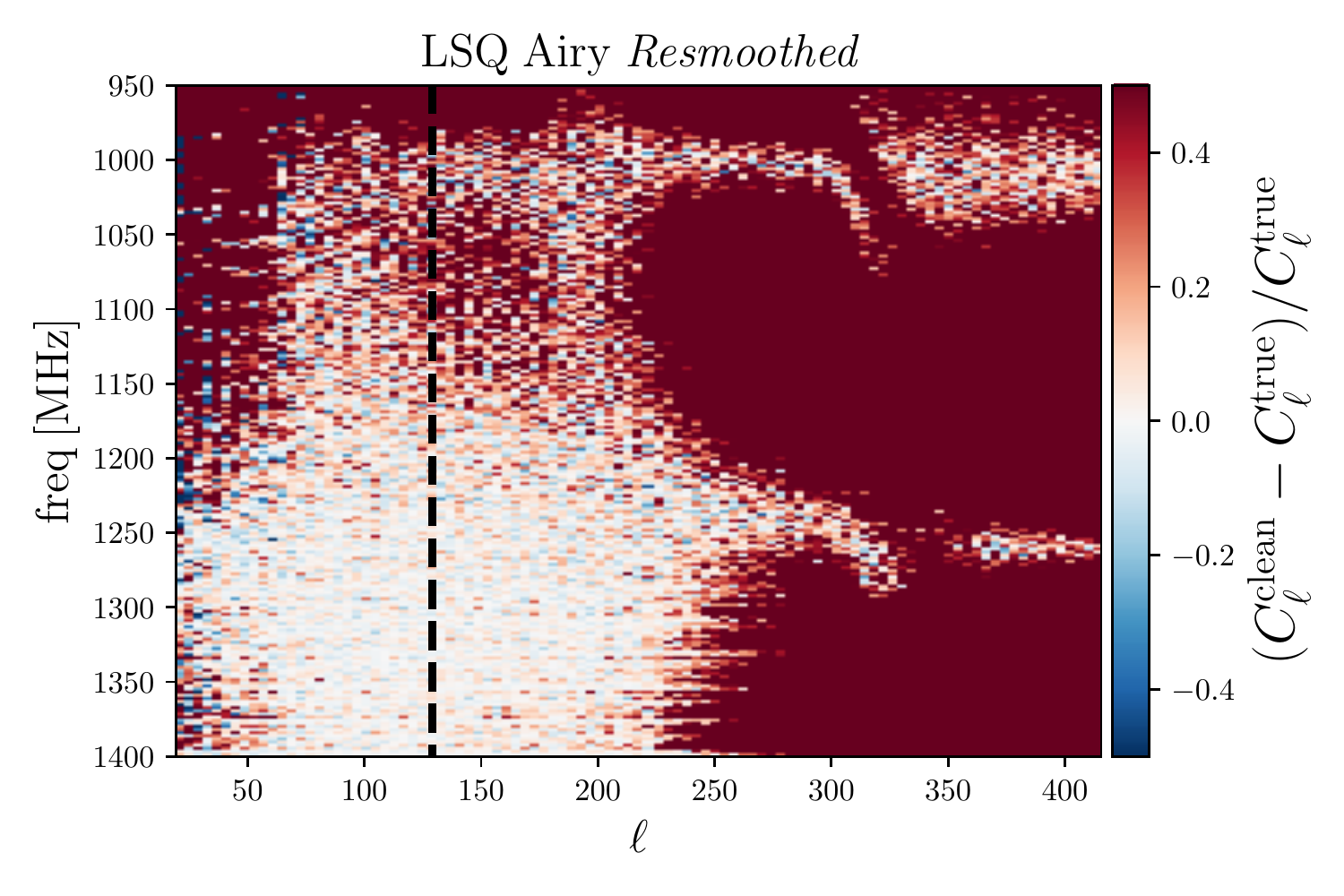}
   \caption{Same as \autoref{fig:unblind_res_cl}, but here we consider the LSQ method.}
   \label{fig:nonblind_gaussian_resmoothing}
\end{figure*}

In this section we report the results of the Blind Challenge.
We present a qualitative overview of the results for the original data-cubes in \secref{sec:original} and for the resmoothed data-cubes in \secref{sec:resmooth}. 
A comprehensive discussion on the relative performances of the various cleaning methods via quantitative metrics is in \secref{sec:spider}.

\subsection{Original data-cube}\label{sec:original}

\paragraph*{Gaussian beam.} In the top panel of \autoref{fig:various_methods_gauss} we show the angular power spectrum $C_\ell$ for a given frequency ($1225$~MHz as an example) for the Gaussian beam case and focusing on the more realistic PSM foreground model. The reconstructed signal is consistent across pipelines,  at least at large scales ($\ell \lesssim 250$), and comparable with the expected input signal. As the beam model starts suppressing the signal, small differences among methods are visible. The effect is slightly stronger for the MeerKAT case, where the noise level and beam suppression are higher.

In the lower panel of \autoref{fig:various_methods_gauss} we plot the line-of-sight power spectrum $P_{\rm los}$, again for the Gaussian beam case and PSM foreground model. At high and intermediate values of $k_\nu$, the cleaned $P_{\rm los}$ show behaviours in good agreement with the true \HI signal. 
At closer inspection, we can see that some of the methods tend to underestimate the signal's amplitude while others slightly over-predict it.
At low $k_\nu$, where most of the foreground power is, all blind methods show some level of over-cleaning, as it is extremely difficult to separate foregrounds from the signal in this region. On the contrary, the LSQ method over-predicts the signal, probably due to the leakage of foreground emission, which is not well isolated and removed by the method, into the \HI plus noise part.

\paragraph*{Airy beam.} The Airy beam model case shown in \autoref{fig:various_methods_airy} presents a more complex scenario. 
At the angular power spectrum level (top panels), there is qualitative agreement among pipelines, except for the LSQ method going astray after $\ell \gtrsim 250$.
Most notably, all cleaning methods consistently display a peak in the $P_{\rm los}$ around $k_\nu \sim 0.045 \, {\rm MHz}^{-1}$ (bottom panels). 
\citet{Matshawule2020} have identified and analysed a similar effect in their simulations. They attributed it to the presence of the $20$~MHz oscillation in the beam width as a function of frequency, which is enforced in their standard modelling of the FWHM of the main lobe in order to reproduce the holographic measurements of the MeerKAT beam by \citet{Asad2021}. 
In our case, the feature is caused by the (oscillating) changing positions of the side-lobes across the frequency band. We believe that the fact that both works find the oscillations at $\sim 20$\,MHz is a coincidence since both oscillations have different origins.

\paragraph*{Beam and foreground structure interaction.} In \autoref{fig:MS5vsPSM} we show the estimator $(P_{\rm clean}-P_{\rm true})/P_{\rm true}$, where $P_{\rm clean}$ is the power spectrum of the residual maps for a specific cleaning methods, while $P_{\rm true}$ is the input signal and noise. 
The {\it peak} feature in the $P_{\rm los}$ of the cleaned data completely disappears using the MS$_{05}$ foreground model, i.e., when the foregrounds are Gaussian. It implies that the more realistic Airy beam alone is not the cause of this effect, but it is instead its combination with the more structured PSM foreground emissions. The latter finding agrees with \citet{Matshawule2020}, and we qualitatively interpret it as follows. Since the sky temperature varies for different lines-of-sight, the frequency behaviour caused by the Airy beam gives rise to oscillations with slightly different amplitude as a function of direction. The different cleaning methods can spot the beam oscillations at the map level, but they tend to miss its exact amplitude in all lines-of-sight. If the sky is just a Gaussian realisation of a foreground-like power spectrum, as for the MS$_{05}$ model, these line-of-sight differences statistically cancel out, and no peak appears in the $P_{\rm los}$. We verified that the above conclusion holds even if the MS$_{05}$ model fluctuations are enhanced by two orders of magnitude. Indeed, running PCA cleaning on these artificial model with strong Gaussian foreground fluctuations (MS$_{05}$ x100) we still find no excess of power in the $P_{\rm los}$ at the scale corresponding to the oscillation in the beam side-lobe.  

On the contrary, due to the realistic sky structures of the PSM foreground model, there is no averaging effect and the $P_{\rm los}$ shows the clear excess at $k_\nu \sim 0.045 \, {\rm MHz}^{-1}$. From \autoref{fig:various_methods_airy} we see that the strongest peak feature in the $P_{\rm los}$ appears for the LSQ and poLOG pipelines, which have less line-of-sight freedom in adapting to the foregrounds. Indeed, even if the peak feature is observed in all methods, they experience it with different severity (see again bottom panels of \autoref{fig:various_methods_airy}). These considerations become important when one tries to mitigate the {\it peak} after the cleaning. For instance, if the contamination is limited to few channels one could flag and remove them from the analysis; on the contrary, artefacts affecting a larger $k_\nu$-range will be harder to handle. %

\medskip
We analyse in more detail the effect of the beam 
on the angular power spectrum. We plot in \autoref{fig:beam_effect} the $C_\ell$ of the cleaned residuals as a function of frequency on the vertical axis. On the left panels we report the PCA(b) method, showing that the cleaning performs differently going from the Gaussian beam case (top) to the Airy beam model (bottom). For comparison, we also plot the $C_\ell$ of the residuals in the Airy beam case for the \fastica(a) (top right panel) showing a similar effect to the PCA(b).
The interaction of the Airy beam with the spatial structure of the PSM foregrounds results in an excess of power at small scales in the residuals that evolves with frequency. As for the peak in the $P_{\rm los}$, the effect in the $C_\ell$ is present only in the cleaned maps and not in the original \HI
convolved with the Airy beam (see the lower panel of \autoref{fig:cl_input}). The poLOG method (lower right panel), which enforces smoothness by construction, is instead free of this small-scale frequency feature. 

We present in \autoref{fig:cellMSvsPSM} the angular power spectrum residuals for the GMCA method, where $C_\ell^{\rm clean}$ is the angular power spectrum of the cleaned maps (as in \autoref{fig:beam_effect}) and $C_\ell^{\rm true}$ is the one for the original foreground free \HI plus noise, convolved with the same beam model. The reconstruction is easier in presence of the simpler MS$_{05}$ foreground model and we find that this conclusion generally holds for all the cleaning methods. As expected from the results of \autoref{fig:MS5vsPSM}, the fringe pattern at small scales appears only for the combined presence of the Airy beam and the PSM foreground model.

\subsection{Resmoothed data-cube}\label{sec:resmooth}

As introduced in \secref{sec:preproc}, the foreground cleaning pipelines have been tested also on pre-processed data-cubes that have been resmoothed with a Gaussian beam to a common lower resolution. Even if this practice has been generally adopted in single-dish experiments to reduce the impact of instrumental systematics contaminating the data \citep[e.g., polarisation leakage as in][]{Switzer2013}, in our particular (and more idealised) simulated setup, 
we instead generally conclude that a simple Gaussian resmoothing does not ease the blind source separation process (especially forcing a simple Gaussian resmoothing in the Airy beam scenarios), although it partially reduces residual foreground contamination for the LSQ method. We now discuss this point in more detail.

\autoref{fig:unblind_res_pk} compares residuals looking at the line-of-sight power spectrum for the SKAO case. We show again the estimator $(P_{\rm clean}-P_{\rm true})/P_{\rm true}$, where now $P_{\rm true}$ is the power spectrum of the resmoothed input signal and noise. 

Top panels refer to the Gaussian beam, bottom to the Airy beam; on the right are the resmoothed cases.
The Gaussian and Airy cases have been already shown in \autoref{fig:various_methods_gauss} and \autoref{fig:various_methods_airy} but the \textit{relative} estimator allows a better quantification of the differences between the original and resmoothed scenarios. 
When maps have been resmoothed, we generally find more signal loss (i.e., a tendency to over-clean) for the blind source separation methods, and a slightly larger $k_{\nu}$-interval affected by the peak feature in the Airy beam case.
The parametric LSQ is an exception and we find that resmoothing helps the reconstruction of the signal. Indeed, the LSQ method performs power-law fits per-pixel across frequency and so relies upon a single pixel to represent the same area of sky across the frequency range. 
Although not leading to signal loss, the resmoothing procedure does slightly enhance the few percent oscillatory pattern arising for the poLOG method, which is probably linked to the specific polynomial truncation.

We show the effect of resmoothing on the angular power spectrum in \autoref{fig:unblind_res_cl}. We present results for mixGMCA while noting that all methods (excluded the particular LSQ case) behave similarly. Resmoothing the data-cube seems to slightly improve the recovery of the $C_\ell$ for the Gaussian beam case, whereas, in the Airy beam case, it only enhances the fringe patterns in the $(C_\ell^{\rm clean} - C_\ell^{\rm true})/C_\ell^{\rm true}$ behaviour (at $\ell \gtrsim 250$).

We report also that, coherently with the line-of-sight power spectrum, resmoothing improves the reconstruction of the angular power spectrum for LSQ method in the Gaussian cases and slightly in the Airy beam case, as shown in \autoref{fig:nonblind_gaussian_resmoothing}.

\medskip

Summarising, we find that for all pipelines the cleaning becomes more difficult in the presence of a more realistic telescope beam. Our simple Gaussian resmoothing does not amend this challenge. In the transverse direction, cleaning methods struggle where the signal clustering is subdominant compared to the noise. In the radial direction, the intermediate range in $k_\nu$ is the less compromised; although, when the spatially structured PSM foregrounds are coupled to the Airy beam, a {\it peak} feature appear for almost all cleaned data-cubes (see \autoref{fig:MS5vsPSM}).

\subsection{Quantitative Comparison}\label{sec:spider}

In this section, we present a set of metrics to allow a \textit{relative} comparison between the various cleaning methods in producing cleaned residual data-cubes whose power spectra reproduce those of the true cosmological signal plus noise.
A comparison in terms of preserved cosmological information is left for future work, whereas these power-spectrum-based metrics allow for an immediate and comprehensive view of the quality of the cleaning.

\subsubsection{Performance metrics}\label{sec:metrics}

\paragraph*{Angular power spectrum.}

The estimator for the accuracy of the recovered angular power spectrum is defined as
\begin{equation*}
    \eta_C(\ell, \nu) \equiv \left((C_\ell^{\rm clean} - C_\ell^{\rm true})/C_\ell^{\rm true}\right)(\nu)     
\end{equation*}
and varies substantially\footnote{The irregularity of our footprint makes the estimation of the angular power spectra also quite noisy, although this is not an issue for the relative comparison among pipelines' results. We leave studies on patch optimisation for future work.} across $\ell$ and frequency $\nu$, as can be seen in  e.g., \autoref{fig:unblind_res_cl}. 
We characterise its overall behaviour with the following metrics.

\begin{enumerate}
    \item[1) $\boldsymbol{\overline{{\rm rms}}}_{C\ell}$.] To have a first estimate of the quality of the cleaning, we compute the the root-mean-square (rms) value of $\eta_C(\ell, \nu)$ for every frequency $\nu$ 
    \begin{equation}
      {\rm rms}_{C\ell}(\nu) = \left(
        \frac{1}{\left(\ell_{max}-\ell_{\rm min}\right)}\sum_{\ell=\ell_{\rm min}}^{\ell_{\rm max}} \eta_C(\ell,\nu)^2 \right)^{1/2}\,,
    \end{equation}
        and define $\overline{{\rm rms}}_{C\ell}$ its mean value across the $N_\nu=512$ channels of our cleaned data-cubes,
    \begin{equation}
       \overline{{\rm rms}}_{C\ell} = \frac{1}{N_\nu} \sum_{i=1}^{N_\nu} {\rm rms}_{C\ell}(\nu_i) \, .
    \end{equation}
    We exclude scales larger than $\ell_{\rm min}=15$ and smaller than $\ell_{\rm max}=500$ 
    to reduce contamination from the mask and the noise, respectively.
    In general, the lower the value of $\overline{{\rm rms}}_{C\ell}$, the better the cleaning, with caveats that we try to track down with the next metrics.
    
    \item[2) $\boldsymbol{\sigma_{\rm rms}}$.] In order to capture the  channel-to-channel variability of the rms value, we also compute its scatter across frequencies  $\sigma_{\rm rms}$
          \begin{equation}
      \sigma_{\rm rms} = \left(\frac{1}{N_\nu} \sum_{i=1}^{N_\nu} \left({\rm rms}_{C\ell}(\nu_i)- \overline{{\rm rms}}_{C\ell}\right)^2\right)^{1/2}.
    \end{equation}
    A smaller value of $\sigma_{\rm rms}$ indicates that a certain method is more consistent in the reconstruction across channels (although it could indicate a consistently biased reconstruction).

    \item[3) $\boldsymbol{\Delta \ell}_{\rm max}$.] 
    A method that perfectly reconstructs large and intermediate scales while getting the noise floor wrong can have a rms higher than a method that is consistently biased at all $\ell$.
  We thus opt for a third metric that quantifies the cumulative number of $\ell$-bins across channels for which the agreement with the expected input signal is better than $30$ per cent, i.e.,
    \begin{equation}
        \Delta \ell_{\rm max} = \sum_{i=1}^{N_\nu} \left(
        \sum_{\ell=\ell_{\rm min}}^{\ell_{\rm max}} f_i(\ell) \right)\,,
    \end{equation}
 with 
    \begin{equation}
    f_i(\ell) = 
    \begin{cases} 
      1 & {\rm if }\, \left| \eta_C(\ell, \nu_i) \right| < 30\% \\
      0 & {\rm else}
   \end{cases} \,.
   \end{equation}

\end{enumerate}

\paragraph*{Line-of-sight power spectrum.}

\begin{figure}
    \centering
    \includegraphics[width=\columnwidth]{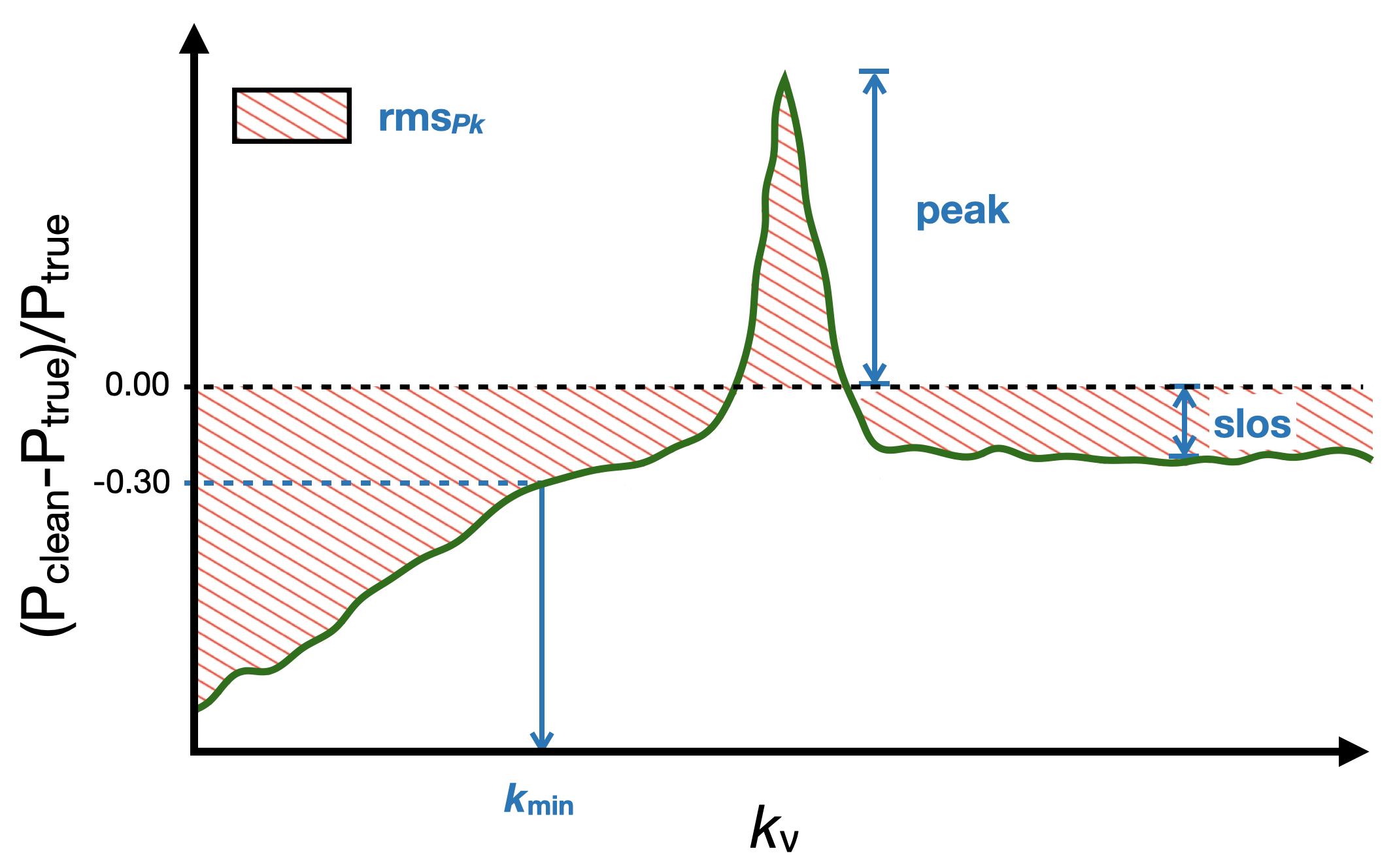} 
    \caption{Sketch summarising the four metrics used to compare the  submitted cleaned residuals against ground truth in terms of the radial $P_{\rm los}(k_\nu)$ power spectrum. The curve is a representative one as in \autoref{fig:unblind_res_pk}. We remind the reader that the {\it peak} feature appears only for the Airy beam model in combination with the non-trivial PSM foregrounds. See main text for definitions.}
    \label{fig:diagram}
\end{figure}

We now consider the radial direction and define
\begin{equation*}
    \eta_P(k) \equiv (P_{\rm clean}(k) - P_{\rm true}(k))/P_{\rm true}(k)\,.
\end{equation*}
Its generic behaviour is more consistent among methods and overall less noisy than the one for the angular power spectrum, as we can see in e.g., \autoref{fig:unblind_res_pk}, also due to the large number of pixels available in our patch. To characterise $\eta_P(k)$, we define the following metrics, also sketched in \autoref{fig:diagram}.

\begin{enumerate}
    \item[1) \textbf{rms$_{Pk}$}.] As for the angular estimator, the first quantity to assess is the distance of the recovered signal from the input one, through the rms value of $\eta_P(k)$,
    \begin{equation}
      {\rm rms}_{Pk} = 
        \left(\frac{1}{N_{k}}\sum_{i=1}^{N_k} \eta_P(k_i)^2 \right)^{1/2},
    \end{equation}
    where $N_k$ is the number of $k$-bins.
    The lower the rms, the more successful the cleaning. 
    
    \item[2) \textbf{slos}.] As noted already in \autoref{fig:unblind_res_pk}, the estimator $\eta_P(k)$ shows a roughly constant bias at small scales for most cleaning methods. Due to over-cleaning, this bias is often negative and an indicator for cosmological signal loss. We thus define slos as the mean value of the estimator for $k_{\nu}>0.1\:{\rm MHz}^{-1}$. Despite the name, residual foregrounds in the cleaned maps may give rise to a positive value for slos. In general, the higher the absolute value for slos, the worse the cleaning performance.
    
    \item[3) $\mathbf{k_{\rm min}}$.] Due to their coherence in frequency, the foreground emission has power predominantly at small $k_\nu$, making these scales the most difficult to recover. To characterise the extent of this contamination, we define $k_{\rm min}$ as the smallest $k_\nu$ 
    at which the residual $P_{\rm los}$ starts deviating more than $\pm 30$ per cent from the expected cosmological signal (i.e., $|\eta_P(k_{\rm min})| = 0.3$).
    A smaller $k_{\rm min}$ indicates a more successful cleaning that extends on a larger range of scales.

    \item[4) \textbf{peak}.] As already mentioned, in the Airy beam case coupled to the PSM foreground model, we observe a spiky feature in the $P_{\rm clean}$ and thus also in $\eta_P$ (see \autoref{fig:unblind_res_pk} and \secref{sec:original} for a more thorough  discussion). 
    The height of this peak is proportional to the extent in $k_\nu$-range in which the spurious artefact appears, and we decide to include it in our set of metrics for the cleaning quality (the higher the peak value, the worse the cleaning performance).
\end{enumerate}

\subsubsection{Method performance ranking}\label{sec:performances}

\begin{figure*}
   \includegraphics[width=\columnwidth]{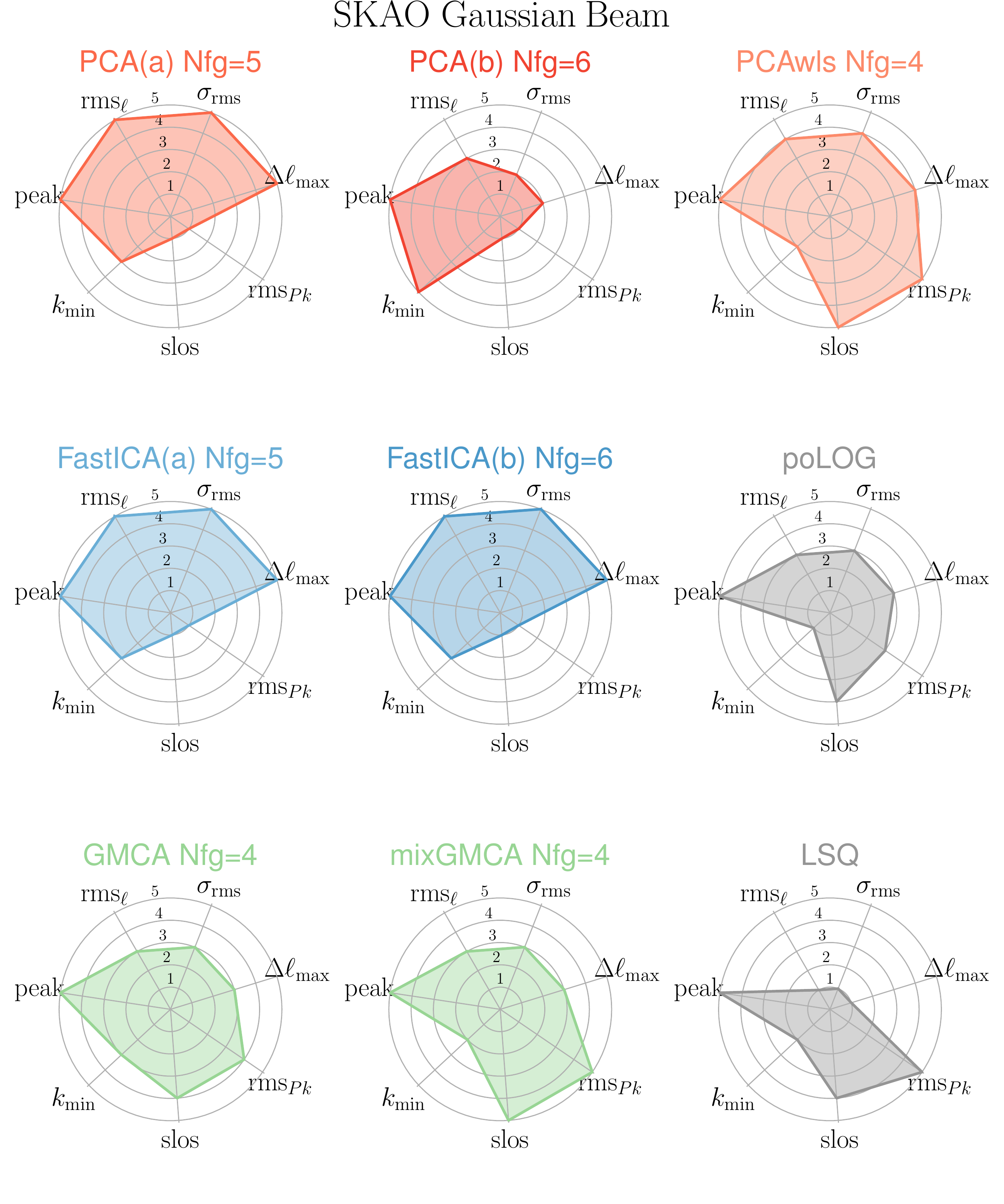}\hspace{0.5cm}
      \includegraphics[width=\columnwidth]{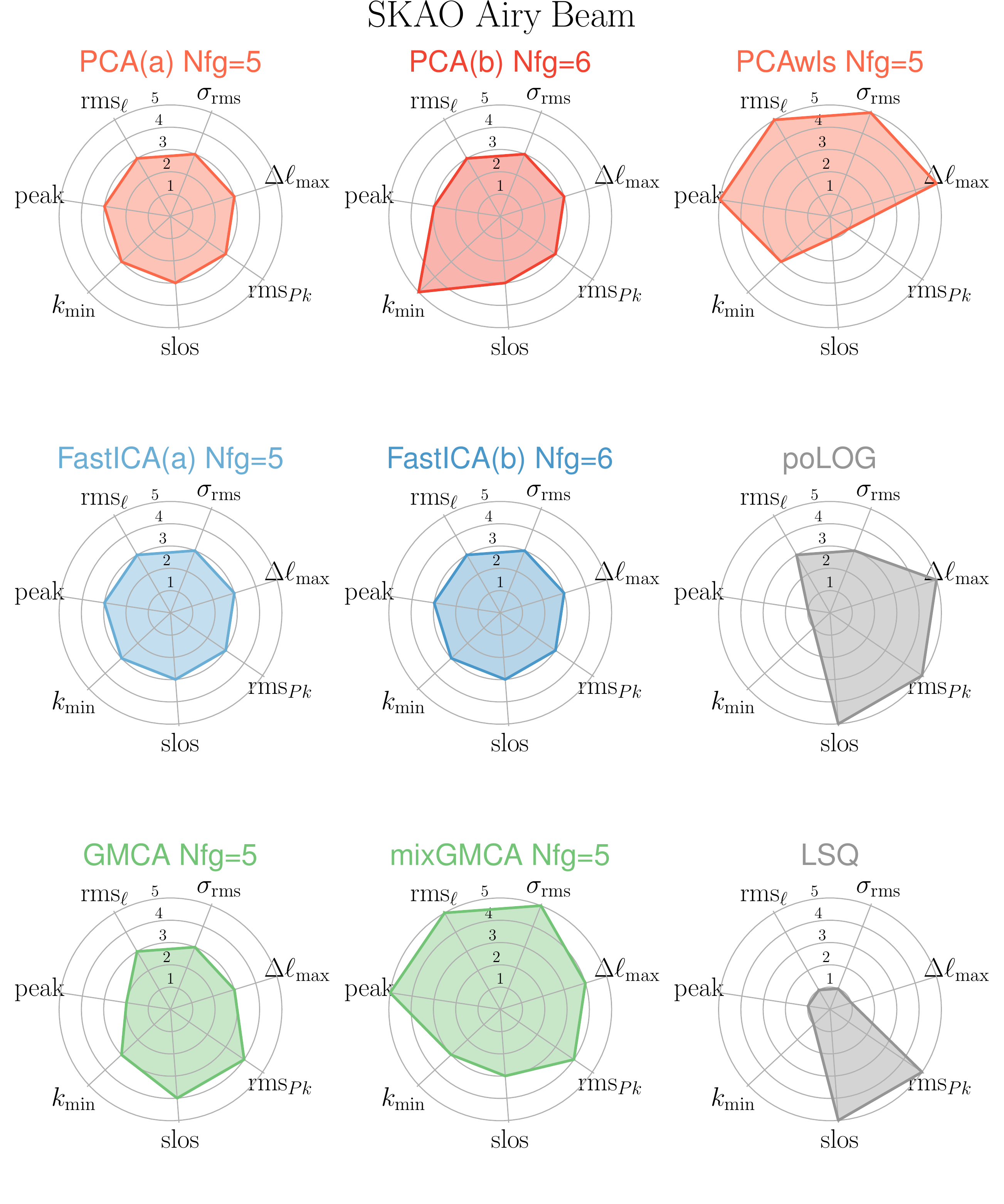}\\
      \medskip \noindent
         \includegraphics[width=\columnwidth]{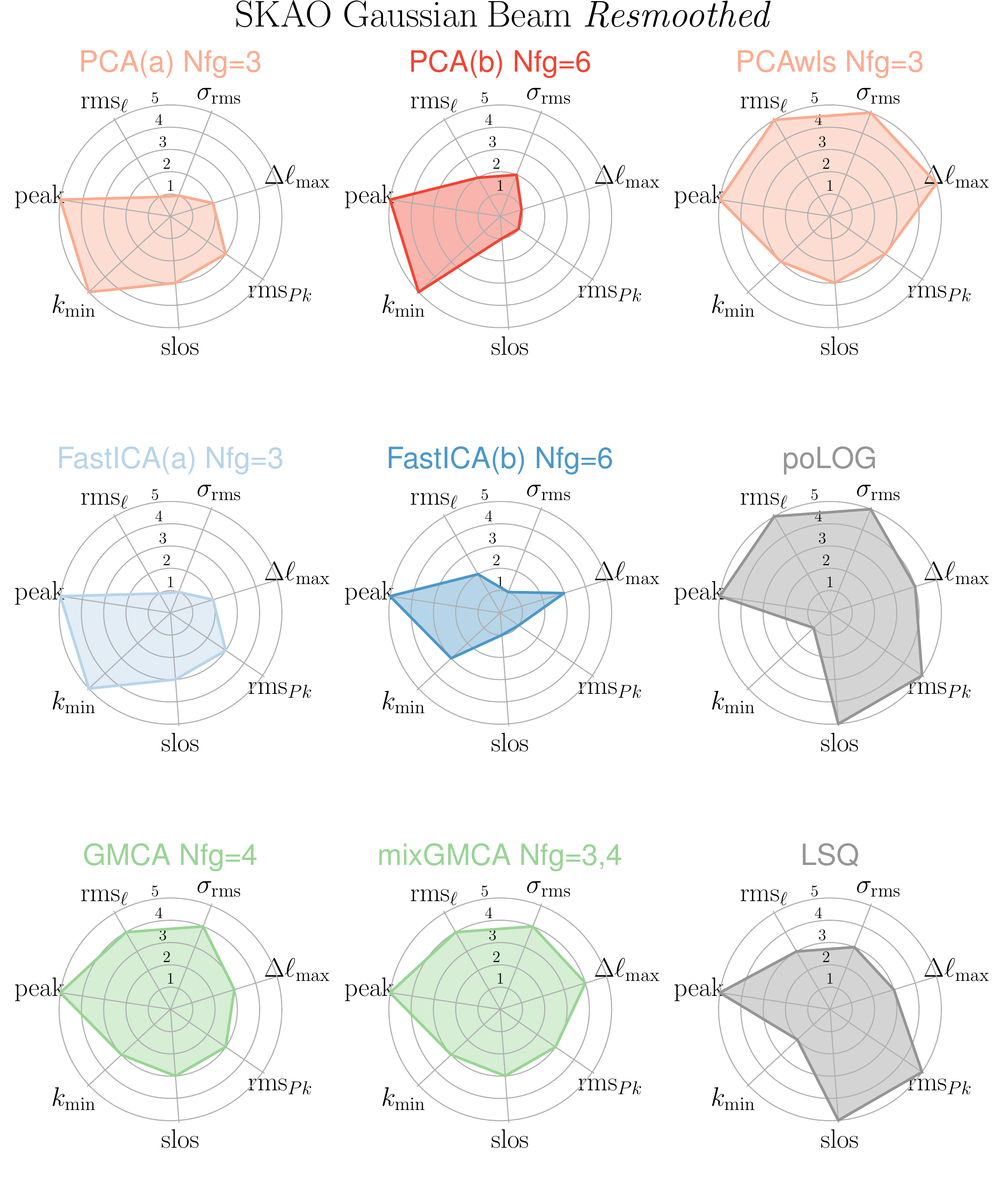}\hspace{0.5cm}
            \includegraphics[width=\columnwidth]{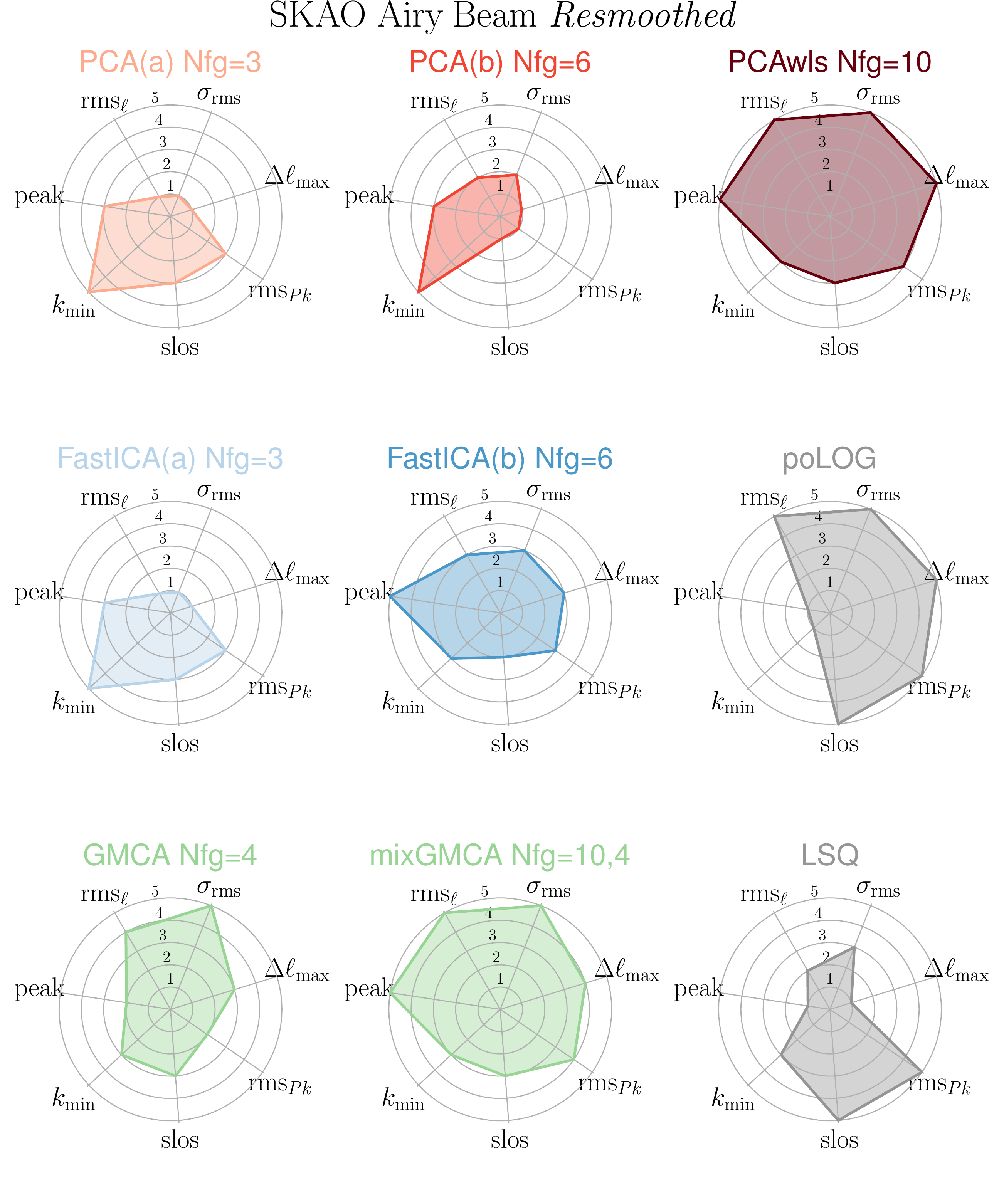}
   \caption{Radar charts showing the performance of the various methods on the different metrics defined in \secref{sec:spider} for a SKAO-MID IM survey and divided in four different panels one for each of the Gaussian/Airy beam model or original/resmoothed combination. For a given metric, we marked each method from 1 to 5, depending on the {\it relative} quality in the cleaning ($1=\,$worst, $5=\,$best); hence, the bigger the area covered by the chart, the better the overall performance.
   In the scenarios with no {\it peak} feature (i.e., Gaussian Beam) we assign a 5 to all pipelines to keep the 7-edge structure for the radar charts.
   Methods are colour-coded: 
   PCA pipelines in red, \fastica in blue, GMCA in green and non-blind methods in grey. For each blind method, we report also the number of subtracted components $N_{\rm fg}$, and the intensity of the colour is scaled proportionally (darker colour corresponds to higher \Nfg) to help the reading. mixGMCA is associated with two different $N_{\rm fg}$: the first for the largest scale PCA and the second for smaller scales GMCA (see also \autoref{table:Nfg}). 
   }
   \label{fig:spiderSKAO}
\end{figure*} 

\begin{figure*}
   \includegraphics[width=\columnwidth]{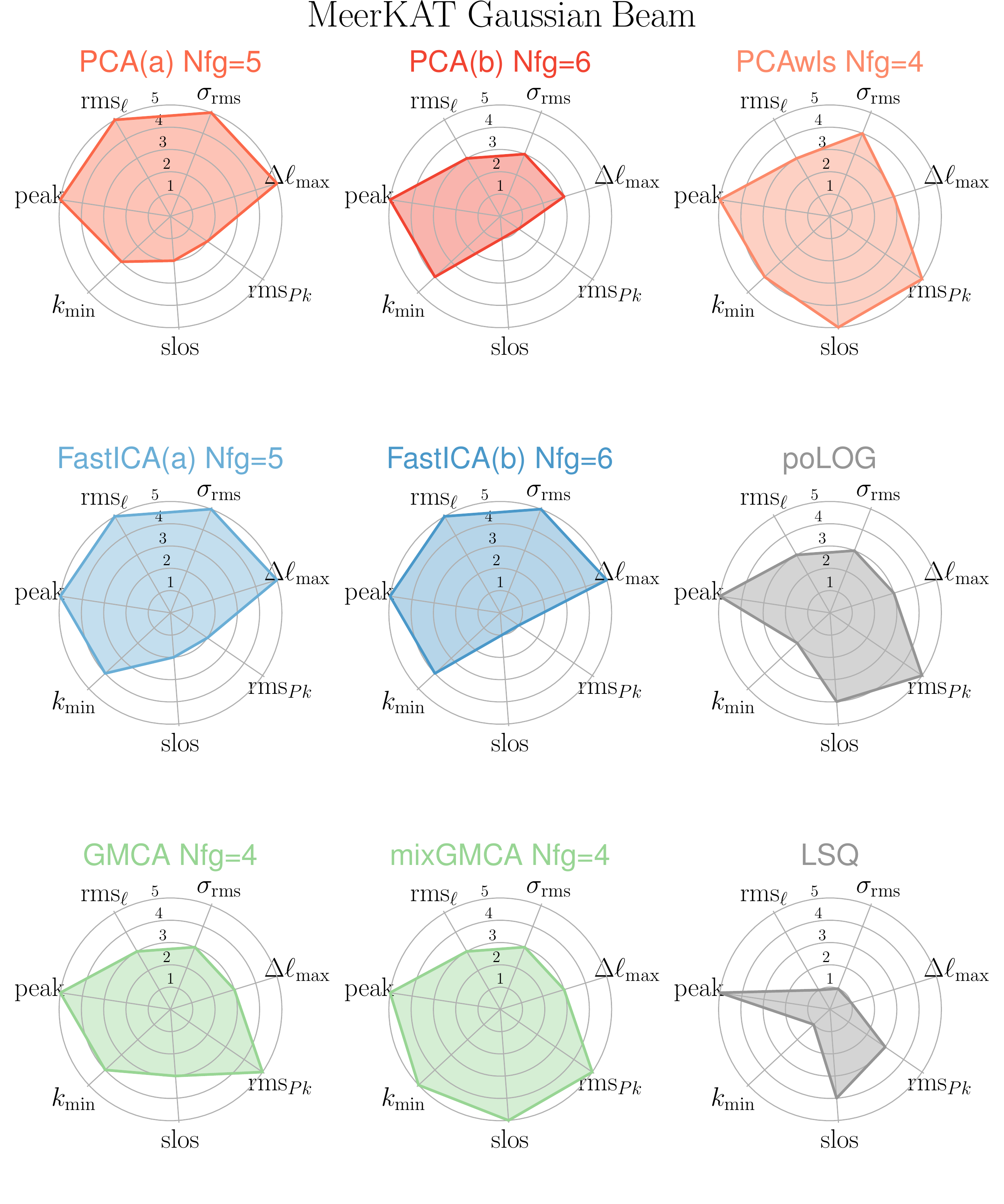}\hspace{0.5cm}
      \includegraphics[width=\columnwidth]{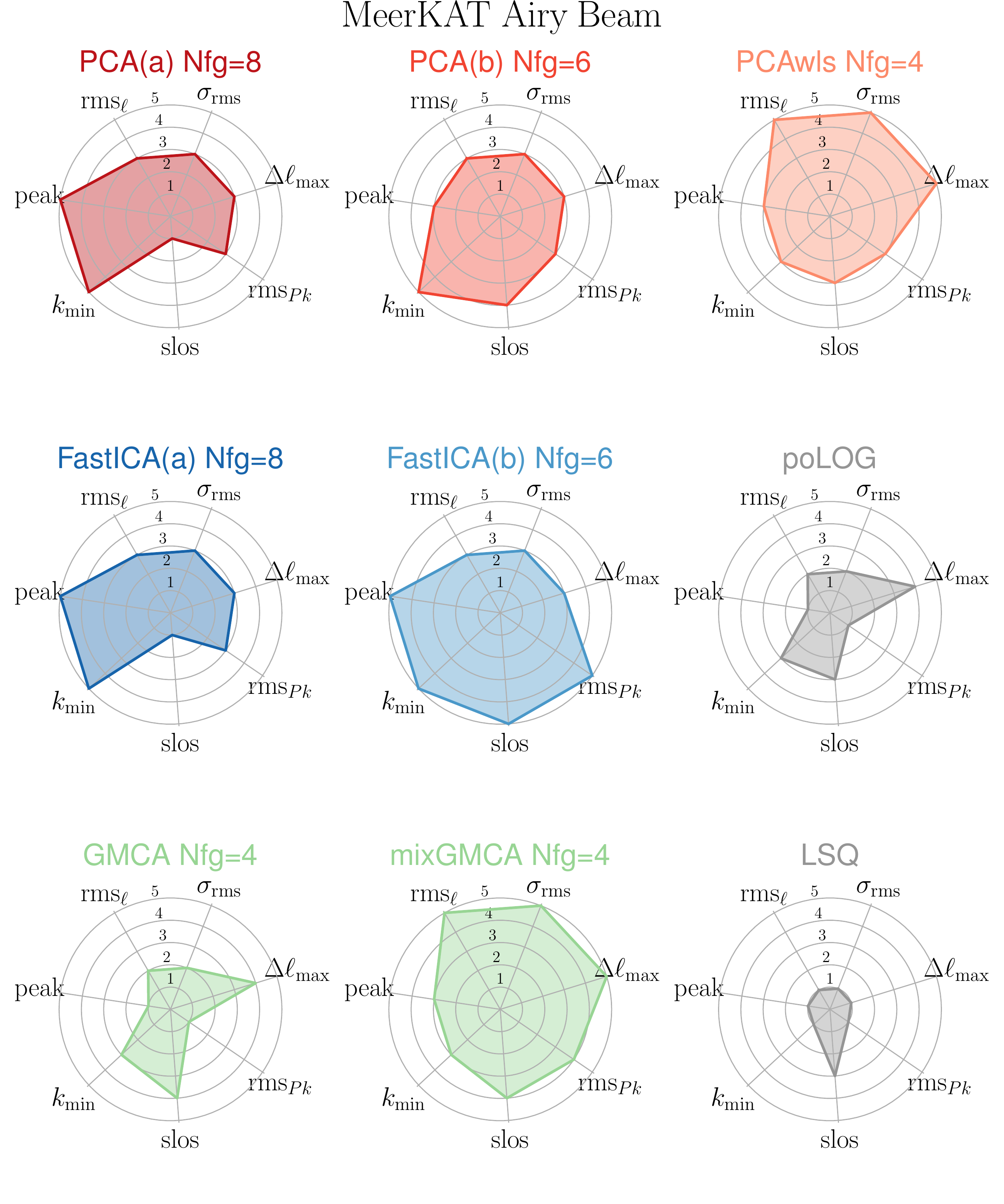}\\
      \medskip \noindent 
         \includegraphics[width=\columnwidth]{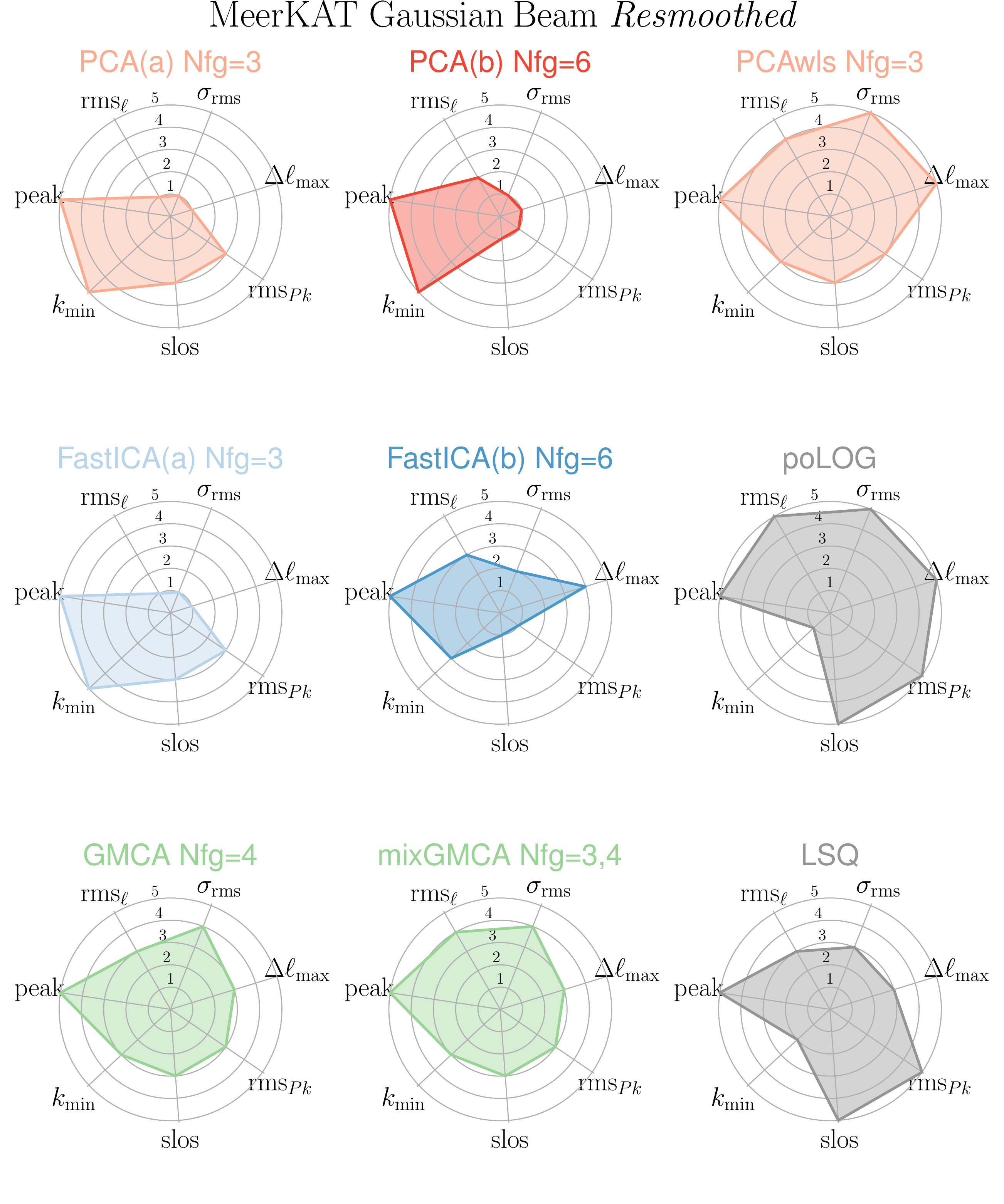}\hspace{0.5cm}
            \includegraphics[width=\columnwidth]{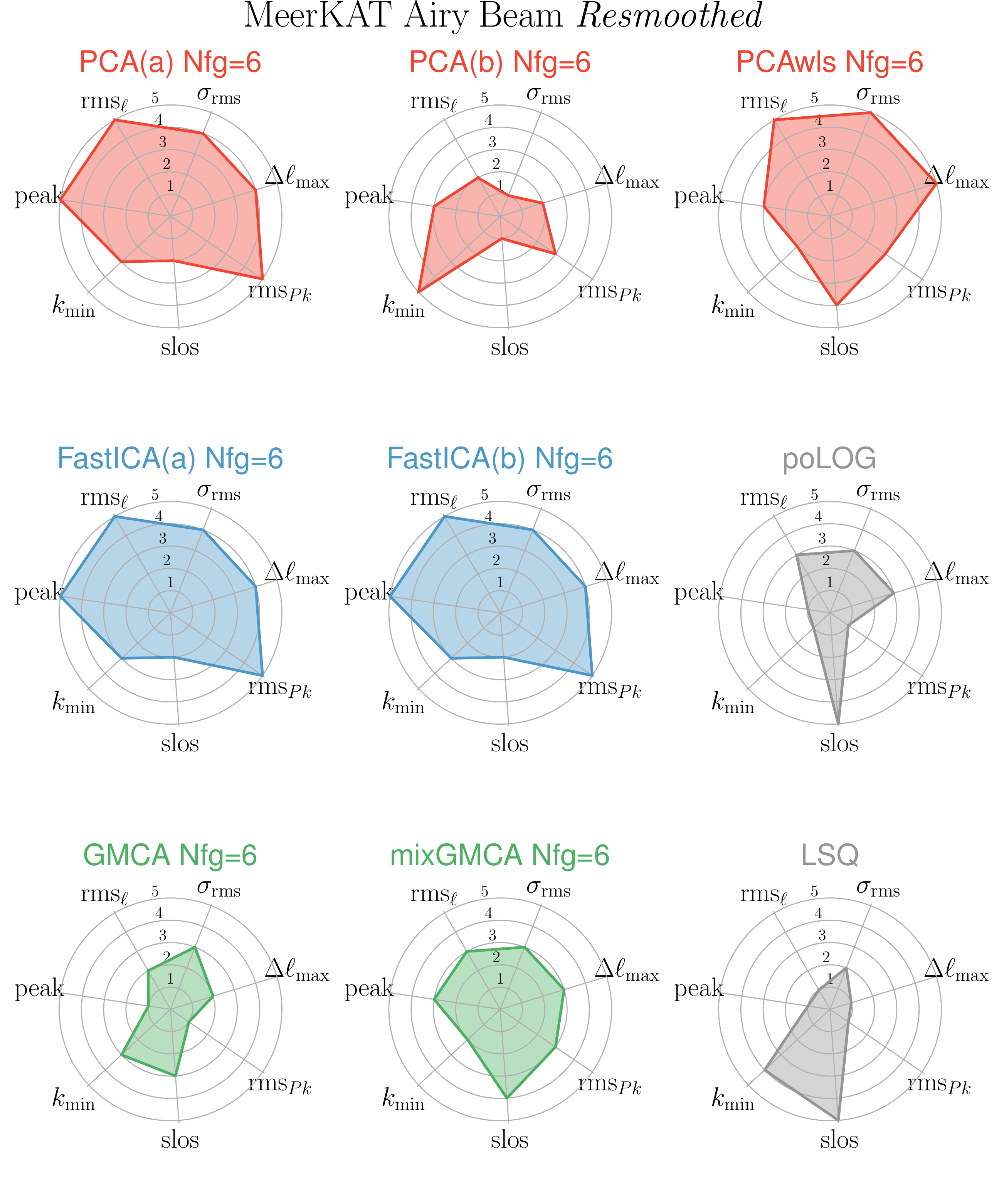}
   \caption{Same as \autoref{fig:spiderSKAO} but for the MeerKAT scenario instead of the SKAO one.
   }
   \label{fig:spiderMK}
\end{figure*}

We evaluate the metrics described above for all submitted residual data-cubes; for each of the sixteen setups and each of the seven metrics, we have a distribution of values (one for each pipeline, see \autoref{table:spiderSKA} and \autoref{table:spiderMeerkat}). We mark each pipeline from 1 to 5 depending on their \textit{relative} performance: the best method scores 5 and the worst method 1, the other marks are assigned binning the interval defined by the two extremes. The binning allows multiple methods to score the same value, including the two extreme ones. 

To visualise the seven metrics together (three for the angular plus four for the line-of-sight power spectra), we compile a radar chart for each submission, where the area covered by the chart relates to the cleaning performance: the larger the area, the more accurate the cleaning. 
We focus on the more realistic PSM foreground model to draw conclusions and present the SKAO cases in \autoref{fig:spiderSKAO} and the MeerKAT cases in \autoref{fig:spiderMK}. Each figure consists of four quadrants: the left column refers to the Gaussian beam cases, the right to the Airy beam, with the corresponding resmoothed scenarios on the second row. We display nine radar charts in each quadrant, one for each method that joined the Challenge. The title of each chart details 1) the method it refers to and 2) the number of sources removed, \Nfg (where applicable). Both are also present in the colour-coding: green for PCA, blue for \fastica, green for GMCA, and grey for poLOG and LSQ, and the intensity of the colour is proportional to \Nfg. We can think of \Nfg as an extra parameter and dimension of the radar charts, since when interpreting the performances of the methods, one should take \Nfg into account. For instance, it is generically true that, in a given observational setup and cleaning method, the higher \Nfg, the more the loss of cosmic signal (e.g., see bottom panel of Figure 4 in \citet{Cunnington:2020njn} and discussion therein).

To preserve the same 7-edge structure for all the radar charts, we decide to show a {\it peak} rating for the cases with no peak (i.e., Gaussian beam), assigning a 5 to all methods.

Looking at \autoref{fig:spiderSKAO} and \autoref{fig:spiderMK}, we can generically conclude that no method clearly outperforms the others and that the efficiency of a given method can vary when facing different types of dirty data-cubes. 
Nevertheless, the richness of these results allows us to understand and highlight different issues related to the contaminants cleaning problem and the methods used to face it. In the next section, we discuss the latter and attempt a comprehensive comparison of the methods' performances for all simulation setups involved in the Challenge.

\section{Discussion}
\label{sec:discussion}

All pipelines show some strengths at different observables and metrics. 
Here, we discuss results by dividing them into component separation method employed.

\paragraph*{PCA performance.}
PCA(a) reconstructs well the angular power spectrum for the Gaussian beam model, in both the SKAO and MeerKAT cases (upper left panel of \autoref{fig:spiderSKAO} and  \autoref{fig:spiderMK}); interestingly, PCA(b), adopting a similar $N_{\rm fg}$, seems to struggle more in the reconstruction of the $C_\ell$. This is particularly true at low frequency, as we notice comparing the upper left panel of \autoref{fig:beam_effect} with \autoref{fig:cl_input}. 
This behaviour is due to the inverse rms weighting used in PCA(b). We have checked this hypothesis after analysing the performance of the submissions and unblinding the results. We re-ran the PCA(b) pipeline (with same parameters) removing the weights when computing the data covariance in \autoref{eq:InverseNoiseCov}. Doing so, and after having the non-weighted PCA(b) go through our performance pipeline, we conclude that the chosen weighting was indeed the reason for a bad reconstruction of the low frequencies. 
Indeed, the data-cubes are characterised by a rms inversely proportional to frequency. I.e., the lower frequency channels are less taken into account by PCA(b), therefore the highest eigenvalues come mainly from the higher frequencies at a fixed value of $N_{\rm fg}$, that forces the shape of the more structured residuals of the higher frequencies to the whole channel range. 
The weighting scheme used in PCA(b) was intended to minimise the influence of noise in the component separation, but down-weighting the lower frequencies actually has proven to be detrimental for the cleaning process. Although this inverse frequency band rms weighting is non-beneficial with these realistic simulations, we cannot discard weighting schemes in general - as for example pixel rms weighting schemes. We will implement these schemes in future work. 

PCAwls shows good performances across all setups and, interestingly, $C_\ell$ are even better reconstructed in presence of the Airy beam model, contrary to what we observe for PCA(a). 
For the original data and in presence of the Gaussian beam model, PCAwls, together with GMCA and mixGMCA, seems to also recover very accurately the radial power spectrum signal (see also the upper left panel of \autoref{fig:unblind_res_pk}). 
The results slightly worsen for the radial power spectrum metrics when moving away from the original data-cube with the Gaussian beam model. This can be seen also in \autoref{fig:unblind_res_pk}, where we notice an increment in the signal loss for all blind methods: the bias at small scales change from less than $\sim20$ per cent to more than $\sim 25$ per cent.
Despite this, PCAwls shows consistently high performances across all metrics and cases. Results are similar for both the original and the resmoothed case, while PCA(a) and (b) typically worsen the quality of the cleaning in the latter case.

\paragraph*{\fastica\ performance.}
\autoref{fig:spiderSKAO} and  \autoref{fig:spiderMK} show that \fastica\ does not improve on PCA, as already discussed in the context of simulations in e.g., \citet{Alonso:2014dhk,Matshawule2020,Cunnington:2020njn}. We recall instead that the application of these two techniques on real data suggests an interesting complementarity and more conservative cleaning results for \fastica\  \citep{Wolz:2015lwa}.
\fastica(b) is more robust than PCA(b) in the low-frequency reconstruction since only the former has been run adopting an rms weighting of the frequency channels.
Moreover, we found good agreement between the two implementations of \fastica, especially for the original data-cubes, where also the \Nfg chosen is similar.

\paragraph*{On \Nfg and resmoothing.} The results for the PCA and \fastica\, residuals presented above highlight two important points: 1) even when using the same $N_{\rm fg}$, the specific implementation of a method and the pre-processing choices (e.g., mean-centring the maps, weighting scheme) play a non-negligible role; 2) the resmoothing of the maps with an extra Gaussian kernel may redistribute information among eigenvectors, suppressing the number of relevant eigenvalues of the frequency-frequency covariance matrix of the data-cube. This may mislead the \Nfg choice. 

\begin{figure}
   \centering
   \includegraphics[width=\columnwidth]{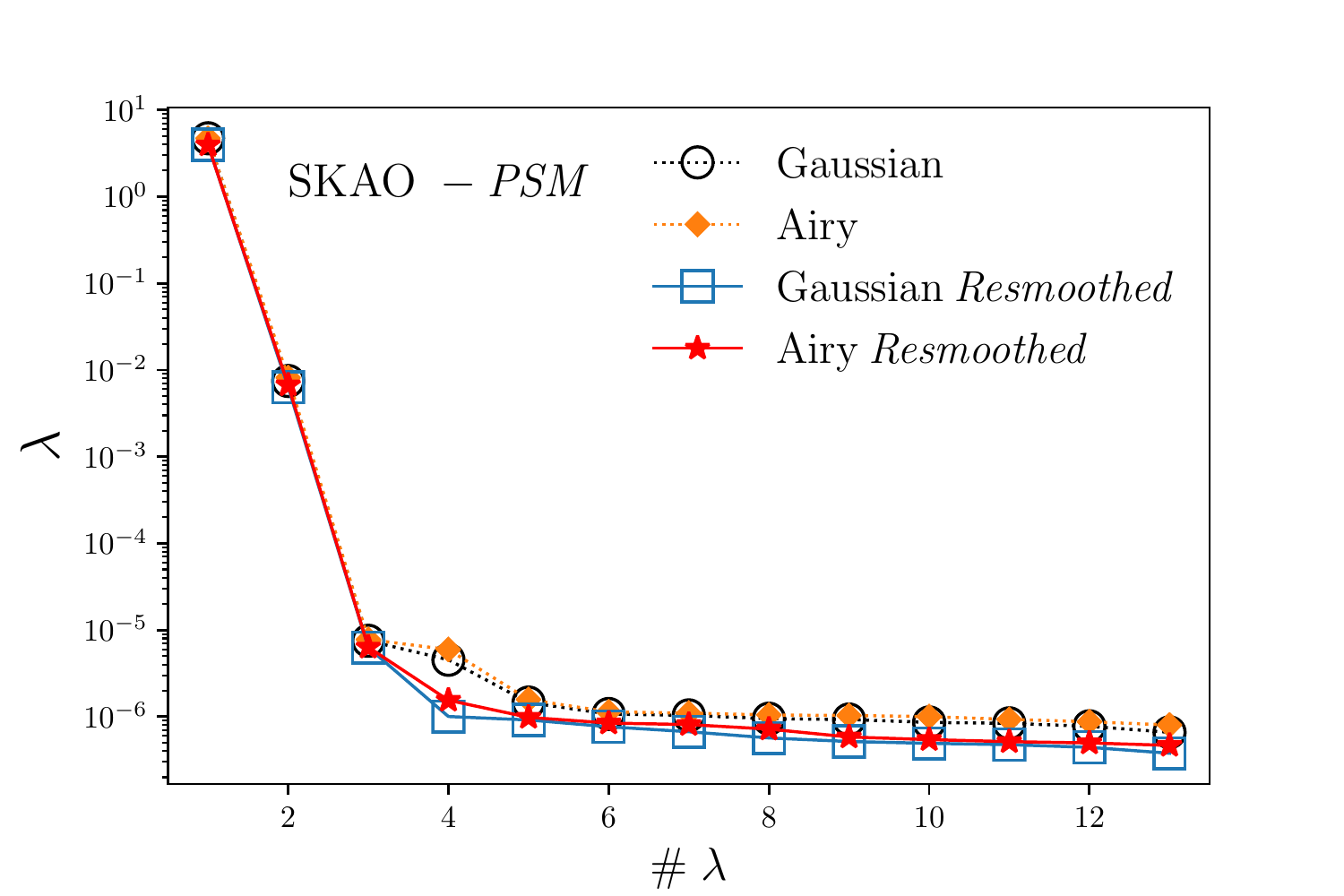} 
   \caption{Ordered eigenvalues of the frequency-frequency covariance of the SKAO - PSM foreground model data-cubes for different beam model and pre-processing options. 
   }
   \label{fig:eigen}   
\end{figure}

We show in \autoref{fig:eigen} the ordered eigenvalues of the frequency-frequency covariance of the SKAO - PSM foreground model data-cubes corresponding to different beam models and resmoothed or not scenarios. As discussed in \secref{sec:PCA}, one criteria for determining \Nfg is to recognise the number of clearly dominant eigenvalues, as the dominant modes are expected to contain most of the foregrounds. Resmoothing redistributes the power of these modes,
potentially suggesting a lower $N_{\rm fg}$.
However, despite the effect on the eigenvalue spectra, our analysis indicates that keeping the same or decreasing \Nfg in the resmoothed cases does not lead to a good cleaning performance: in most cases it led to more under-cleaning in the $C_\ell$ and more over-cleaining in the $P_{\rm los}$.
We stress again that the poor performance of the resmoothing depends on the simulation specifics. Different, more subtle, systematics and real observation contaminants could instead benefit from this type of pre-processing. Moreover, a Gaussian deconvolution was possibly not accurate enough for the Airy beam case (see also \citealt{Matshawule2020}). We postpone a more detailed study of resmoothing to future works.

\paragraph*{GMCA performance.} GMCA and mixGMCA perform similarly in the case of the Gaussian beam model (for both the original and resmoothed cases); indeed, they reconstruct relatively well the radial power spectrum for the original data set (see in particular \autoref{fig:spiderSKAO} and the upper left panel of \autoref{fig:unblind_res_pk}) and the angular power spectrum in the resmoothed cases. Looking at mixGMCA in \autoref{fig:unblind_res_cl} we can get also an idea of the absolute performance of the cleaning on the $C_\ell$: the reconstruction agrees with the input signal better than few percent for a large range of $\ell$ and frequencies. 
As expected, the reconstruction is more difficult for the scales and frequencies more affected by the beam. 

In the more realistic case of the Airy beam model, mixGMCA is better than the GMCA cleaning. Our interpretation is the following. With the Airy beam at play, the morphology of the maps becomes more complex, especially at small scales. The GMCA algorithm seeks and catches those new features and decomposes the signal accordingly, paying most care on those small-scale structures that well satisfy the sparsity assumption. In other words, while performing the source separation process, GMCA decomposes the data-cube in the \Nfg sources that best characterise the small scales, while neglecting the larger (smoother and less sparse) scales. mixGMCA overcomes this problem treating separately the large coarse scale (as decomposed by the wavelet transform) with a PCA cleaning and, moreover, disentangling the \Nfg needed.

\paragraph*{poLOG performance.} The poLOG method 
performs well in presence of a Gaussian beam, and in particular on the resmoothed data-cubes. The mean level of the radial information is correctly reconstructed. However, it is possible that the metrics do not penalise enough the small but present oscillatory pattern shown in \autoref{fig:unblind_res_pk} (see also \citet{Ghosh2011_oscillations,Ghosh2011_resmoothing}).
The results are almost equally good in presence of the Airy beam for the SKAO case, although the quality of the cleaning lowers for the MeerKAT case, possibly due to the more prominent side-lobes of its beam. Interestingly, when the Airy beam is considered, while the peak is clearly visible in the reconstructed radial power spectrum, the angular power spectra, due to the smoothness assumption of poLOG, do not present the typical fringe pattern at high $\ell$ (see \autoref{fig:beam_effect}). 

\paragraph*{LSQ performance.} The LSQ method relies on more physical modelling of the foreground and assumes an a priori knowledge of the monopole of the maps. 
Unsurprisingly, the mean level of the radial information is always correctly reconstructed given the perfect knowledge of the monopole.
LSQ performs satisfactorily in the case of a Gaussian beam model for the resmoothed case, while struggling in the case of the more complex Airy beam model.
Overall, we see that this parametric method is not sophisticated enough to deal with realistic data-cubes. A way forward could be to upgrade it and include a modelling of the specific instrument beam and noise properties.

\paragraph*{On the low frequency channels.} 
From all the angular power spectrum figures presented so far, it is evident that all pipelines 
find more challenging the recovery of the input signal and noise in the lowest frequency channels considered.
We believe that this is due to a combination of effects including 1) the stronger beam suppression and 2) the relative lower intensity of the \HI signal with respect to instrumental noise level (see \autoref{fig:cl_input}). The latter point depends both on the specific \HI model, and on the intrinsic channelisation of the IM experiments (constant channel width corresponds to thicker redshift slices at low frequency, where the clustering of the cosmological signal gets averaged more).

\section{Conclusions}
\label{sec:conclusions}

\subsection*{Summary of the Challenge}
In this work, we presented a Blind Foreground Cleaning Challenge on a realistic set of low-redshift \HI IM simulations for a $\sim 5000$ deg$^2$ single-dish survey with MeerKAT or the SKAO-MID telescope. The simulations, covering the $950-1400$ MHz range, include a \HI signal generated by combining a semi-analytical galaxy formation model with a cosmological halo simulation, and astrophysical foregrounds, generated using two alternative models:
a Gaussian realisation of the foreground 2-point statistic and frequency scaling properties, and a more empirically informed one, based on the Planck Sky Model.
We simulated instrumental effects through a commonly used Gaussian beam and an Airy beam model that includes side-lobes. We modelled a fixed-elevation scanning strategy resulting into a non-homogeneous noise level. 

In summary, the various setup combinations resulted in sixteen {\it dirty} data-cubes to be cleaned, resulting in increasingly realistic scenarios which allow a gradual understanding of the role of individual observational features in the cleaning process.
Nine foreground cleaning pipelines joined this first Blind Challenge, i.e., without prior knowledge of \HI signal, foregrounds, beam model and noise level. Seven of the pipelines (versions of PCA, \fastica\, and GMCA) linearly decompose the given data-cube leveraging statistical properties of the foreground components such as non-Gaussianity or sparsity. The other two methods either impose the foreground smoothness in frequency (polynomial fitting) or make physical assumptions on the foreground properties (least-squares fitting).
Testing many different methods on the same simulation allowed us to quantify their \textit{relative} accuracy on cleaning. We devised a set of criteria to describe the quality of the cleaned residuals in terms of their angular and the line-of-sight power spectra and presented their relative performance using radar charts (see \autoref{fig:spiderSKAO} and \autoref{fig:spiderMK}).

\subsection*{Lessons learned}
Our results suggest that, even among similar methods, subtleties related to each specific implementation can lead to substantial differences in the cleaning performance, and that the choice of \Nfg is not easily deducible and objective without extra prior information of the signal. 
Nevertheless, in presence of a Gaussian beam, all pipelines (with the exception of least-squares fitting) are capable of recovering within $20$ per cent the input power spectra in the frequency range and spatial scales with the least beam suppression.

Interestingly, when the more realistic Airy beam model is considered in combination with the non-Gaussian PSM foregrounds, the cleaning is more complicated and the residuals show 1) a clear spike-like feature in the line-of-sight power spectrum and 2) a fringe pattern in the $C_\ell$ at small angular scales, caused by an oscillation in frequency of the beam side-lobes positions. By enforcing smoothness, the polynomial fitting method is the only exception, not inducing the latter effect on the angular power spectrum.
These systematics are caused by the interaction of spatially structured foregrounds with the far side-lobes of the primary beam. We expect these effects to worsen in the presence of stronger point sources \citep{Matshawule2020} or for observations closer to the Galactic plane. In general, also strong Galactic emission at more than $30\ \deg$ from the line-of-sight could play a role, implying that accurate measurements of the \textit{full} primary beam response will be critical for the success of SKAO, MeerKAT, or any single-dish \HI intensity experiment.

We found that resmoothing with a Gaussian kernel 
does not improve the absolute performance of the cleaning (with the exception of the least-squares fitting method). However, 1) our simulation does not include some challenging systematics --such as polarisation leakage-- that could be mitigated by an aggressive resmoothing\footnote{\citet{McCallum2021} have recently proposed to suppress polarisation leakage at the map-making stage.} and 2) a more accurate deconvolution model including side-lobe structure should be used. 
Most existing cleaning methods do not directly use any beam information during the component separation process, while our results highlight the need for a more accurate treatment of the beam. 
More sophisticated strategies are possible, for example performing component separation and deconvolution simultaneously \citep[e.g.,][]{carloniGertosio2021}.

In general, we conclude that methods based on statistical properties of the data (PCA, \fastica, GMCA, mixGMCA) should be generally preferred to parametric ones, given the current knowledge of foregrounds at the relevant frequencies combined with the systematic effects.

We find that implementing the cleaning in parallel with more than one method is an excellent practice to unveil different data characteristics. 
Indeed, a source separation method is more efficient than another if its assumptions suit the data better, helping develop ad-hoc cleaning strategies.
For instance, we report that mixGMCA, a hybrid PCA-GMCA algorithm,
has shown overall improvement compared to its parent methods and the best consistency among all scenarios (i.e., its performance is satisfactory in all cases). 
Hybrid approaches have the potential to retain the advantages of each of the methods that compose it.
In particular, mixGMCA removes the brightest diffuse astrophysical contamination with PCA on large scales while carefully handling the small-scale instrument-driven defects in the maps with GMCA.

\subsection*{Perspectives}
In this work, we explored several methods available in the literature, making it the most comprehensive study so far for post-reionization \HI IM foreground cleaning. Nevertheless, more methods could be tested on our end-to-end simulations (e.g., GNILC \citep{Olivari2016,Fornazier2021}, GPR \citep{Mertens2018,Soares2021}, KPCA \citep{kpca}).
Known systematics could also be included, such as polarisation leakage \citep{crime,Shaw2015,Spinelli2018}, satellites contamination \citep{Harper2018}, strong RFI-flagging \citep{Carucci:2020enz}, 1/f  noise \citep{2018MNRAS.478.2416H, yichao, chen}, point source masking \citep{masks}, and a more realistic description of the system temperature \citep{Wang2021}.
As more IM data will be available, it will be possible to understand new observational effects and systematics and include them in the modelling, also paving the way to simulation-based learning algorithms for addressing foreground cleaning.

\medskip
This first Challenge is designed as the baseline case to test the ability to recover the \HI cosmological signal, including realistic observational effects. These simulations lay the ground for developing more complex and detailed end-to-end simulations necessary to improve foreground cleaning pipelines leading to robust \HI signal detection in the forthcoming MeerKAT/SKAO era.

\section*{Acknowledgements}

We thank the anonymous referee for useful suggestions that improved the readability of this manuscript.
We warmly thank Jingying Wang, Phil Bull and Keith Grainge for valuable feedback. 
MS would like to thank Tiago Castro for valuable help with the Pinocchio simulations, and Siyambonga Matshawule and Mario Santos for useful discussions. IPC thanks J\'er\^ome Bobin for feedback on the GMCA and mixGMCA implementations. LW would like to thank Clive Dickinson and Keith Grainge for useful discussion in the Challenge set-up.
MS acknowledges funding from the INAF PRIN-SKA 2017 project 1.05.01.88.04 (FORECaST) and support from the INFN INDARK PD51 grant.
IPC acknowledges support from the `Departments of Excellence 2018-2022' Grant (L.\ 232/2016) awarded by the Italian Ministry of University and Research (\textsc{mur}), from the `Ministero degli Affari Esteri della Cooperazione Internazionale - Direzione Generale per la Promozione del Sistema Paese Progetto di Grande Rilevanza ZA18GR02' and, at the early stage of this work, from the European Union through the grant LENA (ERC StG no. 678282). SC is supported by STFC grant ST/S000437/1. MI acknowledges support from the South African Radio Observatory, National Research Foundation (Grant No. 84156) and, at the early stage of this work, from the European Union through the grant LENA (ERC StG no. 678282). JF was supported by the University of Padova under the STARS Grants programme {\em CoGITO: Cosmology beyond Gaussianity, Inference, Theory, and Observations} and by the UK Science \& Technology Facilities Council (STFC) Consolidated Grant ST/P000592/1. AP is a UK Research and Innovation Future Leaders Fellow, grant MR/S016066/1, and also acknowledges support by STFC grant ST/S000437/1.

This research utilised Queen Mary's Apocrita HPC facility, supported by QMUL Research-IT \href{http://doi.org/10.5281/zenodo.438045}{http://doi.org/10.5281/zenodo.438045}. This work made use of the South African Centre for High-Performance Computing, under the project \emph{Cosmology with Radio Telescopes}, ASTRO-0945.
This research made use of \texttt{Numpy} \citep{Numpy2020}, \texttt{Astropy} \citep{Astropy}, \texttt{Scipy} \citep{Scipy2020}, \texttt{healpy} \citep{Zonca2019} and the \textit{HEALPix} \citep{healpix} package.

\smallskip
{\it Author contribution:}
All authors contributed to the design of the Blind Challenge and composing of the article.
MS led the analysis and presentation of the results, performed together with IPC. MS and IPC drafted the first version of the manuscript.  
Simulations: MS (\HI distribution), SC (MS$_{05}$ foregrounds), SH (instrumental effects), MI (PSM foregrounds). 
Participants of the Blind Challenge: IPC (PCAwls, GMCA, mixGMCA), SC (PCA(a), \fastica(a)), MI (LSQ), and JF (PCA(b), \fastica(b), poLOG).
The project was initiated and coordinated by LW and AP as co-chairs of the HI Intensity Mapping Focus Group of the SKA Cosmology SWG. 

\section*{Data Availability}

Simulated data-cubes have been produced in support of this research.
They are publicly available at the \href{https://drive.google.com/drive/folders/1WeBbKHxYtDqntiZo_i54ooPdIQOZh20M?usp=sharing}{UWC-CRC Repository}.
The code used to simulate the instrumental effects can be found on \texttt{github}: \href{https://github.com/SharperJBCA/SWGSimulator}{https://github.com/SharperJBCA/SWGSimulator}.

\bibliographystyle{mnras}
\bibliography{biblio}



\appendix

\section{Pipeline assumptions on the number of foreground components}

We present in \autoref{table:Nfg} the choices made for the various pipelines on the number of foreground sources to subtract in order to clean the different data-cubes (also shown in \autoref{fig:hist_Nfg}). 
In order to assess the most appropriate value of \Nfg to use, one can look for convergence in the power spectra of residuals while increasing the number of removed components (e.g., Figure 10 in \citet{Carucci:2020enz}). The \Nfg can also be estimated by looking at the behaviour of the eigenvalues of the frequency-frequency covariance of data (e.g., Figure 4 in \citet{Cunnington:2020njn}). In the work of \citet{Olivari2016}, an automatised choice of \Nfg is attempted, although highly dependent on prior knowledge of the level of the cosmological signal.
The values reported in \autoref{table:Nfg} do not show a strong consistency across the different methods, neither a clear trend as a function of the cases studied. Subjectivity seems to have played a major role.
Although not reported in the table, the number of foreground components is necessary and crucial for the poLOG method too, since one needs to fix the order of the polynomial that properly describes the foregrounds. \citet{2006ApJ...650..529W} explored different values and concluded the \Nfg$=4$ was sufficient for their $z>6$ simulation.
\citet{2012A&A...540A.129A} considered lower redshifts and truncated their number of components at \Nfg$=2$. 
On the other hand, \citet{Alonso:2014dhk} are more conservative as they concluded that \Nfg$=7$ are needed. 
As a compromise between these previous works, here the poLOG method has always been used with \Nfg$=6$.

\begin{table}
\caption{
The chosen values of the number of subtracted components \Nfg for the different blind cleaning algorithms, as a function of experiment (SKAO/MeerKAT), beam type (Airy/Gaussian), foreground model (PSM/MS$_{05}$) and pre-processing of the data (original data or resmoothed). We remind that the mixGMCA method has two $N_{\rm fg}$, for the large and small scales; here we report both unless the two coincide.}
\resizebox{0.47\textwidth}{!}{%
\begin{tabular}{ccccccccc} 
 \hline
 & \multicolumn{4}{c}{Original data} & \multicolumn{4}{c}{Resmoothed} \\
 \hline
\textit{Beam:} & \multicolumn{2}{c}{Gaussian} & \multicolumn{2}{c}{Airy} & \multicolumn{2}{c}{Gaussian} & \multicolumn{2}{c}{Airy}\\
 \hline
\textit{Fg model:} & MS$_{05}$ & PSM & MS$_{05}$ & PSM &  MS$_{05}$ & PSM & MS$_{05}$ & PSM\\
 \hline \hline
 & \multicolumn{8}{c}{SKAO}\\
\hline
PCA(a) & 5 & 5 & 5 & 5 & 3 & 3 & 3 & 3\\
PCA(b) & 6 & 6 & 6 & 6 & 6 & 6 &6 & 6\\
PCAwls &3 &4  &4 &5 & 3&3 &3 &10\\
\fastica(a)& 5 & 5 & 5 & 5 & 3 & 3 & 3 & 3\\
\fastica(b) & 6 & 6 & 6 & 6 & 6 & 6 &6 & 6\\
GMCA & 3 &4  &4 &5 &4 &4 &3 &4\\
mixGMCA & 3 &4  &4 &5 & 3/4 & 3/4 & 3& 10/4\\
\hline
  & \multicolumn{8}{c}{MeerKAT}\\
  \hline
PCA(a) & 5 & 5 & 5 & 8 & 3 & 3 & 3 & 6 \\
PCA(b) & 6 & 6 & 6 & 6 & 6 & 6 &6 & 6\\
PCAwls  & 3 & 4 & 4 & 4 & 3 &3  &3 &6\\
\fastica(a) & 5 & 5 & 5 & 8 & 3 & 3 & 3 & 6\\
\fastica(b) & 6 & 6 & 6 & 6 & 6 & 6 &6 & 6\\
GMCA & 3 & 4 & 4 & 4 & 4 &4 &3 &6\\
mixGMCA  & 3 & 4 & 4 & 4 & 3/4& 3/4 & 3 & 6\\
\hline
\hline 
\end{tabular}}
\label{table:Nfg}
\end{table}

\section{Performance metrics values}
For completeness, in \autoref{table:spiderSKA} and \autoref{table:spiderMeerkat} we report the values computed for the metrics described in \secref{sec:metrics}, for SKAO and MeerKAT, respectively.
In \secref{sec:performances}, for a given case study (i.e., for a given experiment, a particular beam type and post-processing choice), the performances of the nine different cleaning methods have been ranked and a {\it relative} mark between 1 and 5 has been assigned  (see the radar charts of \autoref{fig:spiderSKAO} and \autoref{fig:spiderMK}).
The values reported in \autoref{table:spiderSKA} and \autoref{table:spiderMeerkat} carry further information. For example, it is possible to see the (negative) effect of resmoothing on both the slos and $\Delta \ell_{\rm max}$ (expressed as the percentage of reconstructed $C_\ell$ values with a precision better than 30 per cent). Moreover, while for the Gaussian beam both the SKAO-MID and MeerKAT setups lead to similar results, the smaller side-lobes of the SKAO-MID dishes ease the cleaning performances.

\begin{table}
\caption{The values of the seven metrics described in \secref{sec:spider} used for ranking the various cleaning methods for the SKAO case. For simplicity, we express $\Delta \ell_{\rm max}$ as a percentage. The metric $k_{\rm min}$ is expressed in MHz$^{-1}$. The peak feature is present only when considering the Airy beam and is thus not reported for the Gaussian beam case.}
\resizebox{0.48\textwidth}{!}{\begin{tabular}{llllllll}
\hline
Method & ${\rm rms}_\ell$ &  $\sigma_{\rm rms}$ & $\Delta \ell_{\rm max}$ & ${\rm rms}_{Pk}$ & ${\rm slos}$ & $k_{\rm min}$ & ${\rm peak}$\\
 & & & (\%) & & & [MHz$^{-1}$] &  \\
\hline \hline
\multicolumn{8}{c}{Gaussian}\\
\hline
PCA(a) & 0.15 & 0.12 & 94.5 & 0.14 & 0.14 & 0.0044 & - \\
PCA(b) & 0.54 & 1.34 & 83.6 & 0.16 & 0.16 & 0.0022 & - \\
PCAwls & 0.15 & 0.13 & 93.7 & 0.03 & 0.03 & 0.0067 & - \\
FastICA(a) & 0.15 & 0.12 & 94.5 & 0.14 & 0.14 & 0.0044 & - \\
FastICA(b) & 0.15 & 0.12 & 94.3 & 0.16 & 0.16 & 0.0044 & - \\
GMCA   & 0.18 & 0.23 & 92.7 & 0.03 & 0.03 & 0.0044 & - \\
mixGMCA & 0.16 & 0.14 & 93.4 & 0.03 & 0.03 & 0.0067 & - \\
poLOG  & 0.19 & 0.32 & 92.9 & 0.04 & 0.03 & 0.0089 & - \\
LSQ    & 2.23 & 4.87 & 54.4 & 0.03 & 0.03 & 0.0067 & -\\
\hline 
\multicolumn{8}{c}{Airy}\\
\hline
PCA(a) & 0.67 & 1.14 & 82.5 & 0.16 & 0.13 & 0.0044 & 1.09  \\
PCA(b) & 0.89 & 1.74 & 76.5 & 0.18 & 0.16 & 0.0022 & 1.10  \\
PCAwls & 0.28 & 0.38 & 88.3 & 0.23 & 0.24 & 0.0044 & -0.02 \\
FastICA(a) & 0.67 & 1.14 & 82.5 & 0.16 & 0.13 & 0.0044 & 1.09  \\
FastICA(b) & 0.66 & 1.11 & 82.4 & 0.18 & 0.15 & 0.0044 & 1.08  \\
GMCA   & 0.73 & 1.24 & 83.0 & 0.15 & 0.11 & 0.0044 & 1.24  \\
mixGMCA & 0.34 & 0.42 & 84.7 & 0.15 & 0.15 & 0.0044 & 0.04  \\
poLOG  & 0.69 & 1.28 & 88.8 & 0.14 & 0.03 & 0.0089 & 1.44  \\
LSQ    & 4.15 & 8.72 & 49.5 & 0.14 & 0.03 & 0.0089 & 1.44 \\
\hline
\multicolumn{8}{c}{Gaussian \textit{Resmoothed}}\\
\hline
PCA(a) & 36.16 & 88.13 & 56.8 & 0.26 & 0.26 & 0.0044 & - \\
PCA(b) & 24.64 & 55.49 & 52.2 & 0.35 & 0.36 & 0.0044 & - \\
PCAwls & 10.72 & 34.88 & 67.8 & 0.26 & 0.26 & 0.0067 & - \\
FastICA(a) & 36.16 & 88.13 & 56.8 & 0.26 & 0.26 & 0.0044 & - \\
FastICA(b) & 19.39 & 95.85 & 64.7 & 0.34 & 0.34 & 0.0067 & - \\
GMCA   & 12.32 & 36.37 & 64.8 & 0.25 & 0.26 & 0.0067 & - \\
mixGMCA & 11.58 & 36.04 & 66.0 & 0.26 & 0.26 & 0.0067 & - \\
poLOG  & 10.22 & 33.90 & 66.6 & 0.05 & 0.001 & 0.0133 & - \\
LSQ    & 13.23 & 40.25 & 63.9 & 0.03 & 0.0004 & 0.0089 & -\\
\hline
\multicolumn{8}{c}{Airy  \textit{Resmoothed}}\\
\hline
PCA(a) & 41.74 & 94.85 & 47.3 & 0.32 & 0.26 & 0.0044 & 2.35 \\
PCA(b) & 26.23 & 79.55 & 46.8 & 0.39 & 0.35 & 0.0044 & 2.22 \\
PCAwls & 12.62 & 40.10 & 61.5 & 0.28 & 0.29 & 0.0067 & 0.20 \\
FastICA(a) & 41.74 & 94.85 & 47.3 & 0.32 & 0.26 & 0.0044 & 2.35 \\
FastICA(b) & 17.16 & 57.05 & 54.0 & 0.31 & 0.31 & 0.0067 & 0.61 \\
GMCA   & 15.56 & 42.33 & 56.0 & 0.36 & 0.29 & 0.0067 & 2.56 \\
mixGMCA & 13.59 & 41.29 & 58.2 & 0.28 & 0.29 & 0.0067 & 0.20 \\
poLOG  & 13.50 & 41.23 & 62.5 & 0.27 & 0.001 & 0.0111 & 2.96 \\
LSQ    & 28.78 & 62.10 & 46.8 & 0.27 & 0.001 & 0.0067 & 2.93\\
\hline
\hline 
\end{tabular}}
\label{table:spiderSKA}
\end{table}

\begin{table}
\caption{Same as \autoref{table:spiderSKA} but for the MeerKAT case. For simplicity, we express $\Delta \ell_{\rm max}$ as a percentage. The metric $k_{\rm min}$ is expressed in MHz$^{-1}$. The peak feature is present only when considering the Airy beam and is thus not reported for the Gaussian beam case.}
\resizebox{0.48\textwidth}{!}{\begin{tabular}{llllllll}
\hline
Method & ${\rm rms}_\ell$ &  $\sigma_{\rm rms}$ & $\Delta \ell_{\rm max}$ & ${\rm rms}_{Pk}$ & ${\rm slos}$ & $k_{\rm min}$ & ${\rm peak}$\\
 & & & (\%) & & & [MHz$^{-1}$] &  \\
\hline \hline
\multicolumn{8}{c}{Gaussian}\\
\hline
PCA(a)     & 0.14 & 0.09 & 95.0 & 0.06 & 0.06 & 0.0044 & - \\
PCA(b)     & 0.24 & 0.39 & 90.3 & 0.07 & 0.07 & 0.0044 & - \\
PCAwls     & 0.15 & 0.10 & 94.2 & 0.04 & 0.04 & 0.0044 & - \\
FastICA(a) & 0.14 & 0.09 & 95.0 & 0.06 & 0.06 & 0.0044 & - \\
FastICA(b) & 0.15 & 0.09 & 94.8 & 0.07 & 0.07 & 0.0044 & - \\
GMCA       & 0.19 & 0.24 & 93.3 & 0.05 & 0.05 & 0.0044 & - \\
mixGMCA    & 0.15 & 0.11 & 94.0 & 0.04 & 0.04 & 0.0044 & - \\
poLOG      & 0.18 & 0.22 & 93.9 & 0.05 & 0.04 & 0.0089 & - \\
LSQ        & 2.27 & 3.82 & 50.2 & 0.05 & 0.04 & 0.0111 & -\\
\hline 
\multicolumn{8}{c}{Airy}\\
\hline
PCA(a)     & 0.53 & 0.43  & 75.6 & 0.06 & 0.06 & 0.0022 & 0.10  \\
PCA(b)     & 1.86 & 3.36  & 70.2 & 0.16 & 0.04 & 0.0022 & 2.27  \\
PCAwls     & 0.34 & 0.36  & 84.4 & 0.06 & 0.05 & 0.0044 & 0.29  \\
FastICA(a) & 0.53 & 0.43  & 75.6 & 0.06 & 0.06 & 0.0022 & 0.10  \\
FastICA(b) & 0.54 & 0.45  & 75.3 & 0.03 & 0.03 & 0.0022 & 0.15  \\
GMCA       & 4.40 & 7.98  & 78.7 & 1.14 & 0.04 & 0.0044 & 17.50 \\
mixGMCA    & 0.34 & 0.36  & 83.8 & 0.05 & 0.04 & 0.0044 & 0.30  \\
poLOG      & 4.46 & 8.20  & 81.7 & 1.19 & 0.05 & 0.0067 & 18.24 \\
LSQ        & 8.47 & 12.84 & 47.0 & 1.21 & 0.05 & 0.0133 & 18.42\\
\hline
\multicolumn{8}{c}{Gaussian \textit{Resmoothed}}\\
\hline
PCA(a)     & 102.29 & 255.73 & 56.5 & 0.26 & 0.27 & 0.0022 & - \\
PCA(b)     & 54.23  & 214.19 & 53.8 & 0.35 & 0.36 & 0.0022 & - \\
PCAwls     & 20.79  & 66.59  & 67.1 & 0.26 & 0.27 & 0.0067 & - \\
FastICA(a) & 102.29 & 255.73 & 56.5 & 0.26 & 0.27 & 0.0022 & - \\
FastICA(b) & 29.72  & 134.30 & 66.4 & 0.34 & 0.34 & 0.0067 & - \\
GMCA       & 25.89  & 71.83  & 64.0 & 0.25 & 0.26 & 0.0067 & - \\
mixGMCA    & 23.29  & 69.53  & 65.3 & 0.26 & 0.27 & 0.0067 & - \\
poLOG      & 18.82  & 63.22  & 66.9 & 0.06 & 0.002 & 0.0155 & - \\
LSQ        & 25.95  & 79.45  & 64.8 & 0.03 & 0.001 & 0.0111 & -\\
\hline
\multicolumn{8}{c}{Airy  \textit{Resmoothed}}\\
\hline
PCA(a)     & 25.07 & 67.71  & 52.3 & 0.27 & 0.28 & 0.0067 & 0.26  \\
PCA(b)     & 48.93 & 223.23 & 43.7 & 0.60 & 0.32 & 0.0022 & 8.00  \\
PCAwls     & 24.87 & 65.96  & 53.0 & 0.58 & 0.13 & 0.0089 & 8.74  \\
FastICA(a) & 25.07 & 67.71  & 52.3 & 0.27 & 0.28 & 0.0067 & 0.26  \\
FastICA(b) & 25.07 & 67.71  & 52.3 & 0.27 & 0.28 & 0.0067 & 0.26  \\
GMCA       & 40.85 & 84.91  & 45.1 & 4.30 & 0.26 & 0.0067 & 65.60 \\
mixGMCA    & 28.53 & 70.24  & 49.8 & 0.58 & 0.14 & 0.0089 & 8.76  \\
poLOG      & 32.42 & 77.99  & 51.1 & 4.41 & 0.004 & 0.0133 & 67.36 \\
LSQ        & 62.52 & 131.25 & 38.1 & 4.44 & 0.003 & 0.0044 & 67.90\\
\hline
\hline 
\end{tabular}}
\label{table:spiderMeerkat}
\end{table}


\bsp	
\label{lastpage}
\end{document}